\documentclass[latin9]{aa}

\usepackage{graphicx}
\usepackage{adjustbox}
\usepackage{txfonts}
\usepackage{url}
\usepackage{relsize}
\usepackage{color}
\usepackage{amstext}
\usepackage{amsmath}

\begin{document}

\def\simlt{\mathrel{\rlap{\lower 3pt\hbox{$\sim$}}\raise 2.0pt\hbox{$<$}}}
\def\simgt{\mathrel{\rlap{\lower 3pt\hbox{$\sim$}} \raise 2.0pt\hbox{$>$}}}

\def\lsim{\,\lower2truept\hbox{${< \atop\hbox{\raise4truept\hbox{$\sim$}}}$}\,}
\def\gsim{\,\lower2truept\hbox{${> \atop\hbox{\raise4truept\hbox{$\sim$}}}$}\,}

\def\beq{\begin{equation}}
\def\eeq{\end{equation}}

      \title{Effects of observer peculiar motion on the isotropic background frequency spectrum: \\ From the monopole to higher multipoles}

    \author{T. Trombetti\inst{1,2}\fnmsep\thanks{e-mail:trombetti@ira.inaf.it}
                     \and
           C. Burigana\inst{1,3,4}
                               \and
           F. Chierici\inst{1,2,5}
           }

    \institute{INAF, Istituto di Radioastronomia, Via Piero Gobetti 101, I-40129 Bologna, Italy
    \and
    CNR, Istituto di Scienze Marine, Via Piero Gobetti 101, I-40129 Bologna, Italy
    \and
    Dipartimento di Fisica e Scienze della Terra, Universit\`a di Ferrara, Via Giuseppe Saragat 1, I-44122 Ferrara, Italy
    \and
    INFN, Sezione di Bologna, Via Irnerio 46, I-40127 Bologna, Italy
     \and
    INGV, Sezione Roma 2, Via di Vigna Murata 605, I-00143 Roma, Italy}

   \date{Received ...; accepted ...}

 \abstract
 {The observer peculiar motion produces boosting effects in the anisotropy pattern of the considered background with frequency spectral behaviours related to its frequency spectrum.}
{We study how the frequency spectrum of the background isotropic monopole emission is modified and transferred to the frequency spectra
at higher multipoles,  $\ell$. 
We performed the analysis in terms of spherical harmonic expansion 
up to a certain value of $\ell_{\rm max}$, for various models of background radiation, spanning the range between the radio and the far-infrared.} 
{We derived a system of linear equations to obtain the spherical harmonic coefficients and provide the explicit solutions up to $\ell_{\rm max} = 6$.
These are written as linear combinations of the signals at $N = \ell_{\rm max} +1$ colatitudes.
We take advantage of the symmetry property of the associated Legendre polynomials with respect to $\pi/2,$ which allows for the separation of the system into two subsystems: 
1) for $\ell=0$ and even multipoles and 2) for odd multipoles. This improves the accuracy of the solutions with respect to an arbitrary choice of the adopted colatitudes.} 
{We applied the method to different types of monopole spectra represented in terms of analytical or semi-analytical functions, that is, four types of distortions of the photon distribution function of the cosmic microwave background and four types of extragalactic background signals superimposed 
onto the cosmic microwave background's Planckian spectrum, along with several different combinations of these types.
We present our results in terms of the spherical harmonic coefficients and of the relationships between the observed and the intrinsic monopole spectra,
as well as in terms of the corresponding all-sky maps and angular power spectra.
For certain representative cases, we compare the results of the proposed method with those obtained using more computationally demanding numerical integrations or map generation and inversion.
The method is generalized to the case of an average map composed by accumulating data taken with sets of different observer velocities, as is necessary when including the effect of the observer motion relative to the Solar System barycentre.}
{The simplicity and efficiency of the proposed method can significantly alleviate the computational effort required for accurate theoretical predictions 
and for the analysis of data derived by future projects across a variety of cases of interest. 
Finally, we discuss the superposition of the cosmic microwave background intrinsic anisotropies and of the effects induced by the observer peculiar motion, exploring 
the possibility of constraining the intrinsic dipole embedded in the kinematic dipole in the presence of background spectral distortions.} 

      
      \titlerunning{Observer motion and background spectrum from monopole to higher multipoles}

   \maketitle

%-------------------------------------------------------------------

\section{Introduction}
\label{sec:intro}

The peculiar motion of an observer relative to an ideal reference frame at rest with respect to the cosmic background in a given frequency band,
produces boosting effects in the anisotropy patterns at low multipoles with frequency spectral behaviours related to the spectrum of the isotropic monopole emission. 
The largest effect is on the dipole, that is, on the anisotropy at the $\ell=1$ multipole, which is mainly attributed to the solar system
barycentre motion.
The study of the dipole anisotropy spectrum is an alternative way to the achievement of absolute measurements for extracting information about the background monopole spectrum.
This approach was originally proposed by \cite{1981A&A....94L..33D} within the framework of cosmic microwave background (CMB) spectral distortions that possibly occurred in the cosmic plasma at different epochs.
This method has been exploited by \cite{2015ApJ...810..131B} in the context of future CMB anisotropy missions and, in particular, 
by \cite{2016JCAP...03..047D}  with the aim of improving the characterization of the cosmic infrared background (CIB) spectrum.
Numerical simulations have been performed to assess the impact of instrumental performance, 
potential residuals from imperfect foreground subtraction, and relative calibration uncertainties in the reconstruction of the types of signals  described
above \citep{2018JCAP...04..021B}.
This differential approach has been investigated to be applied to the analysis of the redshifted 21cm line 
\citep{2017PhRvL.118o1301S} and of its diurnal pattern in drift-scan observations \citep{2018ApJ...866L...7D}. 
Recent predictions of the cosmic dipole from four types of imprints that are expected from (or associated with) cosmological reionization -- the diffuse free-free (FF) emission, the Comptonization distortion, the redshifted 21cm line, and the 
radio extragalactic background, along with combinations of these types -- have been presented by \cite{2019A&A...631A..61T}. 

In this work, we carry out an analysis of the effect of the observer peculiar motion on the 
frequency spectra of the monopole and of the anisotropy patterns at higher multipoles for the monopole component of various types of background radiation, ranging from the radio to the far-infrared (far-IR). 
For a blackbody spectrum, the amplitude of this effect decreases as $\beta^\ell$ at increasing $\ell$, where $\beta \ll 1$ is the module of the dimensionless peculiar velocity of the observer,
which is defined by the vector $\vec{\beta} = \vec{{\rm v}} / c$, $c$ being the speed of light.

In general, the vector $\vec{\beta}$ is the sum of an almost constant component, $\vec{\beta_{\rm C}}$, that is due to the motion of the Solar System barycentre 
with respect to the cosmic background 
and of a time varying component, $\vec{\beta_{\rm V}} = \vec{\beta_{\rm V}} (t)$,
which is due to the motion of the observer relative to the Solar System barycentre reference frame. 
For ground-based or sub-orbital experiments, $\vec{\beta_{\rm V}}$ is given by the combination of the motions 
of the Earth around the Solar System barycentre
($\beta_{\rm ES} \simeq 10^{-4}$)
and of the experimental equipment on the Earth's surface
($\beta_{\rm EE} \simeq 1.7 \times 10^{-6}$ for an experiment located at the Earth equator).
For past, planned, and proposed CMB space missions, $\vec{\beta_{\rm V}}$ is given by the combination of the motion of the
Earth or of the second Lagrangian point of the Sun-Earth system (L2) around the Solar System barycentre ($\beta_{\rm L2} \simeq 1.01\, \beta_{\rm ES}$)
and of the motion of spacecraft around the Earth or around L2, according to the adopted trajectory. For example, 
the typical spacecraft velocity around L2 is $\beta_{\rm L} \simeq 1.7 \times 10^{-6}$ for a Lissajous 'orbit' with a 'radius' of $\simeq 3 \times 10^{5}$\,km described over a timescale of around six months.
We note that the Solar System motion around the Galactic centre and the presence of local Universe gravitational fields imply 
time variations of 
$\vec{\beta}$, 
but they 
can be neglected
within the typical duration, $\tau_{\rm S}$, of a survey. For example, in the approximation of a uniform circular motion of the Solar System around the Galactic centre, with
a rotation period $P = 2\pi / \omega \simeq 2.4 \times 10^8$\,yr and a velocity ${\rm v}_{\rm SG} \simeq 220$\,km/s, that is, ${\beta}_{\rm SG} \simeq 7.3 \times 10^{-4}$,
the relative variation of the velocity at a timescale of $\tau_{\rm S} \simeq 5$\,yr is
$\Delta {\rm v}_{\rm SG} / {\rm v}_{\rm SG} = \Delta {\beta}_{\rm SG} / {\beta}_{\rm SG} \simeq \omega^2 \cdot ({\rm v}_{\rm SG} / \omega) \cdot \tau_{\rm S} / \,{\rm v}_{\rm SG} = \omega \, \tau_{\rm S} \simeq 1.3 \times 10^{-7}$.

For numerical estimates, we 
will typically assume
that the CMB dipole is due to velocity effects only and we neglect the modulation of $\vec{\beta}$ introduced by the contribution of $\vec{\beta_{\rm V}}$. 
After the correction for the spacecraft motion around the Solar System barycentre,  
the nominal CMB dipole amplitude according to the {\it Planck} 2015 results 
\citep{2016A&A...594A...1P,2016A&A...594A...5P,2016A&A...594A...8P} is $A_{\rm dip} = (3.3645 \pm 0.002)$ mK.  
When using the Low Frequency Instrument alone
\citep{2020A&A...641A...2P}, 
the most recent analysis of the {\it Planck} 2018 results gives 
an almost identical value of $A_{\rm dip}$ as in the 2015 release.
On the other hand, when including, again, the High Frequency Instrument,
the analysis of the {\it Planck} 2018 results gives $A_{\rm dip} = (3.36208 \pm 0.00099)$ mK \citep{2020A&A...641A...3P,2020A&A...641A...1P},
which is a slightly lower value. These results are clearly compatible within the errors. 
Based on a joint analysis \citep{2009ApJ...707..916F} of the data from the Far Infrared Absolute Spectrophotometer (FIRAS)
on board the Cosmic Background Explorer (COBE) and from the Wilkinson Microwave Anisotropy Probe (WMAP),
we adopt $T_0 = (2.72548 \pm 0.00057)$\,K for the current CMB effective temperature in the blackbody spectrum approximation,
where $aT_0^4$ gives the current CMB energy density with $a = 8\pi I_3 k^4 / (hc)^3$, 
$I_3 = \pi^4/15$, $k$ and $h$ the Boltzmann and Planck constants.
We use the velocity ${\rm v}_{\rm C} = (369.82\pm0.11)$\,km/s given in Table 3 of \cite{2020A&A...641A...1P},
that is, $\beta = \beta_{\rm C} = {\rm v}_{\rm C} / c \simeq 1.2336 \times 10^{-3} \simeq A_{\rm dip} / T_0$,
to characterise 
the velocity of the Solar System barycentre
with respect to the cosmic background. 

The main contribution to the modulation of $\vec{\beta}$ coming from $\vec{\beta_{\rm V}}(t)$,
which produces an effect amounting to $\simeq 8.1$\,\% of the global signal,
is derived from the component of the observer motion 
due to the revolution of the Earth or of L2 around the Solar System barycentre (see above estimates).

We study the effect of peculiar motion in terms of spherical harmonic expansion up to a certain value of $\ell_{\rm max}$. In this way, we introduce a relative error in the prediction of the effect 
at a given $\ell$
that strongly decreases with $\ell_{\rm max}$. 
Neglecting the contributions from higher orders, the dipole anisotropy spectrum was estimated as the difference between the signal measured in the direction of motion and in its perpendicular direction
\citep{1981A&A....94L..33D}, namely, in terms of a very simple linear combination of the signals in two specific directions.
In this work, we show how this concept can be generalized to derive the frequency behaviour of the anisotropy pattern up to higher multipoles. We provide both a recipe and explicit solutions that can be directly used
for accurate and swift theoretical predictions of the individual multipole patterns and of the global pattern, allowing us to bypass the need for more computationally demanding approaches that are based on
delicate numerical integrations or on map generation and inversion. 

In Sect. \ref{sec:theoframe}, we introduce the adopted formalism and 
the set of equations we aimed to solve.  
We provide the explicit equations and solutions up to $\ell_{\rm max} = 6$ in Sect. \ref{sec:sol7} and $\ell_{\rm max} = 4$ in Sect. \ref{sec:sol5} in order to point out some general properties of the solutions. 
In Sects. \ref{sec:sol7BB} and \ref{sec:sol5BB}, we work out these solutions for the particular case of a blackbody spectrum to make clear their simple link with the contributions coming from the various orders of $\beta$.
In some cases, we show their equivalence with the 
corresponding explicit exact solution at the order of $\beta$ corresponding to $\ell_{\rm max}$.
In Sects. \ref{sec:dip_3and2col} and \ref{sec:dip_2col}, we discuss the solutions up to low $\ell_{\rm max}$, that is, $\ell_{\rm max} = 1$ and 2, derived using just two or three colatitudes. 
Some remarks linking the general properties of the found solutions at various $\ell_{\rm max}$
with the monopole spectrum integration and differentiation are given in Sect. \ref{sec:highl_derivs}. 
The main applications and results of the proposed method are presented in Sect. \ref{sec:monmod} for eight specific types of background: we give concise presentations of the monopole spectrum models adopted in this study, 
we describe the main features of the found solutions and, for two very different cases, we compare them with the results based on a numerical integration. 
Some applications to combinations of signals are discussed in Sect. \ref{sec:combi}.
In Sect. \ref{sec:compa}, we briefly present a set of results related to all-sky maps and angular power spectra, also for the purpose of comparison with previous analyses based on map generation and inversion. 
In Sect. \ref{sec:varybeta}, we discuss how the developed method can be directly generalized 
from the case of maps obtained with a constant observer velocity to the case of maps 
derived from the average of data taken with a set of different observer velocities,
as, for example, in the case of $\vec{\beta}$ modulated by $\vec{\beta_{\rm V}}(t)$.
In Sect. \ref{sec:all}, we focus on the global pattern at microwave frequencies,
discussing the superposition of the CMB intrinsic anisotropies and of the effects induced by the observer peculiar motion.
The possibility of constraining the intrinsic dipole embedded in the kinematic dipole in the presence of CMB spectral distortions
is then discussed in Sect. \ref{sec:dip_all}.
Finally, in Sect. \ref{sec:conclu}, we draw our main conclusions.
Some technical aspects are provided in the three sections of the appendix.
\section{Theoretical framework and formalism}
\label{sec:theoframe}

The peculiar velocity effect on the frequency spectrum can be evaluated on the whole sky
using the complete description of the Compton-Getting effect \citep{1970PSS1825F}. This is based
on the Lorentz invariance of the photon distribution function.
In this work, we are interested in the effects induced on the monopole (or global) signal that, by definition, is isotropic in an ideal reference frame at rest with respect to the CMB or,   
more generally, 
to the cosmic background under consideration. In principle, the CMB and the other cosmic backgrounds provide information on 
processes that possibly occurred at different epochs or that are differently weighted for a range of redshift shells. Thus, the above ideal reference frame should correctly refer to the corresponding cosmic phase. 

At a given $\nu$, the photon distribution function, $\eta^{\rm BB/dist}$,
for the considered type of spectrum needs to be computed with the frequency multiplied by the product $(1 - \hat{n} \cdot \vec{\beta})/(1 - {\beta}^2)^{1/2}$.
The notation `BB/dist' stands for a blackbody spectrum or for any type of non-blackbody signal
(or for combinations of signals).
This accounts for all the possible sky directions, which are defined by the unit vector ${\hat{n}}$,
relative
to the peculiar velocity of the observer,
which is defined by the vector $\vec{\beta}$ in the reference frame at rest with respect to the considered cosmic background. This includes all the
orders in $\vec{\beta}$ and the link with the geometrical properties induced at each multipole. 
We study the effect in terms of equivalent thermodynamic temperature, $T_{\rm th} (\nu)$, defined as the temperature of the blackbody having the same $\eta(\nu)$ at the frequency $\nu$,
\begin{equation}
      T_{\rm th} (\nu) ={h\nu\over k} {1 \over \ln(1+1/\eta(\nu))} \, .
    \label{eq:t_therm}
\end{equation}

\noindent
The observed signal map is then given by \citep{2018JCAP...04..021B}
\begin{equation}\label{eq:eta_boost}
T_{\rm th}^{\rm BB/dist} (\nu, {\hat{n}}, \vec{\beta}) =
\frac{xT_{0}} {{\rm{ln}}(1 + 1 / (\eta(\nu, {\hat{n}}, \vec{\beta}))^{\rm BB/dist}) } \, ,
\end{equation}
\noindent where $\eta(\nu, {\hat{n}}, \vec{\beta}) = \eta(\nu')$ with
$\nu' = \nu (1 - {\hat{n}} \cdot \vec{\beta})/(1 - \beta^2)^{1/2}$,
$x=h\nu/(kT_r)$ and $T_r=T_0(1+z)$ are the redshift invariant dimensionless frequency and the redshift-dependent effective temperature of the CMB.

 The unit vector ${\hat{n}}$ is associated to the polar coordinates $\theta$ (colatitude) and $\phi$ (longitude). The function 
$T_{\rm th}^{\rm BB/dist} (\nu, {\hat{n}}, \vec{\beta}) = T_{\rm th}^{\rm BB/dist} (\nu, \theta, \phi, \vec{\beta})$ can be expanded in spherical harmonics. 
We adopt a reference system with the $z$ axis parallel to the observer velocity and we can then simply replace $\vec{\beta}$ with $\beta$ in the above dependencies. Thus,

\begin{equation}\label{eq:harm}
T_{\rm th}^{\rm BB/dist} (\nu, \theta, \phi, \beta) =
\sum_{\ell=0}^{\ell_{\rm max}} \sum_{m=-\ell}^{\ell} a_{\ell,m} (\nu, \beta) Y_{\ell,m}(\theta, \phi) \, ,
\end{equation}

\noindent
where $Y_{\ell,m}(\theta, \phi)$ are the spherical harmonics related to the associated Legendre polynomials, ${P}_\ell^m ({\rm cos} \, \theta)$, and the coefficients
$a_{\ell,m} (\nu,\beta)$ contain information on the background spectrum and the observer velocity.

In the adopted reference system, the isotropy of the background monopole, or, equivalently, of $\eta$, implies that $T_{\rm th}^{\rm BB/dist}$ depends on 
$\theta$ but not on $\phi$. Thus, in Eq. \eqref{eq:harm}, we can take only the terms with $m=0$ and in this case, $Y_{\ell,m}(\theta, \phi) = {\tilde P}_\ell^m ({\rm cos} \, \theta)$, where ${\tilde P}_\ell^m ({\rm cos} \, \theta)$ are
the renormalized associated Legendre polynomials: 

\begin{equation}\label{eq:legendre}
{\tilde P}_\ell^m ({\rm cos} \theta) = \sqrt{\frac{2\ell+1}{4\pi} \frac{(\ell -m)!}{(\ell+m)!}} {P}_\ell^m ({\rm cos} \, \theta) \, .
\end{equation}

In general, for a real function, the coefficients of the spherical harmonics expansion with $m>0$ are related to the coefficients with $m<0$ by the relation
$a_{\ell,m}^* = (-1)^m \, a_{\ell,-m}$, where the index $*$ denotes the complex conjugation. 
We note that for this problem and with the adopted reference system
with the $z$ axis parallel (or antiparallel) to the observer velocity,
we are interested only in the non-vanishing coefficients with $m=0$, but, in general, we can also see that the coefficients $a_{\ell,m}$ with $m \ne 0$ do not vanish. The publicly available tools allow us to efficiently 
compute the $a_{\ell,m}$ passing from a reference system to another (see \cite{2005ApJ...622..759G}).

Formally, the coefficients $a_{\ell,m} (\nu,\beta)$ can be computed through an inversion of Eq. \eqref{eq:harm}:

\begin{align}\label{eq:harminv}
& a_{\ell,m} (\nu, \beta) = \int_{\theta=0}^\pi \int_{\phi=0}^{2\pi} T_{\rm th}^{\rm BB/dist} (\nu, \theta, \phi, \beta) e^{-im\phi} {\tilde P}_\ell^m ({\rm cos} \, \theta) \, {\rm sin} \, \theta \, d\theta \, d\phi \, \nonumber
\\ & = \int_{\theta=0}^\pi \int_{\phi=0}^{2\pi} \left[{ T_{\rm th}^{\rm BB/dist} (\nu, \theta, \phi, \beta) - T_{\rm th}^{\rm BB/dist,rest} (\nu) }\right] 
\\ & \cdot e^{-im\phi} {\tilde P}_\ell^m ({\rm cos} \, \theta) \, {\rm sin} \, \theta \, d\theta \, d\phi 
+ a_{\ell,m}^{\rm rest} (\nu) \, , \nonumber
\end{align}

\noindent
where $T_{\rm th}^{\rm BB/dist}$ is evaluated through Eq. \eqref{eq:eta_boost} and
$m=0$. In the last part of Eq. \eqref{eq:harminv},
$a_{\ell,m}^{\rm rest} (\nu) = \sqrt{4\pi}  T_{\rm th}^{\rm BB/dist,rest} (\nu)$ and $T_{\rm th}^{\rm BB/dist,rest} (\nu)$ are 
evaluated in the background rest frame, that is, these are
the intrinsic spherical harmonics expansion coefficients and the intrinsic (isotropic) background monopole expressed in equivalent thermodynamic temperature.
For this problem, $a_{\ell,m}^{\rm rest} (\nu)$ does not vanish only for $\ell = 0$. 
This form in Eq. \eqref{eq:harminv} is useful in the numerical computation (see also Sect. \ref{sec:NE}) because the integrand function becomes
the difference between the equivalent thermodynamic temperatures in the reference frames in motion and at rest with respect to the background. 
For a general background spectrum, this approach requires a delicate
and computationally demanding integration over $\theta$. For a small $\beta$, it could be difficult to achieve the extreme precision needed to characterize the fine and small details of spectral features.
We can instead consider Eq. \eqref{eq:harm} with $m=0$ for a set of $N$ directions, namely, of colatitudes $\theta_i$ with $i=0,N-1$, 
to construct a linear system of $N$ equations in the $N$ unknowns $a_{\ell,0} (\nu,\beta)$, with $\ell=0,N-1$, that can be solved given the corresponding $N$ values of $T_{\rm th}^{\rm BB/dist} (\nu, \theta_i, \phi, \beta)$, provided that the determinant of the system coefficient matrix does not vanish.
The solutions for the unknowns $a_{\ell,0} (\nu,\beta)$ can be then written as linear combinations of $N$ signals, $T_{\rm th}^{\rm BB/dist} (\nu, \theta_i, \phi, \beta)$,
that are evaluated for a given background monopole at $N$ colatitudes. 

With this simple scheme, we can fully characterize the observed signal map $T_{\rm th}^{\rm BB/dist} (\nu, {\hat{n}}, \vec{\beta})$ up to the desired multipole component $\ell_{\rm max} = N - 1$.
Let us assume, as a rule of thumb, that the amplitude of this effect decreases at increasing multipole as 
$\beta^{\ell \cdot p}$, with $p \approx 1$. The value appropriate to the case of a blackbody spectrum is $p=1$, as mentioned in Sect. \ref{sec:intro}, while, in general, the effective scaling with $\ell$ is frequency-dependent 
and related to the monopole spectrum shape, as discussed in next sections.   
Considering a spherical harmonic expansion up to $\ell_{\rm max}$ leads to neglect the contributions from $\ell > \ell_{\rm max}$.
Thus, given the above scaling, the relative error in the computation of the effect at a given $\ell \le \ell_{\rm max}$ is, at most, on the order of $\beta^{(\ell_{\rm max}-\ell+j) \cdot p}$.
For a generic choice of the $N$ colatitudes, we simply have $j = 1$.
Since $\beta$ is on the order of $10^{-3}$, adopting $\ell_{\rm max} = 6,$ we expect to achieve an extremely high numerical accuracy that is sufficient for any application even in the very distant future; whereas, 
setting $\ell_{\rm max} = 4$ can be adequate for predicting the corresponding multipole patterns as part of the analysis 
in forthcoming and planned (or proposed) surveys 
since no relevant error is introduced by neglecting the contributions at higher multipoles.
In general, an accuracy up to any desired order can be then achieved with this approach by just computing $T_{\rm th}^{\rm BB/dist}$  in only a relatively small number of sky directions, $N =  \ell_{\rm max} + 1$.

We note that $Y_{0,0} = \sqrt{1/(4\pi)}$ and that for $\ell=0,  $ and for even  $\ell,$ the associated Legendre polynomials ${P}_\ell^0 ({\rm cos} \theta)$ are symmetric with respect to $\theta = \pi / 2$, whereas for odd $\ell,$ they vanish at $\theta = \pi / 2$ and are antisymmetric with respect to $\theta = \pi / 2$.
This suggests that the linear system of $N$ equations using $\theta = \pi / 2$ and pairs of colatitudes symmetric with respect to $\theta = \pi / 2$ are expected 
satisfy the following properties:
(i) for $\theta = \pi / 2$, all the coefficients multiplying $a_{\ell,0}$ are null for odd $\ell$; 
(ii) for each pair of colatitudes -- if $\ell$ is even, 
the coefficient multiplying $a_{\ell,0}$ is the same, whereas, it is the opposite result for odd $\ell$.

As made evident in the next sections, these properties
can be used to significantly simplify the explicit solution of the system because they allow us to combine the equations into two separate subsystems: 1) $(N-1)/2+1$ equations for 
$a_{\ell,0}$, with $\ell = 0$ and even $\ell $  and 2)  $(N-1)/2$ equations for $a_{\ell,0}$ with odd $\ell$. For even $\ell_{\rm max}$, a choice of odd $N = \ell_{\rm max} + 1$ colatitudes $\theta_i$ 
that satisfies the above symmetry implies $j=2$ (instead of 1) for even $\ell$ in the scaling, $\beta^{(\ell_{\rm max}-\ell+j) \cdot p}$,
of the relative error of the method (see also the discussion at the end of Appendix \ref{app:beta_amplified}).
For odd $\ell_{\rm max}$, the system can be built with $N = \ell_{\rm max} + 1$ colatitudes $\theta_i$ as above, but avoiding the inclusion of $\pi / 2$. 
The system can be split into two separate subsystems of $N/2$ equations:\ 1) for  
$\ell = 0$ and even $\ell$ and 2) for  
odd $\ell$; and in this case, $j=2$ for odd $\ell$ (see also Sect. \ref{sec:dip_3and2col} and the discussion at the end of Sect. \ref{sec:dip_2col}).
This property allows us to achieve a significant improvement in accuracy with respect to a generic choice of the $N$ colatitudes $\theta_i$.

\section{Explicit solutions up to $\ell_{\rm max} = 6$}
\label{sec:sol7}

Explicitly expanding $T_{\rm th}^{\rm BB/dist}$ in spherical harmonics up to $\ell_{\rm max} = 6,$ we get

\begin{align}
\label{eq:harm6}
& T_{\rm th}^{\rm BB/dist} = a_{0,0} \sqrt{\frac{1}{4\pi}} \nonumber
\\ & + a_{1,0} \sqrt{\frac{3}{4\pi}} {\rm cos} \, \theta \nonumber
\\ & + a_{2,0} \sqrt{\frac{5}{4\pi}} \left({ \frac{3}{2} {\rm cos}^2 \, \theta - \frac{1}{2} }\right) \nonumber
\\ & + a_{3,0} \sqrt{\frac{7}{4\pi}} \left({ \frac{5}{2} {\rm cos}^3 \, \theta - \frac{3}{2} {\rm cos} \, \theta }\right) 
\\ & + a_{4,0} \sqrt{\frac{9}{4\pi}} \left({ \frac{35}{8} {\rm cos}^4 \, \theta - \frac{15}{4} {\rm cos}^2 \, \theta + \frac{3}{8} }\right) \nonumber
\\ & + a_{5,0} \sqrt{\frac{11}{4\pi}} \left({ \frac{63}{8} {\rm cos}^5 \, \theta - \frac{35}{4} {\rm cos}^3 \, \theta + \frac{15}{8} {\rm cos} \, \theta }\right) \nonumber 
\\ & + a_{6,0} \sqrt{\frac{13}{4\pi}} \left({ \frac{231}{16} {\rm cos}^6 \, \theta - \frac{315}{16} {\rm cos}^4 \, \theta + \frac{105}{16} {\rm cos}^2 \, \theta - \frac{5}{16} }\right) \, , \nonumber
\end{align}

\noindent
where we omit, for simplicity, the dependence of $T_{\rm th}^{\rm BB/dist}$ on $\nu$, $\theta,$ and $\beta$ and the dependencies of $a_{\ell,0}$ on $\nu$ and $\beta$.

To write the linear system of seven equations, we are able to choose among infinite possibilities, and the explicit form
of the system (but not the solution up to the adopted maximum multipole) depends on the specific choice.
Among the possible choices satisfying the symmetry properties described above, we selected a 
set of colatitudes $\theta_i$ such that the values of ${\rm cos} \, \theta_i$ are rational numbers or just involve $\sqrt{2}$ in order to simplify the algebra:
$\theta_i = 0, \pi/4, \pi/3, \pi/2, (2/3)\pi, (3/4)\pi,$ and $\pi$. 

After a series of calculations, we derived the corresponding linear system.
We obtain

\begin{align}\label{eq:7rings_0}
T_{\rm th}^{\rm BB/dist} & (\theta=0) = \sqrt{\frac{1}{4\pi}} a_{0,0} + \sqrt{\frac{3}{4\pi}} a_{1,0}   
\\ & + \sqrt{\frac{5}{4\pi}} a_{2,0}  + \sqrt{\frac{7}{4\pi}} a_{3,0} + \sqrt{\frac{9}{4\pi}} a_{4,0} \nonumber
\\ & + \sqrt{\frac{11}{4\pi}} a_{5,0} + \sqrt{\frac{13}{4\pi}} a_{6,0} \, , \nonumber
\end{align}

\begin{align}\label{eq:7rings_1}
T_{\rm th}^{\rm BB/dist} & (\theta=\pi/4) = \sqrt{\frac{1}{4\pi}} a_{0,0} + \frac{\sqrt{2}}{2} \sqrt{\frac{3}{4\pi}} a_{1,0} 
\\ & + \frac{1}{4} \sqrt{\frac{5}{4\pi}} a_{2,0}  - \frac{\sqrt{2}}{8} \sqrt{\frac{7}{4\pi}} a_{3,0} - \frac{13}{32} \sqrt{\frac{9}{4\pi}} a_{4,0} \nonumber
\\ & - \frac{17\sqrt{2}}{64} \sqrt{\frac{11}{4\pi}} a_{5,0} -\frac{19}{128} \sqrt{\frac{13}{4\pi}} a_{6,0} \, , \nonumber
\end{align}

\begin{align}\label{eq:7rings_2}
T_{\rm th}^{\rm BB/dist} & (\theta=\pi/3) = \sqrt{\frac{1}{4\pi}} a_{0,0} + \frac{1}{2} \sqrt{\frac{3}{4\pi}} a_{1,0} 
\\ & - \frac{1}{8} \sqrt{\frac{5}{4\pi}} a_{2,0}  - \frac{7}{16} \sqrt{\frac{7}{4\pi}} a_{3,0} - \frac{37}{128} \sqrt{\frac{9}{4\pi}} a_{4,0} \nonumber
\\ & + \frac{23}{256} \sqrt{\frac{11}{4\pi}} a_{5,0} + \frac{331}{1024} \sqrt{\frac{13}{4\pi}} a_{6,0} \, , \nonumber
\end{align}

\begin{align}\label{eq:7rings_3}
T_{\rm th}^{\rm BB/dist} & (\theta=\pi/2) = \sqrt{\frac{1}{4\pi}} a_{0,0} + 0 \cdot a_{1,0}   
\\ & - \frac{1}{2} \sqrt{\frac{5}{4\pi}} a_{2,0}  + 0 \cdot a_{3,0} + \frac{3}{8} \sqrt{\frac{9}{4\pi}} a_{4,0} \, , \nonumber
\\ & + 0 \cdot a_{5,0} -\frac{5}{16} \sqrt{\frac{13}{4\pi}} a_{6,0} \nonumber
\end{align}

\begin{align}\label{eq:7rings_4}
T_{\rm th}^{\rm BB/dist} & (\theta=(2/3)\pi) = \sqrt{\frac{1}{4\pi}} a_{0,0} - \frac{1}{2} \sqrt{\frac{3}{4\pi}} a_{1,0} 
\\ & - \frac{1}{8} \sqrt{\frac{5}{4\pi}} a_{2,0}  + \frac{7}{16} \sqrt{\frac{7}{4\pi}} a_{3,0} - \frac{37}{128} \sqrt{\frac{9}{4\pi}} a_{4,0} \nonumber
\\ & - \frac{23}{256} \sqrt{\frac{11}{4\pi}} a_{5,0} + \frac{331}{1024} \sqrt{\frac{13}{4\pi}} a_{6,0} \nonumber
\end{align}

\begin{align}\label{eq:7rings_5}
T_{\rm th}^{\rm BB/dist} & (\theta=(3/4)\pi) = \sqrt{\frac{1}{4\pi}} a_{0,0} - \frac{\sqrt{2}}{2} \sqrt{\frac{3}{4\pi}} a_{1,0} 
\\ & + \frac{1}{4} \sqrt{\frac{5}{4\pi}} a_{2,0}  + \frac{\sqrt{2}}{8} \sqrt{\frac{7}{4\pi}} a_{3,0} - \frac{13}{32} \sqrt{\frac{9}{4\pi}} a_{4,0} \nonumber
\\ & + \frac{17\sqrt{2}}{64} \sqrt{\frac{11}{4\pi}} a_{5,0} -\frac{19}{128} \sqrt{\frac{13}{4\pi}} a_{6,0} \, , \nonumber
\end{align}

\begin{align}\label{eq:7rings_6}
T_{\rm th}^{\rm BB/dist} & (\theta=\pi) = \sqrt{\frac{1}{4\pi}} a_{0,0} - \sqrt{\frac{3}{4\pi}} a_{1,0}  
\\ & + \sqrt{\frac{5}{4\pi}} a_{2,0} - \sqrt{\frac{7}{4\pi}} a_{3,0} + \sqrt{\frac{9}{4\pi}} a_{4,0} \nonumber
\\ & - \sqrt{\frac{11}{4\pi}} a_{5,0} + \sqrt{\frac{13}{4\pi}} a_{6,0} \, . \nonumber 
\end{align}

Equations \eqref{eq:7rings_0}--\eqref{eq:7rings_6} constitute the linear system that is to be solved; and the 
determinant of the coefficients of the system matrix, $\simeq -0.303$, does not vanish.
We can solve the system using the methods of elimination and substitution.

As anticipated, we can combine the above equations to split the system into two subsystems.
By adding the left and right sides of Eq. \eqref{eq:7rings_0} and Eq. \eqref{eq:7rings_6}, of Eq. \eqref{eq:7rings_1} and  Eq. \eqref{eq:7rings_5}, and of Eq. \eqref{eq:7rings_2} and Eq. \eqref{eq:7rings_4}, 
we get three equations which, complemented using Eq. \eqref{eq:7rings_3}, 
form a linear system involving only the four unknowns $a_{\ell,0}$ with $\ell=0$ and even $\ell$. We can solve it by substitution. Equation \eqref{eq:7rings_3} allows us to express $2 \sqrt{1/(4\pi)} a_{0,0}$
as a combination of $T_{\rm th}^{\rm BB/dist} (\theta=\pi/2)$, $a_{2,0}$, $a_{4,0}$, 
and
$a_{6,0}$ to be put in the other three equations. From the first one, we then express $a_{2,0}$ as a combination 
of $T_{\rm th}^{\rm BB/dist} (\theta=0)$, $T_{\rm th}^{\rm BB/dist} (\theta=\pi/2)$,
$T_{\rm th}^{\rm BB/dist} (\theta=\pi)$, $a_{4,0}$, 
and
$a_{6,0}$ to be put in the two other remaining equations. 
We then represent $a_{4,0}$ as a combination of $T_{\rm th}^{\rm BB/dist} (\theta=0)$, $T_{\rm th}^{\rm BB/dist} (\theta=\pi/4)$, 
$T_{\rm th}^{\rm BB/dist} (\theta=\pi/2)$, $T_{\rm th}^{\rm BB/dist} (\theta=(3/4)\pi)$, $T_{\rm th}^{\rm BB/dist} (\theta=\pi)$,
and
$a_{6,0}$,
which allows us to derive
first the solution for $a_{6,0}$:

\begin{align}
\label{eq:sol7rings_a6}
a_{6,0} & = \frac{64}{693} \sqrt{\frac{4\pi}{13}} 
\Bigl[  \left({T_{\rm th}^{\rm BB/dist} (\theta=0) + T_{\rm th}^{\rm BB/dist} (\theta=\pi) }\right) 
\\ &  - 6 \left({T_{\rm th}^{\rm BB/dist} (\theta=\pi/4) + T_{\rm th}^{\rm BB/dist} (\theta=(3/4)\pi) }\right)  \nonumber
\\ &  + 8 \left({T_{\rm th}^{\rm BB/dist} (\theta=\pi/3) + T_{\rm th}^{\rm BB/dist} (\theta=(2/3)\pi) }\right)  \nonumber
\\ &  - 6 T_{\rm th}^{\rm BB/dist} (\theta=\pi/2)  \Bigr] \,. \nonumber
\end{align}

\noindent 
As suggested in the introduction, 
$a_{6,0}$ is written in terms of a linear combination of the set of values of $T_{\rm th}^{\rm BB/dist}$ computed for the seven adopted colatitudes.
With substitution, we subsequently derive the solution for $a_{4,0}$, $a_{2,0}$ and, finally, for $a_{0,0}$:

\begin{align}\label{eq:sol7rings_a4}
a_{4,0} & = \frac{8}{385} \sqrt{\frac{4\pi}{9}} 
\Bigl[ { 9 \left({T_{\rm th}^{\rm BB/dist} (\theta=0) + T_{\rm th}^{\rm BB/dist} (\theta=\pi) }\right) } 
\\ & { - 10 \left({T_{\rm th}^{\rm BB/dist} (\theta=\pi/4) + T_{\rm th}^{\rm BB/dist} (\theta=(3/4)\pi) }\right) } \nonumber
\\ & { - 16 \left({T_{\rm th}^{\rm BB/dist} (\theta=\pi/3) + T_{\rm th}^{\rm BB/dist} (\theta=(2/3)\pi) }\right) } \nonumber
\\ & { + 34T_{\rm th}^{\rm BB/dist} (\theta=\pi/2) } \Bigr]  \, , \nonumber 
\end{align}

\begin{align}\label{eq:sol7rings_a2}
a_{2,0} & = \frac{1}{693} \sqrt{\frac{4\pi}{5}} 
\Bigl[ { 121 \left({T_{\rm th}^{\rm BB/dist} (\theta=0) + T_{\rm th}^{\rm BB/dist} (\theta=\pi) }\right) } 
\\ & { + 396 \left({T_{\rm th}^{\rm BB/dist} (\theta=\pi/4) + T_{\rm th}^{\rm BB/dist} (\theta=(3/4)\pi) }\right) } \nonumber
\\ & { - 352 \left({T_{\rm th}^{\rm BB/dist} (\theta=\pi/3) + T_{\rm th}^{\rm BB/dist} (\theta=(2/3)\pi) }\right) } \nonumber
\\ & { - 330 T_{\rm th}^{\rm BB/dist} (\theta=\pi/2) } \Bigr]  \, , \nonumber 
\end{align}

\begin{align}\label{eq:sol7rings_a0}
a_{0,0} & = \frac{\sqrt{4\pi}}{630} 
\Bigl[ { 29 \left({T_{\rm th}^{\rm BB/dist} (\theta=0) + T_{\rm th}^{\rm BB/dist} (\theta=\pi) }\right) } 
\\ & { + 120 \left({T_{\rm th}^{\rm BB/dist} (\theta=\pi/4) + T_{\rm th}^{\rm BB/dist} (\theta=(3/4)\pi) }\right) } \nonumber
\\ & { + 64 \left({T_{\rm th}^{\rm BB/dist} (\theta=\pi/3) + T_{\rm th}^{\rm BB/dist} (\theta=(2/3)\pi) }\right) } \nonumber
\\ & { + 204 T_{\rm th}^{\rm BB/dist} (\theta=\pi/2) } \Bigr] \, . \nonumber
\end{align}

Subtracting left and right sides of Eq. \eqref{eq:7rings_0} and Eq. \eqref{eq:7rings_6}, of Eq. \eqref{eq:7rings_1} and  Eq. \eqref{eq:7rings_5}, and of Eq. \eqref{eq:7rings_2} and Eq. \eqref{eq:7rings_4} 
we get three equations that form a linear system involving only the three unknowns $a_{\ell,0}$ with odd $\ell$. From the difference between the first of these equations and the second equation multiplied by $\sqrt{2}$ 
and the difference between the first equation and the third equation multiplied by $2$, we can write a system for $a_{3,0}$ and $a_{5,0}$. We first derive $a_{5,0}$

\begin{align}\label{eq:sol7rings_a5}
a_{5,0} & = \frac{32}{189} \sqrt{\frac{4\pi}{11}}
\Bigl[ { \left({T_{\rm th}^{\rm BB/dist} (\theta=0) - T_{\rm th}^{\rm BB/dist} (\theta=\pi) }\right) } 
\\ & { - 3\sqrt{2} \left({T_{\rm th}^{\rm BB/dist} (\theta=\pi/4) - T_{\rm th}^{\rm BB/dist} (\theta=(3/4)\pi) }\right) } \nonumber
\\ & { + 4 \left({T_{\rm th}^{\rm BB/dist} (\theta=\pi/3) - T_{\rm th}^{\rm BB/dist} (\theta=(2/3)\pi) }\right) } \Bigr] \nonumber
\end{align}

\noindent
and then, by substitution, $a_{3,0}$ and $a_{1,0}$

\begin{align}\label{eq:sol7rings_a3}
a_{3,0} & = \frac{2}{135} \sqrt{\frac{4\pi}{7}} 
\Bigl[ { 13 \left({T_{\rm th}^{\rm BB/dist} (\theta=0) - T_{\rm th}^{\rm BB/dist} (\theta=\pi) }\right) } 
\\ & { + 15\sqrt{2} \left({T_{\rm th}^{\rm BB/dist} (\theta=\pi/4) - T_{\rm th}^{\rm BB/dist} (\theta=(3/4)\pi) }\right) } \nonumber
\\ & { - 56 \left({T_{\rm th}^{\rm BB/dist} (\theta=\pi/3) - T_{\rm th}^{\rm BB/dist} (\theta=(2/3)\pi) }\right) } \Bigr]  \, , \nonumber
\end{align}

\begin{align}\label{eq:sol7rings_a1}
a_{1,0} & = \frac{1}{210} \sqrt{\frac{4\pi}{3}} 
\Bigl[ { 29 \left({T_{\rm th}^{\rm BB/dist} (\theta=0) - T_{\rm th}^{\rm BB/dist} (\theta=\pi) }\right) }
\\ & { + 60\sqrt{2} \left({T_{\rm th}^{\rm BB/dist} (\theta=\pi/4) - T_{\rm th}^{\rm BB/dist} (\theta=(3/4)\pi) }\right) } \nonumber
\\ & { + 32 \left({T_{\rm th}^{\rm BB/dist} (\theta=\pi/3) - T_{\rm th}^{\rm BB/dist} (\theta=(2/3)\pi) }\right) } \Bigr] \, . \nonumber
\end{align}

We note that the structure of the solutions for $a_{\ell,0}$ for $\ell = 0,$ and even $\ell$, involving the sums of $T_{\rm th}^{\rm BB/dist}$ at pairs of colatitudes symmetric
with respect to $\pi/2$ and $T_{\rm th}^{\rm BB/dist}$ at $\pi/2$, as well as the structure of the solutions for
odd $\ell$, involving the differences of $T_{\rm th}^{\rm BB/dist}$ at pairs of colatitudes symmetric with respect to $\pi/2$, reflect the 
symmetry and antisymmetry properties that are 
discussed at the end of Sect. \ref{sec:theoframe}, together with the corresponding implications for the system solution accuracy. 

The solutions expressed in Eqs. \eqref{eq:sol7rings_a6}-\eqref{eq:sol7rings_a1} can be compared with each of the Eqs. \eqref{eq:7rings_0}-\eqref{eq:7rings_6} for a given colatitude $\theta_i$: 
as expected, the sum of products of the various coefficients that multiply $a_{\ell,0}$, for $\ell = 0, 6$,
in the equation for $T_{\rm th}^{\rm BB/dist}(\theta_i)$ with the coefficients in Eqs. \eqref{eq:sol7rings_a6}-\eqref{eq:sol7rings_a1} that multiply $T_{\rm th}^{\rm BB/dist}(\theta_i)$ gives exactly one.
Remarkably, except for $\theta_i = \pi/2$, where only $\ell = 0$ and the even multipoles contribute to $T_{\rm th}^{\rm BB/dist}$, for all the other colatitudes $\theta_i$ the above sum 
is equally contributed for one half by $\ell = 0$ and by the even multipoles and for one half by the odd multipoles.
This is another property related to the symmetry with respect to $\pi/2$ of the set of colatitudes adopted.

\section{Solutions for a blackbody up to $\ell_{\rm max} = 6$}
\label{sec:sol7BB}

Let us consider the specific case of the CMB, assumed to ideally exhibit a blackbody monopole spectrum 
with an effective temperature, $T_0$, in the CMB rest frame. In this case, the photon distribution function is 
\begin{equation}\label{eq:etaBB}
\eta^{\rm BB} (\nu) = \eta^{\rm BB} (x) = \frac{1}{e^{x} -1} \, .
\end{equation}
\noindent
Equation \eqref{eq:eta_boost} then gives the well-known expression

\begin{equation}\label{eq:eta_boostBB}
T_{\rm th}^{\rm BB} (\nu, \theta, \phi, \beta) = \frac{x T_0}{x'} = \frac{T_0 (1-\beta^2)^{1/2}}{1-\beta {\rm cos} \theta} \, ,
\end{equation}

\noindent
with $x'=h\nu'/(kT_r)$, highlighting that $T_{\rm th}^{\rm BB} (\nu, \theta, \phi, \beta)$ does not depend on $\phi$ nor on $\nu$. 

The observed CMB effective temperature averaged over the full sky, $T_{\rm 0,obs}$, is given by 

\begin{equation}\label{eq:T0obsDef}
T_{\rm 0,obs} = \frac{1}{4\pi} \int_{\theta=0}^{\pi} \int_{\phi=0}^{2\pi} \frac{T_0 (1-\beta^2)^{1/2}}{1-\beta {\rm cos} \theta} \, {\rm sin} \theta \, d\theta d\phi \, .
\end{equation}

For an observer at rest with respect to the CMB, $\beta = 0$
and then the substitution of the integration variable, $\theta,$ with a new variable, $w={\rm cos} \, \theta$, 
obviously gives $T_{\rm 0,obs} = T_0$ and implies that in the expansion represented by Eq. \eqref{eq:harm}, as specified by Eq. \eqref{eq:harm6}, 
the only non-vanishing contribution to $T_{\rm 0,obs}$ comes from a term associated to the multipole coefficient $a_{0,0}$.
Since
$Y_{0,0} = \sqrt{1/(4\pi)}$, $a_{0,0} = T_0 \sqrt{4\pi}$. 

For an observer in motion with respect to the CMB, $\beta \ne 0$ and $T_{\rm 0,obs}$ can be calculated by simply substituting the integration variable, $\theta,$ with $w=1-\beta {\rm cos} \, \theta$. We get

\begin{equation}\label{eq:T0obs}
T_{\rm 0,obs} = \frac{1}{2} (1-\beta^2)^{1/2} T_0 \frac{1}{\beta} {\rm ln} \frac{1+\beta}{1-\beta} \, ,
\end{equation}

\noindent
as already reported in \cite{2020JCAP...02..026L} (see also \cite{2011MNRAS.415.3227C}, \cite{2014PhRvD..89l3504D}). 
By replacing ${\rm ln} [(1+\beta)/(1-\beta)]$ with its expansion in Taylor's series up to $\beta^7$, that is, with $2 [\beta + (1/3)\beta^3 + (1/5)\beta^5 + (1/7)\beta^7]$,
we find:

\begin{equation}\label{eq:T0obsTay}
T_{\rm 0,obs} = \frac{1}{2} (1-\beta^2)^{1/2} T_0 \cdot 2 \cdot [1 + (1/3)\beta^2 + (1/5)\beta^4 + (1/7)\beta^6] \, .
\end{equation}

\noindent
We now specify the coefficients, $a_{\ell,0}$, given by Eqs. \eqref{eq:sol7rings_a6}--\eqref{eq:sol7rings_a1} to the blackbody case using Eq. \eqref{eq:eta_boostBB} to compute 
$T_{\rm th}^{\rm BB}$ at the seven considered colatitudes. After solving the algebra, we get

\begin{equation}\label{eq:sol7rings_a6BB}
a_{6,0}^{\rm BB} = \frac{128}{231} \sqrt{\frac{4\pi}{13}} (1-\beta^2)^{1/2} T_0 \cdot \frac{\beta^6}{(1-\beta^2)(2-\beta^2)(4-\beta^2)} \, ,
\end{equation}

\begin{equation}\label{eq:sol7rings_a4BB}
a_{4,0}^{\rm BB} = \frac{16}{385} \sqrt{\frac{4\pi}{9}} (1-\beta^2)^{1/2} T_0 \cdot \frac{44\beta^4-17\beta^6}{(1-\beta^2)(2-\beta^2)(4-\beta^2)} \, , 
\end{equation}

\begin{equation}\label{eq:sol7rings_a2BB}
a_{2,0}^{\rm BB} = \frac{2}{21} \sqrt{\frac{4\pi}{5}} (1-\beta^2)^{1/2} T_0 \cdot \frac{56\beta^2-50\beta^4+5\beta^6}{(1-\beta^2)(2-\beta^2)(4-\beta^2)} \, ,
\end{equation}

\begin{align}\label{eq:sol7rings_a0BB}
a_{0,0}^{\rm BB} & = \frac{1}{2} \sqrt{4\pi} (1-\beta^2)^{1/2} T_0 
\\ & \cdot \frac{16-(68/3)\beta^2+(118/15)\beta^4-(204/315)\beta^6}{(1-\beta^2)(2-\beta^2)(4-\beta^2)} \, , \nonumber
\end{align}

\noindent
and

\begin{equation}\label{eq:sol7rings_a5BB}
a_{5,0}^{\rm BB} = \frac{64}{63} \sqrt{\frac{4\pi}{11}}  (1-\beta^2)^{1/2} T_0 \cdot \frac{\beta^5}{(1-\beta^2)(2-\beta^2)(4-\beta^2)} \, ,
\end{equation}

\begin{equation}\label{eq:sol7rings_a3BB}
a_{3,0}^{\rm BB} = \frac{4}{45} \sqrt{\frac{4\pi}{7}}  (1-\beta^2)^{1/2} T_0 \cdot \frac{36\beta^3-23\beta^5}{(1-\beta^2)(2-\beta^2)(4-\beta^2)} \, ,
\end{equation}

\begin{equation}\label{eq:sol7rings_a1BB}
a_{1,0}^{\rm BB} = \sqrt{\frac{4\pi}{3}}  (1-\beta^2)^{1/2} T_0 \cdot \frac{8\beta-(46/5)\beta^3+(71/35)\beta^5}{(1-\beta^2)(2-\beta^2)(4-\beta^2)} \, .
\end{equation}

In the above expressions, the factor $(1-\beta^2)^{1/2} T_0$ clearly comes from Eq. \eqref{eq:eta_boostBB} while the three factors in the denominator come from the choice of the pairs of colatitudes
$\theta$ symmetric to $\pi/2$, that have been set to $0$ and $\pi$, $\pi/4$ and $(3/4)\pi$, $\pi/3,$ and $(2/3)\pi$. We observe also that, because of the adopted $\ell_{\rm max} = 6$ and the separation of the system 
into two subsystems, 
the solutions for $a_{6,0}^{\rm BB}$ and $a_{5,0}^{\rm BB}$ (see Eqs. \eqref{eq:sol7rings_a6BB} and \eqref{eq:sol7rings_a5BB}) do not show at numerator additional terms coming from higher multipoles,
while they appear in the solutions for $a_{\ell,0}^{\rm BB}$ for $\ell \le 4$ (see Eqs. \eqref{eq:sol7rings_a4BB}-\eqref{eq:sol7rings_a0BB} and Eqs. \eqref{eq:sol7rings_a3BB}-\eqref{eq:sol7rings_a1BB}).

It is evident that the coefficients, $a_{\ell,0}^{\rm BB}$, do not depend on $\theta$. Thus, 
the substitution of the integration variable $\theta$ with $w={\rm cos} \, \theta$, 
in the expansion represented by Eq. \eqref{eq:harm}
again
gives $T_{\rm 0,obs}$ and implies that the only non-vanishing contribution
comes from a term associated to the multipole coefficient $a_{0,0}^{\rm BB}$.
An expansion in Taylor's series up to $\beta^6$, gives $1/(1-\beta^2) = 1 + \beta^2+\beta^4+\beta^6$, 
$1/(2-\beta^2) = (1 + \beta^2/2+\beta^4/4+\beta^6/8)/2,$
and $1/(4-\beta^2) = (1 + \beta^2/4+\beta^4/16+\beta^6/64)/4$. 
It is then simple 
to verify
that, at the same order in $\beta$, Eq. \eqref{eq:sol7rings_a0BB} 
gives exactly the result expressed by Eq. \eqref{eq:T0obsTay}, as is required, in principle.

Equation \eqref{eq:harminv} allows us to analytically derive the $a_{\ell,0}(\nu,\beta)$ for any $\ell$ for relatively simple dependencies of
$T_{\rm th}^{\rm BB} (\nu, \theta, \phi, \beta)$, 
as in the case of the blackbody spectrum, namely, for Eq. \eqref{eq:eta_boostBB}. 
The form of the integrand in $\theta$ involves only the function ${\rm sin} \, \theta / (1-\beta {\rm cos} \, \theta)$ 
multiplied by polynomials in ${\rm cos} \, \theta$ and when substituting
the integration variable $\theta$ with $w=1-\beta {\rm cos} \, \theta$, 
the integrand consists only of functions as $1/w$ and powers of $w$. We omit the tedious calculation at $\ell \ge 2$. Instead, for $\ell = 1,$ we get:
\begin{align}\label{eq:exact_a1BB}
a_{1,0}^{\rm BB} & = 2\pi (1-\beta^2)^{1/2} T_0 \int_{\theta=0}^\pi \frac{{\rm sin} \, \theta}{1-\beta {\rm cos} \, \theta} \sqrt{\frac{3}{4\pi}} {\rm cos} \, \theta \, d\theta
\\ & = 2\pi \sqrt{\frac{3}{4\pi}} (1-\beta^2)^{1/2} T_0 \frac{1}{\beta^2} \left[{ {\rm ln} \frac{1+\beta}{1-\beta} -2\beta }\right] \nonumber \, ,
\end{align}

\noindent
where, by replacing
${\rm ln} [(1+\beta)/(1-\beta)]$ with its expansion in Taylor's series up to $\beta^7$, gives:

\begin{align}\label{eq:exact_a1BB7ord}
a_{1,0}^{\rm BB} = \sqrt{4\pi} \sqrt{3} (1-\beta^2)^{1/2} T_0 [\beta/3+\beta^3/5+\beta^5/7] \, .
\end{align}

\noindent
Performing a Taylor's series expansion up to $\beta^6$ for $1/(1-\beta^2)$, $1/(2-\beta^2)$ and $1/(4-\beta^2)$, 
it is simple to verify that at the same order in $\beta$, Eq. \eqref{eq:sol7rings_a1BB} for $a_{1,0}^{\rm BB}$ 
gives precisely the result expressed by Eq. \eqref{eq:exact_a1BB7ord}.

Finally, we remember that in Sect. \ref{sec:sol7} we discussed deriving both $a_{0,0}$ and $a_{1,0}$ by substitution 
in the last step 
of the calculation to solve
the corresponding linear subsystem. Thus,
the consistencies discussed above for $T_{\rm 0,obs}$ and $a_{1,0}^{\rm BB}$ at the adopted order also represent a further verification of the derived algebraic solutions.

\section{Explicit solutions up to $\ell_{\rm max} = 4$}
\label{sec:sol5}

For many applications, a computation up to $\ell_{\rm max} = 4$ suffice to get the relevant information. We then provide here simpler solutions based on $N=5$ equations
using the set of colatitudes $0, \pi/4, \pi/2, (3/4)\pi$ and $\pi$. This also allows us to explicitly focus on some of the mentioned properties of the proposed method.

We construct a system formed only by Eqs. \eqref{eq:7rings_0}, \eqref{eq:7rings_1}, \eqref{eq:7rings_3}, \eqref{eq:7rings_5} and \eqref{eq:7rings_6} ignoring
the terms associated to $a_{5,0}$ and  $a_{6,0}$, and we again solve it with the methods of elimination and substitution in a way similar to that described in Sect. \ref{sec:sol7}. The solutions are

\begin{align}\label{eq:sol5rings_a4}
a_{4,0} & = \frac{8}{35} \sqrt{\frac{4\pi}{9}} 
\Bigl[ { \left({T_{\rm th}^{\rm BB/dist} (\theta=0) + T_{\rm th}^{\rm BB/dist} (\theta=\pi) }\right) }
\\ & { - 2 \left({T_{\rm th}^{\rm BB/dist} (\theta=\pi/4) + T_{\rm th}^{\rm BB/dist} (\theta=(3/4)\pi) }\right) } \nonumber
\\ & { + 2 T_{\rm th}^{\rm BB/dist} (\theta=\pi/2) } \Bigr]  \nonumber \, ,
\end{align}

\begin{align}\label{eq:sol5rings_a2}
a_{2,0} & = \frac{1}{21} \sqrt{\frac{4\pi}{5}} 
\Bigl[ { 5 \left({T_{\rm th}^{\rm BB/dist} (\theta=0) + T_{\rm th}^{\rm BB/dist} (\theta=\pi) }\right) } 
\\ & { + 4 \left({T_{\rm th}^{\rm BB/dist} (\theta=\pi/4) + T_{\rm th}^{\rm BB/dist} (\theta=(3/4)\pi) }\right) } \nonumber
\\ & { - 18 T_{\rm th}^{\rm BB/dist} (\theta=\pi/2) } \Bigr] \nonumber \, ,
\end{align}

\begin{align}\label{eq:sol5rings_a0}
a_{0,0} & = \frac{\sqrt{4\pi}}{30} 
\Bigl[ { \left({T_{\rm th}^{\rm BB/dist} (\theta=0) + T_{\rm th}^{\rm BB/dist} (\theta=\pi) }\right) } 
\\ & { + 8 \left({T_{\rm th}^{\rm BB/dist} (\theta=\pi/4) + T_{\rm th}^{\rm BB/dist} (\theta=(3/4)\pi) }\right) } \nonumber
\\ & { +12 T_{\rm th}^{\rm BB/dist} (\theta=\pi/2) } \Bigr] \, . \nonumber
\end{align}

\noindent
Next, we have

\begin{align}\label{eq:sol5rings_a3}
a_{3,0} & = \frac{2}{5} \sqrt{\frac{4\pi}{7}} 
\Bigl[ { \left({T_{\rm th}^{\rm BB/dist} (\theta=0) - T_{\rm th}^{\rm BB/dist} (\theta=\pi) }\right) } 
\\ & { - \sqrt{2} \left({T_{\rm th}^{\rm BB/dist} (\theta=\pi/4) - T_{\rm th}^{\rm BB/dist} (\theta=(3/4)\pi) }\right) } \Bigr] \nonumber \, ,
\end{align}

\begin{align}\label{eq:sol5rings_a1}
a_{1,0} & = \frac{1}{10} \sqrt{\frac{4\pi}{3}} 
\Bigl[ { \left({T_{\rm th}^{\rm BB/dist} (\theta=0) - T_{\rm th}^{\rm BB/dist} (\theta=\pi) }\right) } 
\\ & { + 4 \sqrt{2} \left({T_{\rm th}^{\rm BB/dist} (\theta=\pi/4) - T_{\rm th}^{\rm BB/dist} (\theta=(3/4)\pi) }\right) } \Bigr] \,  \nonumber .
\end{align}

While the structure of the solutions expressed by Eqs. \eqref{eq:sol5rings_a4}--\eqref{eq:sol5rings_a1} is analogous to the structure of Eqs. \eqref{eq:sol7rings_a4}--\eqref{eq:sol7rings_a0}
and Eqs. \eqref{eq:sol7rings_a3}--\eqref{eq:sol7rings_a1}, 
the different algebraic coefficients reflect the different choice adopted for the set of colatitudes.

\section{Solutions for a blackbody up to $\ell_{\rm max} = 4$}
\label{sec:sol5BB}

Specifying the coefficients, $a_{\ell,0}$, given by Eqs. \eqref{eq:sol5rings_a4}--\eqref{eq:sol5rings_a1} to the blackbody case (see Eq. \eqref{eq:eta_boostBB}), we compute 
$T_{\rm th}^{\rm BB}$ at the five considered colatitudes
and we obtain

\begin{equation}\label{eq:sol5rings_a4BB}
a_{4,0}^{\rm BB} = \frac{16}{35} \sqrt{\frac{4\pi}{9}} (1-\beta^2)^{1/2} T_0 \cdot \frac{\beta^4}{(1-\beta^2)(2-\beta^2)} \, ,
\end{equation}

\begin{equation}\label{eq:sol5rings_a2BB}
a_{2,0}^{\rm BB} = \frac{2}{21} \sqrt{\frac{4\pi}{5}} (1-\beta^2)^{1/2} T_0 \cdot \frac{14\beta^2-9\beta^4}{(1-\beta^2)(2-\beta^2)} \, ,
\end{equation}

\begin{equation}\label{eq:sol5rings_a0BB}
a_{0,0}^{\rm BB} = \frac{\sqrt{4\pi}}{30} (1-\beta^2)^{1/2} T_0 \cdot \frac{60-70\beta^2+12\beta^4}{(1-\beta^2)(2-\beta^2)} \, ,
\end{equation}

\noindent
and

\begin{equation}\label{eq:sol5rings_a3BB}
a_{3,0}^{\rm BB} = \frac{4}{5} \sqrt{\frac{4\pi}{7}}  (1-\beta^2)^{1/2} T_0 \cdot \frac{\beta^3}{(1-\beta^2)(2-\beta^2)} \, ,
\end{equation}

\begin{equation}\label{eq:sol5rings_a1BB}
a_{1,0}^{\rm BB} = \sqrt{\frac{4\pi}{3}}  (1-\beta^2)^{1/2} T_0 \cdot \frac{2\beta-(9/5)\beta^3}{(1-\beta^2)(2-\beta^2)} \, .
\end{equation}

In this case ($\ell_{\rm max} = 4$), the solutions for $a_{4,0}^{\rm BB}$ and $a_{3,0}^{\rm BB}$ do not show, in terms of the numerator, additional 
higher multipoles terms,
as they appear 
at
$\ell \le 2$. The algebraic coefficients appearing in Eqs. \eqref{eq:sol5rings_a4BB}--\eqref{eq:sol5rings_a1BB} 
and in Eqs. \eqref{eq:sol7rings_a4BB}--\eqref{eq:sol7rings_a0BB} and \eqref{eq:sol7rings_a3BB}--\eqref{eq:sol7rings_a1BB}  are different, but these sets of equations give exactly the same solutions
when the ratios of their polynomials in $\beta$ are computed up to the order of $\beta^4$. Analogously, Eq. \eqref{eq:sol5rings_a0BB} gives for $T_{\rm 0,obs}$
the same result of Eq. \eqref{eq:T0obsTay}, and Eq. \eqref{eq:sol5rings_a1BB} is equivalent to Eq. \eqref{eq:exact_a1BB7ord}
when they are computed up to the same order of power in $\beta$.

\section{Explicit solutions up to $\ell_{\rm max} = 2$ and 1}
\label{sec:dip_3and2col}

It is helpful to write the solutions for low values of $\ell_{\rm max}$.

From the $N=3$ equations at the colatitudes $0, \pi/2$ and $\pi$ (Eqs. \eqref{eq:7rings_0}, \eqref{eq:7rings_3}, and \eqref{eq:7rings_6}), ignoring
the terms associated to $a_{\ell,0}$ for $\ell > 2$, we get

\begin{align}\label{eq:sol3rings_a2}
a_{2,0} & = \frac{1}{3} \sqrt{\frac{4\pi}{5}} 
\Bigl[ { \left({T_{\rm th}^{\rm BB/dist} (\theta=0) + T_{\rm th}^{\rm BB/dist} (\theta=\pi) }\right) } 
\\ & { - 2 T_{\rm th}^{\rm BB/dist} (\theta=\pi/2) } \Bigr] \nonumber \, ,
\end{align}

\begin{align}\label{eq:sol3rings_a1}
a_{1,0} & = \frac{1}{2} \sqrt{\frac{4\pi}{3}} 
\Bigl[ { \left({T_{\rm th}^{\rm BB/dist} (\theta=0) - T_{\rm th}^{\rm BB/dist} (\theta=\pi) }\right) } \Bigr] \, , 
\end{align}

\begin{align}\label{eq:sol3rings_a0}
a_{0,0} & = \frac{\sqrt{4\pi}}{6} 
\Bigl[ { \left({T_{\rm th}^{\rm BB/dist} (\theta=0) + T_{\rm th}^{\rm BB/dist} (\theta=\pi) }\right) } 
\\ & { + 4 T_{\rm th}^{\rm BB/dist} (\theta=\pi/2) } \Bigr] \, . \nonumber
\end{align}

Using only $N=2$ equations at the colatitudes $0$ and $\pi$ (Eqs. \eqref{eq:7rings_0} and \eqref{eq:7rings_6}), neglecting
the terms 
at
$\ell > 1$, we have

\begin{align}\label{eq:sol2rings_a1}
a_{1,0} & = \frac{1}{2} \sqrt{\frac{4\pi}{3}} 
\Bigl[ { \left({T_{\rm th}^{\rm BB/dist} (\theta=0) - T_{\rm th}^{\rm BB/dist} (\theta=\pi) }\right) } \Bigr] \, , 
\end{align}

\begin{align}\label{eq:sol2rings_a0}
a_{0,0} & = \frac{\sqrt{4\pi}}{2} \Bigl[ { \left({T_{\rm th}^{\rm BB/dist} (\theta=0) + T_{\rm th}^{\rm BB/dist} (\theta=\pi) }\right) } \Bigr] \, . 
\end{align}

With $N=2$ equations but at the colatitudes $0$ and $\pi/2$ (Eqs. \eqref{eq:7rings_0} and \eqref{eq:7rings_3}), which is clearly not symmetric with respect to $\pi/2$, 
instead we get:
\begin{align}\label{eq:sol2asimmrings_a1}
a_{1,0} & = \sqrt{\frac{4\pi}{3}} 
\Bigl[ { \left({T_{\rm th}^{\rm BB/dist} (\theta=0) - T_{\rm th}^{\rm BB/dist} (\theta=\pi/2) }\right) } \Bigr] \, , 
\end{align}

\begin{equation}\label{eq:sol2asimmrings_a0}
a_{0,0} = \sqrt{4\pi} \Bigl[ { \left({T_{\rm th}^{\rm BB/dist} (\theta=\pi/2) } \right) } \Bigr] \, . 
\end{equation}

\section{On dipole estimations based on two colatitudes}
\label{sec:dip_2col}

As proposed by \cite{1981A&A....94L..33D}, a suitable and observationally intuitive approximation for the dipole spectrum can be expressed in terms of the difference 
of $T_{\rm th}^{\rm BB/dist}$ in the direction of motion and in its perpendicular direction.

Equations \eqref{eq:7rings_0} and \eqref{eq:7rings_3} allow us to express this difference in terms of a combination of 
the coefficients $a_{\ell,0} (\nu,\beta)$ up to $\ell_{\rm max} = 6:$ 

\begin{align}\label{eq:DeltaTtherm_alm}
    \Delta_{0,\pi/2} T_{\rm th}^{\rm BB/dist} & = \sqrt{\frac{3}{4\pi}} a_{1,0} + \frac{3}{2} \sqrt{\frac{5}{4\pi}} a_{2,0}  + \sqrt{\frac{7}{4\pi}} a_{3,0} 
\\ & + \frac{5}{8} \sqrt{\frac{9}{4\pi}} a_{4,0} + \sqrt{\frac{11}{4\pi}} a_{5,0} + \frac{21}{16} \sqrt{\frac{13}{4\pi}} a_{6,0} \nonumber \, .
\end{align}

\noindent
Neglecting the contributions from $\ell > 1$, Eqs. \eqref{eq:sol2asimmrings_a1} and \eqref{eq:DeltaTtherm_alm} are equivalent.
We can also express the semi-difference in $T_{\rm th}^{\rm BB/dist}$ measured in the direction of motion and in its opposite direction using Eqs. \eqref{eq:7rings_0} and \eqref{eq:7rings_6}

\begin{align}\label{eq:DeltaTtherm_alm_line}
   \frac{1}{2} \Delta_{0,\pi} T_{\rm th}^{\rm BB/dist} & = \sqrt{\frac{3}{4\pi}} a_{1,0} + \sqrt{\frac{7}{4\pi}} a_{3,0} + \sqrt{\frac{11}{4\pi}} a_{5,0} \, 
\end{align}

\noindent
and in leaving out the terms at $\ell > 1$, Eq. \eqref{eq:DeltaTtherm_alm_line} is equivalent to Eq. \eqref{eq:sol2rings_a1}.

The estimation of $a_{1,0} (\nu,\beta)$ through the simple difference of $T_{\rm th}^{\rm BB/dist}$ in only two directions 
can be performed using the two colatitudes $\theta = 0$ and $\theta = \pi$ to automatically suppress the contributions from $\ell = 2$ (and from higher even $\ell$), as
discussed at the end of Sect. \ref{sec:theoframe}.
The same holds for any other pair of colatitudes symmetric with respect to $\pi/2$ (as can be derived, for example, by combining Eqs. \eqref{eq:7rings_1} and \eqref{eq:7rings_5} or Eqs. \eqref{eq:7rings_2} and \eqref{eq:7rings_4}). The solutions presented in Sect. \ref{sec:sol7} can instead be used to correct for the contributions from the odd terms at $\ell = 3$ and 5
(and from the terms at even $\ell$, when using Eq. \eqref{eq:DeltaTtherm_alm}).

\section{Solutions for $a_{\ell,0}$ and spectrum integration or differentiation}
\label{sec:highl_derivs}

Equation \eqref{eq:harminv} shows that the solution for a given $a_{\ell,0}$ is an integral of $T_{\rm th}^{\rm BB/dist}$ over $\theta$. Since $\beta$ is very small, when $\theta$ spans in the interval $[0,\pi]$, around $\pi/2$, 
the values of $T_{\rm th}^{\rm BB/dist}$ in the integrand come from frequency values in a small interval around $\nu / (1-\beta^2)^{1/2}$
(see Eq. \eqref{eq:eta_boost} and the relation between $\nu$ and $\nu'$), making the integral sensitive to the local variation of $T_{\rm th}^{\rm BB/dist}$.
Formally, the solutions expressed by Eqs. \eqref{eq:sol7rings_a6}--\eqref{eq:sol7rings_a1} and \eqref{eq:sol5rings_a4}--\eqref{eq:sol5rings_a1} can be regarded 
as
definitions of sets of weights 
assigned to a small number of values of
function $T_{\rm th}^{\rm BB/dist}$ in a given set of colatitudes, or corresponding frequencies, 
to compute the integrals that give the coefficients $a_{\ell,m}$ in Eq. \eqref{eq:harminv}. 

To a first-order approximation, the dipole spectrum induced by the observer peculiar velocity is directly proportional to the first logarithmic derivative of the photon occupation number, $\eta(\nu)$, with respect to the frequency, $\nu$ \citep{1981A&A....94L..33D}.
This concept can be generalized to higher multipoles. 
Let us consider the partial derivative of Eq. \eqref{eq:harminv} with respect to the frequency, $\nu$. 
According to Leibniz's rule, when performing the differentiation under the integral sign,
a further multiplicative factor involving the product, $\beta \, {\rm cos}\, \theta,$ enters in the integral over $\theta$, other than factors depending on the form of $\eta(\nu)$. 
As is evident from Eq. \eqref{eq:harm6}, a further power of ${\rm cos}\, \theta$ appears passing from $\ell$ to $\ell +1$ in the associated Legendre polynomials and, consequently, in
the integrand function of $a_{\ell,0}$ (see Eq. \eqref{eq:harminv}). 
Thus, the subsequent $a_{\ell,0}(\nu)$ at increasing $\ell$ is tightly related to the subsequent derivatives of $\eta(\nu)$ with respect to $\nu$, or in other words,
their frequency behaviours are particular sensitive to the local (in frequency space) monopole spectrum variation up to increasing derivative order. 

It is interesting to note certain properties of the coefficients (or weights) in Eqs. \eqref{eq:sol7rings_a6}--\eqref{eq:sol7rings_a1}, \eqref{eq:sol5rings_a4}--\eqref{eq:sol5rings_a1},
\eqref{eq:sol3rings_a2}-\eqref{eq:sol3rings_a0}, and \eqref{eq:sol2rings_a1}-\eqref{eq:sol2rings_a0}
that are related to the separation of odd and even multipoles in the system solution.
As already discussed, this separation appears when we adopt sets of colatitudes $\theta$ symmetrically located around $\pi/2$.
The central weight, applied to $\theta = \pi/2$, is zero for odd $\ell$ but not for even $\ell$. For angles $\theta$ symmetric with respect to $\pi/2,$ the weights are 
opposite for odd $\ell$ and equal for even $\ell$. The sum of the weights vanishes, except for $\ell = 0$: in this case, the sum is exactly unit, when divided by 
the 'normalization' factor, $\sqrt{4\pi}$ (see also Eq. \eqref{eq:harm6}).
These properties are identical to those satisfied by the weights for the centred approximations at a grid point for the generation of finite difference formulas
on arbitrarily spaced grids for any order of derivative \citep{1988MC,1998SIAMR..40..685F}.
Furthermore, we note that the relative weights in Eqs. \eqref{eq:sol3rings_a2}, \eqref{eq:sol3rings_a1}, and \eqref{eq:sol2rings_a1}
are equivalent to the relative weights for the centred approximations at a grid point for the second and first order of derivative, the relative weights in Eq. \eqref{eq:sol2rings_a0} are equivalent to
the relative weights for the centred approximations at the halfway point for the zero order of derivative, while the relative weights in Eqs. \eqref{eq:sol2asimmrings_a1} and \eqref{eq:sol2asimmrings_a0}
are equivalent to the relative weights for the one-sided approximations at a grid point for the first and zero order of derivative. The different level of approximation in the estimate of $a_{1,0}$ via
Eq. \eqref{eq:DeltaTtherm_alm} and Eq. \eqref{eq:DeltaTtherm_alm_line}, neglecting terms at $\ell > 1$, is clearly related to the different accuracies of the one-sided and centred scheme for numerical differentiation.
Finally, the relative weights for $a_{0,0}$ in Eq. \eqref{eq:sol3rings_a0} are not equivalent to relative weights for the zero order of derivative of the schemes mentioned above.
This is of increasing evidence in the weights of the solutions at $\ell_{max} > 2$.
Remarkably, they do not satisfy the sign alternation appearing in the weights of the centred approximations at a grid point of finite difference formulas moving from the central node to the more external nodes.
Indeed, they store the relations between the $a_{\ell,0}$ at different $\ell$ and the temperatures at the adopted set of colatitudes that originates from the system solution at the corresponding $\ell_{max}$
(this is analogous to the 'mixing' of derivatives discussed above).

\section{Monopole spectrum models and single signal results}
\label{sec:monmod}

The method described 
can be applied to any type of signal and to combinations of signals, provided that they are summed in terms of additive quantities, such as the photon distribution function, $\eta$, or the antenna temperature,
\begin{equation}
        T_{\rm ant} (\nu) ={h\nu\over k} \eta(\nu) \, .
    \label{eq:t_ant}
\end{equation}

\noindent
In this work, we consider eight different types of monopole spectrum that can be represented in terms of analytical or semi-analytical functions. 

We first focus on four types of signals
characterized by a CMB-distorted photon distribution function, $\eta^{\rm dist} (\nu)$ that is different from the 
blackbody, $\eta^{\rm BB} (\nu)$, at the present temperature $T_0$.
We then consider four types of extragalactic background superimposed onto the CMB blackbody spectrum.
We give only a concise description of the various models, referring to the literature for further information.
On the other hand, we report the equations relevant for a clear connection with Sect. \ref{sec:all}.

We first consider 
the
signals 
that are more relevant (or  essentially relevant) at low frequencies
(radio domain) 
and then 
those
that are relevant 
over a very wide frequency range 
(up to the far-IR) 
or more 
important
at increasing frequency. 
We compare the results based on the proposed method
(the solutions 
in Sect. \ref{sec:sol7}) with the computation based 
on 
direct numerical integration (see Eq. \eqref{eq:harminv} and the discussion in Sect. \ref{sec:NE}).
For simplicity, we perform the comparison (see also Appendix \ref{app:beta_amplified}) only for two representative cases, which were chosen because they are very different with regard to the spectrum features.

The results are presented in terms of the following quantities:

\begin{itemize}

\item The difference, $\Delta T_{th}$, between the equivalent thermodynamic temperature 
of the intrinsic monopole spectrum for the adopted model and the CMB present temperature $T_0$. 

\item The ratio, $R = (a_{0,0} (\nu,\beta) / \sqrt{4\pi}) / T_{th} (\nu)$, between the equivalent thermodynamic temperature of observed (see Eq. \eqref{eq:harm6}) 
and intrinsic monopole, expressed in terms of the difference $\Delta R = R - R^{\rm BB}$, where 
$R^{\rm BB} \simeq (1 - 2.5362 \times 10^{-7})$ is the same ratio but for the case of the blackbody, $\eta^{\rm BB} (\nu)$, at the present temperature $T_0$ 
(see Eqs. \eqref{eq:T0obsDef}--\eqref{eq:T0obsTay} and the discussion in Sect. \ref{sec:sol7BB}). 

\item The coefficients $a_{\ell,0} (\nu,\beta)$ from $\ell = 1$ to $\ell_{\rm max} = 6$ 
(see Eqs. \eqref{eq:sol7rings_a6}--\eqref{eq:sol7rings_a2} and \eqref{eq:sol7rings_a5}--\eqref{eq:sol7rings_a1})
expressed in terms of their difference
with  
the blackbody case (see Eqs. \eqref{eq:sol7rings_a6BB}--\eqref{eq:sol7rings_a2BB} and \eqref{eq:sol7rings_a5BB}--\eqref{eq:sol7rings_a1BB}).

\end{itemize}

\begin{figure*}[ht!]
\centering
                  \includegraphics[width=18.5cm]{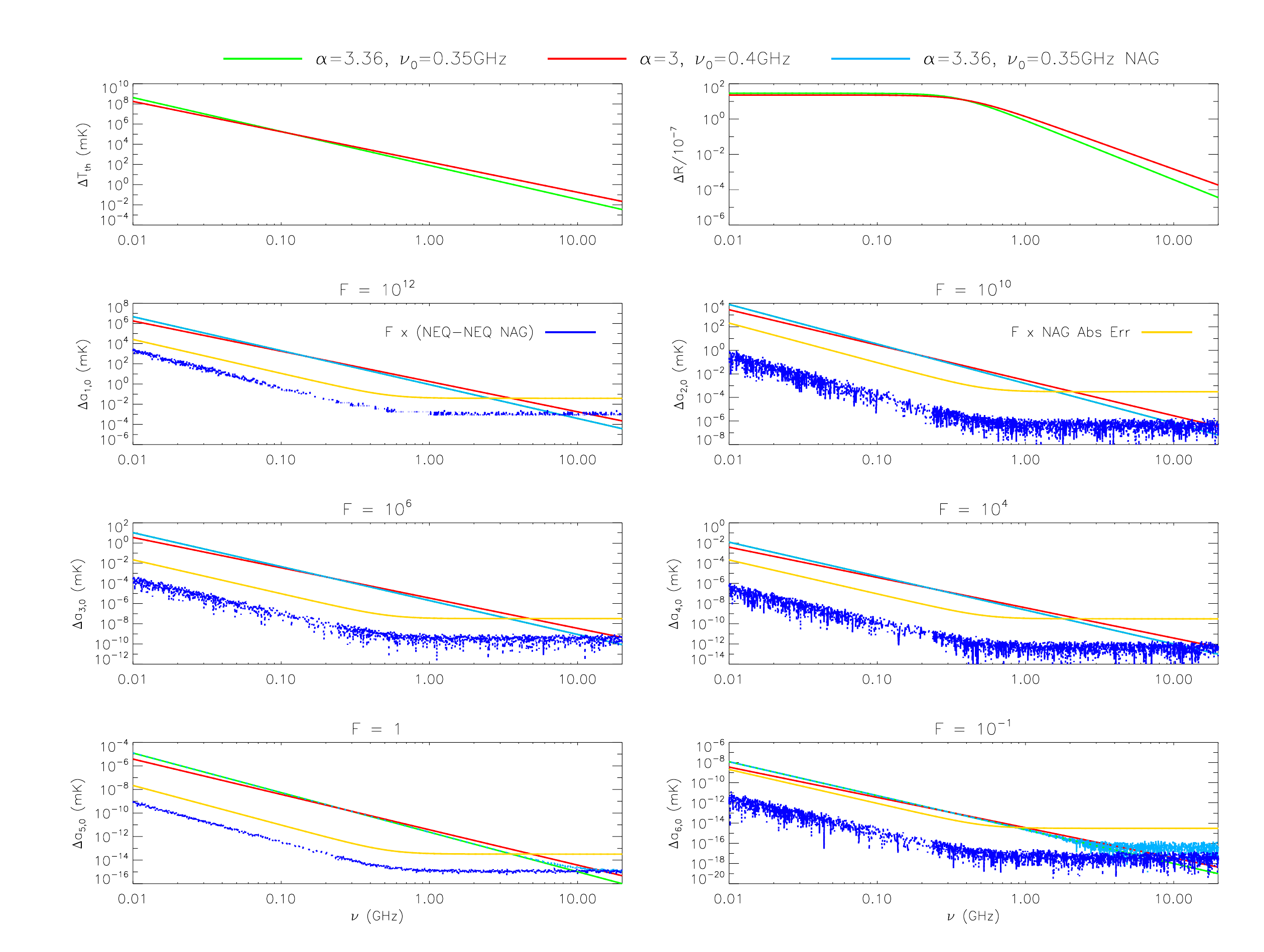}
    \caption{$\Delta T_{th}$, $\Delta R$ and $\Delta a_{\ell,0}$ from $\ell = 1$ to $\ell_{\rm max} = 6$ for the considered non-equilibrium models.
    Solid lines (or dots) correspond to positive (or negative) values. Green and light blue lines are essentially superimposed up to $\ell =4$, 
    where only one of the two lines can be appreciated.
    Their difference multiplied by a factor F to have values compatible with the adopted range is displayed by the blue lines.
    Yellow lines refer to the nominal integration error quoted by the routine {\it D01AJF}, again multiplied by the factor F. See also the legend and the text.}
    \label{fig:NEQ}
\end{figure*}

\subsection{Possible non-equilibrium imprint at low frequencies}
\label{sec:NE}

An important extragalactic background signal that is much larger than the CMB background predicted for a blackbody spectrum at an equilibrium temperature 
in agreement with FIRAS results is observed at radio frequencies, in particular, below a few GHz (see e.g. \cite{Dowell:2018}). A signal excess could be also present 
at $\simeq 3.3$ GHz, as claimed by \citep{2011ApJ...730..138S} on the basis of the second generation of the
Absolute Radiometer for Cosmology, Astrophysics, and Diffuse Emission (ARCADE 2)
data.
Models based on contributions by faint astrophysical sources, on interactions between dark matter (DM) and baryons, or 
on their combinations
have been invoked to explain this background, 
possibly together with the pronounced absorption profile of the 21cm redshifted line signal (see also Sect. \ref{sec:21cm}), which has also been claimed by
\cite{EDGESobs2018Nature};
see, for example, \cite{2011ApJ...734....6S}, \cite{Barkana2018Nature}, \cite{2018arXiv180210094M}, \cite{2018ApJ...868...63E} and \cite{2019IAUS..346..365M} (see also \cite{Subrahmanyan:2013}, \cite{2018Natur.564E..32H} and \cite{2018MNRAS.481L...6S}).

\cite{2020PhRvR...2a3210B} proposed an alternative explanation of the signal excess in the low frequency background, involving a mechanism of stochastic frequency diffusion in the perspective of non-equilibrium statistical mechanics. The model implies a modification of the standard Kompaneets equation \citep{1957JETP....4..730K}, explicitly considered by the authors in the limit that includes only the scattering, and
a relaxation of the Einstein detailed balance relation. The resulting abundance of low frequency photons can be described by a stationary solution of the photon distribution function in the form: 

\begin{equation}\label{eq:etaNE}
        \eta(\nu) = \frac{1}{{\rm exp} \left[{\int^\nu d\nu' \gamma(\nu')/D(\nu')}\right] - 1} = \frac{1}{e^{\psi(\nu)} - 1} \, ,
\end{equation}

\noindent
where $\gamma$ and $D$ are frequency dependent friction and diffusion terms and 
the function $\psi(\nu)$ can be approximated by 

\begin{equation}\label{eq:psiNE}
       \psi(\nu) = \frac{h\nu}{kT^\star} \frac{(\nu/\nu_0)^\alpha}{1+(\nu/\nu_0)^\alpha}  \, .
\end{equation}

\noindent
Subtracting from the global extragalactic background signal the contribution by extragalactic radio sources, for instance assuming the model by \cite{2008ApJ...682..223G} with an amplification
factor of $\simeq 1.3$ in the resulting background 
(see also Sect. \ref{sec:exback_radio}), 
and comparing the residual background
with their almost complete collection
of cosmic background absolute 
temperature data, 
they found: $\nu_0 \simeq 0.4$\,GHz for $\alpha = 3$ (and $T^\star = T_0$ to 
fit high frequency data), $\nu_0 \simeq 0.35$\,GHz, and $\alpha \simeq 3.36$ using both $\nu_0$ and $\alpha$ as fit variables.

In Fig. \ref{fig:NEQ}, we show $\Delta T_{th}$, $\Delta R$ and the coefficients $a_{\ell,0} (\nu,\beta)$ for $\ell = 1, 6$, expressed in terms of $\Delta a_{\ell,0}$, 
derived
for the two sets of best-fit parameters
according to the solutions in Sect. \ref{sec:sol7}
and 
in one case,
also on the basis of Eq. \eqref{eq:harminv}. 
The computations were performed in quadruple precision. 
We first carried out some tests
with a simple Gaussian quadrature scheme \citep{press1992numerical}, 
using various accuracy parameter values and point numbers (e.g.
with the accuracy parameter ({\it EPS}) set to $10^{-9}$ and 2048 points), and compare the result with the explicit analytical solution for a blackbody: we find unreliable results above $\ell = 3$
(or above $\ell = 4$ provided that Eq. \eqref{eq:harminv} is written as in the last equality). We then performed the numerical integration using the 
very accurate and efficient routine {\it D01AJF} of the Numerical Algorithms Group (NAG) Numerical Library,
available
only 
in double precision,
setting integration accuracy parameters to the smallest values 
and increasing the number of sub-intervals used by the routine and the related workspace allocation. However, we
verified that
splitting the integral in terms of sums of integrals over subsets of the integration intervals does not improve the accuracy at all.

There is
very good agreement between the results found with the routine {\it D01AJF} and the solutions in Sect. \ref{sec:sol7} with $\ell_{\rm max} = 6$ (particularly at lower multipoles, where the lines are 
superimposed and indistinguishable).
Their differences are compatible with a combination of higher order terms, that is, beyond $\ell = 6$, and integration errors, that are missing
in the solutions in Sect. \ref{sec:sol7} and present in the numerical results, respectively.
The two types of differences clearly appear, respectively, at lower frequencies, where the signal is higher and the relative integration error is lower, 
and at higher frequencies, where the signal is lower and the relative integration error is higher.    
We report also the nominal integration error quoted by the routine {\it D01AJF}: the comparison with the above differences suggests that this error is likely very conservative.
In Appendix \ref{app:beta_amplified}, we provide some results derived adopting a much larger value of $\beta$ that implies much larger signals, relatively higher contributions from higher multipoles, as well as
relatively lower numerical integration errors: the analysis clearly supports the above interpretation.

Figure \ref{fig:NEQ} shows that the typical power law shape of the intrinsic monopole spectrum, subsequent to the subtraction of the blackbody at the present temperature $T_0$,
is displayed also at higher multipoles, as already discussed in \cite{2019A&A...631A..61T} for the dipole. Remarkably, we find that the ratio, $R$, between observed and intrinsic monopole, 
$\Delta R$ in Fig. \ref{fig:NEQ}, 
is not frequency independent, as in the case of a blackbody, but exhibits a frequency dependence
related to the assumed intrinsic monopole spectrum.
At low frequencies, below $\sim 1$\,GHz, the values of $\Delta R$ are positive and with amplitudes comparable to 
$|R^{\rm BB} - 1|$ or even larger.

\subsection{Comptonization distortion and free-free diffuse emission}
\label{sec:CFF}

Many types of sources of photon and energy injections in cosmic plasma generates Comptonization distortions \citep{1972JETP...35..643Z} via electron heating, and in ionizing the matter, they also produce FF distortions.
Although these signatures can be generated both before (see e.g. \cite{2012MNRAS.419.1294C}) and after the cosmological recombination epoch (see e.g. \cite{1986ApJ...300....1S} and \cite{1994LNP...429...28D}),
the cosmological reionization 
associated with the early formation phases of bound structures is the most remarkable source of these distortions. 
Two key parameters quantify the amplitudes of these imprints that for a given model, are tightly coupled. They are the Comptonization parameter, $u$,
proportional to the global fractional energy exchange 
between matter and radiation in the cosmic plasma
(for small distortions $u \simeq (1/4) \Delta \varepsilon/ \varepsilon_{\rm i}$), 
and the 
FF distortion parameter, $y_{B} (x)$,
defined by integrals over the relevant redshift interval.
On the other hand, even in the context of the reionization process, 
a variety of astrophysical mechanisms can contribute to determine the final distortion levels (see e.g. \cite{2016JCAP...03..047D}, \cite{2018JCAP...04..021B} and references therein). 
The resulting distorted photon distribution function is well approximated by
\begin{equation}\label{eq:etaC}
\eta^{\rm FF+C} \simeq \eta_{\rm i} + u {x / \phi_{\rm i} \,e^{x/\phi_{\rm i}} \over (e^{x/\phi_{\rm i}} - 1)^{2}}  \left ( {x/\phi_{\rm i} \over { {\rm tanh}[x/(2\phi_{\rm i})] }} - 4 \right) + \frac{y_{B}(x)}{x^{3}} \, , 
\end{equation}
\noindent where $\eta_{\rm i}$ is the photon occupation number at the dissipation process initial time 
denoted with the subscript ${\rm i}$. Neglecting other processes, $\eta_{\rm i}$ can be assumed to have a Planckian distribution at the
initial temperature defined by $\phi_{\rm i} = \phi(z_{\rm i}) = (1+ \Delta \varepsilon/ \varepsilon_{\rm i})^{-1/4} \simeq 1-u$, that is, 
$\eta_{\rm i} = 1 / (e^{x/\phi_{\rm i}} - 1)$. 

The global Comptonization distortion depends linearly on matter density, 
thus, assuming a uniform medium is not critical for computing $u$. Conversely, bremsstrahlung depends quadratically on matter density
and, in the presence of a substantial intergalactic medium (IGM) matter density contrast,
the FF distortion is amplified with respect to the case of a homogeneous medium (\cite{2004ApJ...606L...5C,2011MNRAS.410.2353P,2014MNRAS.437.2507T})
by a factor  of $\simeq 1 + \sigma^{2} (z)$, that is, $\Omega_{b}^{2} (z) \rightarrow \Omega_{b,homog}^{2}  (1+\sigma^{2}(z))$,
 with $\sigma^{2} (z)$ as the baryonic matter variance
related to the thermal properties of the DM particles. 

\begin{figure*}[ht!]
\centering
         \includegraphics[width=18.5cm]{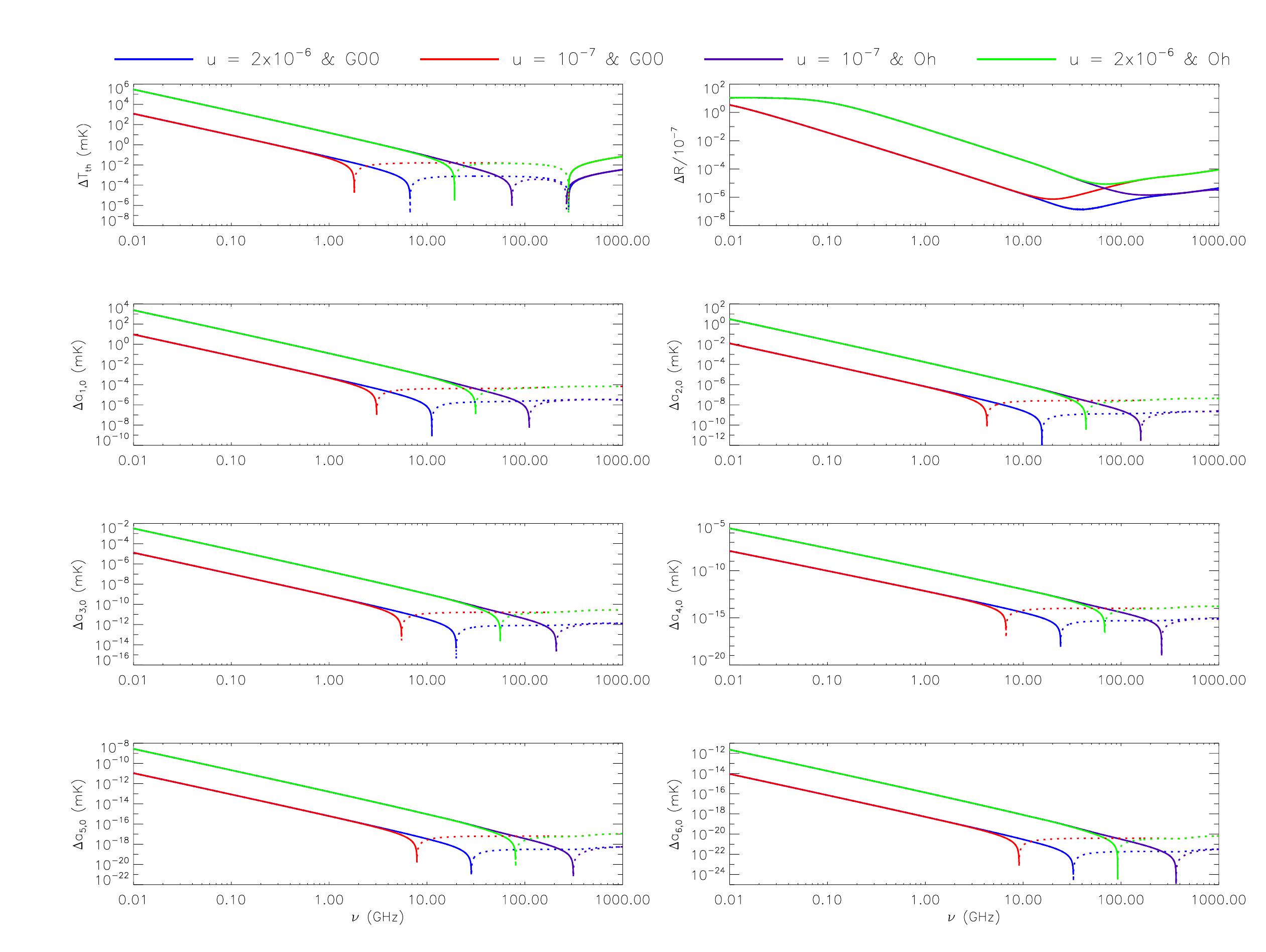}
    \caption{$\Delta T_{th}$, $\Delta R$ and $\Delta a_{\ell,0}$ from $\ell = 1$ to $\ell_{\rm max} = 6$ for the considered combined Comptonization and diffuse FF distortion models. 
    Solid lines (or dots) correspond to positive (or negative) values. See also the legend and the text.}
    \label{fig:FFeC}
\end{figure*}

Following \cite{2019A&A...631A..61T}, we consider two pairs of different FF and Componization distortion models
to identify a plausible range of possible distortions. We first consider the ionization history of \cite{2000ApJ...542..535G},
resulting in a Thomson optical depth $\tau$
that is fully consistent with 
recent {\it Planck} results, along with a fixed cut-off value $k_{max} = 100$. We coupled it with two different levels of Comptonization distortion, characterized by $u = 10^{-7}$, which is very close to that derived in 
\cite{2008MNRAS.385..404B} for the \cite{2000ApJ...542..535G} model and corresponds to an almost minimal energy injection
consistent with the current constraints on $\tau$, 
and by $u = 2 \times 10^{-6}$, a value that accounts for possible additional energy injections by a broad set of astrophysical phenomena. 

At long wavelengths, $\lambda = c / \nu  \gsim \lambda_0 = 1.5$\,cm, $y_{B}$ is well-described by a linear dependence on ${\rm log} \lambda$, $y_{B}^{\rm lin} \simeq a^{\rm lin} {\rm log} \lambda + b^{\rm lin}$,
while at $\lambda = c / \nu  \lsim \lambda_0$ a quadratic dependence, $y_{B}^{\rm quad} \simeq a^{\rm quad} {\rm log}^2 \lambda + b^{\rm quad} {\rm log} \lambda + c^{\rm quad}$,
works better. The coefficients $a^{\rm lin}$ and $b^{\rm lin}$ are given in appendix C of \cite{2014MNRAS.437.2507T}:
for the adopted model $a^{\rm lin} \simeq 3.292 \times 10^{-9}$, $b^{\rm lin} \simeq 2.070 \times 10^{-9}$.
To allow for continuous 
derivatives of $y_{B}$ also at frequencies around the transition 
between the two regimes, thus avoiding to introduce spurious oscillations in the resulting $a_{\ell,0}$, we need to properly join the two representations.
Combining them with exponential weights, 
\begin{equation}\label{eq:expweight}
y_{B} = y_{B}^{\rm lin} (1 - e^{ -(\lambda/\lambda_0)^d}) + y_{B}^{\rm quad} e^{ -(\lambda/\lambda_0)^d} \, ,
\end{equation}
\noindent
with $d=3/2$, is suitable to this purpose.
A best-fit (see table C1 of \cite{2014MNRAS.437.2507T})
gives 
$a^{\rm quad} \simeq -4.657 \times 10^{-10}$, $b^{\rm quad} \simeq 1.210 \times 10^{-9}$, $c^{\rm quad} \simeq 2.841 \times 10^{-9}$.

Larger FF distortions are expected from the integrated contribution of an ensemble of ionized halos at substantial redshifts,   
as in the model by \cite{1999ApJ...527...16O} that predicts a value of $y_{B} \sim 1.5 \times 10^{-6}$ at $\nu \sim 2$ GHz.
We then consider a second pair of models rescaling the above FF representation to $y_{B}({\rm 2\,GHz}) = 1.5 \times 10^{-6}$,
coupled with Comptonization distortions with $u = 10^{-7}$ or $2 \times 10^{-6}$.

In general, the frequency behaviour of $y_{B}$ is significantly less model dependent than its overall amplitude.
A power law representation of $y_{B}$ with a single amplitude parameter is\ adopted, for simplicity, in Sect. \ref{sec:all}.
In this approximation
\begin{equation}
y_{B} (\nu)  \simeq A_{\rm FF} \, (\nu/{\rm GHz})^{-\zeta} \, ,
\label{eq:yBpowlaw}
\end{equation}
\noindent 
where,  
assuming
$\zeta \simeq 0.15$, suitable values of $A_{\rm FF}$ at low frequencies (several\,MHz\,$\lsim \nu \lsim$\,some\,GHz)
are respectively $A_{\rm FF} \simeq 7.012 \times 10^{-9}$ and 
$1.664\times 10^{-6}$ \citep{2019A&A...631A..61T}.
Assuming the same slope but values of $A_{\rm FF}$ multiplied by a proper factor, $f$, offers a reasonable approximation 
also at 30\,GHz\,$\lsim \nu \lsim$\,100\,GHz
(on average we find $f \sim 0.578$, while it ranges between $\simeq 0.550$ and $\simeq 0.596$).
Of course, a better power law fit can be found jointly varying $A_{\rm FF}$ and $\zeta$ according to the considered frequency range.
As examples, for the first model 
at 10\,MHz\,$\lsim \nu \lsim$\,1\,GHz 
(or at 30\,GHz\,$\lsim \nu \lsim$\,100\,GHz ) 
we find $A_{\rm FF} \simeq 7.225 \times 10^{-9}$ and $\zeta \simeq 0.143$ 
(or $A_{\rm FF} \simeq 5.236 \times 10^{-9}$ and $\zeta \simeq 0.214$).
The adopted intrinsic monopole models are shown in Fig. \ref{fig:FFeC} (top-left panel) in terms of $\Delta T_{th}$. 

The results
in Fig. \ref{fig:FFeC} are derived multiplying $y_{B}$ in Eq. \eqref{eq:expweight} by a damping function, 
${\rm exp} (-(\lambda/\lambda_{1})^d) {\rm exp} (-(\lambda/\lambda_{2})^d)$, which is relevant 
at very short wavelengths ($\lambda_{1} = 0.09$\,cm and $\lambda_{2} = 0.05$\,cm, corresponding to $\simeq 333$\,GHz and 600\,GHz)
to make the results at $\nu \gsim 400$\,GHz dependent essentially only on the Comptonization term.
This does not appreciably affect the results shown in the various panels of Fig. \ref{fig:FFeC} at $\nu \lsim 400$\,GHz.
On the other hand, while a better theoretical characterization of the FF emission at very high frequencies is required for a proper estimate in this context, 
we note that at $\nu \gsim 400$\,GHz the signal associated to the CIB, discussed in Sect. \ref{sec:cib}, dominates over the other terms at any multipole.

The differences, $\Delta a_{\ell,0}$,
derived for these models
are 
positive at low frequencies, where the FF term dominates, and negative at high frequencies, where the Comptonization prevails.
The transition from the FF to the Comptonization regime, which ranges from about 3 GHz to about 300 GHz,
depends on the relative amplitude of the two parameters $y_B$ and $u$. This generalizes the result already found by \cite{2019A&A...631A..61T} for the dipole:
in particular, the transition frequency between the two regimes clearly 
increases with 
$\ell$, increasing from a maximum value of $\sim 100$\,GHz for $\ell = 1$ to 
a maximum value 
around 
$\sim 350$\,GHz
for the highest values of $\ell$.
Again, the approximate power law shape of the intrinsic monopole spectrum at low frequencies is maintained at higher multipoles. 

The ratio
between observed and intrinsic monopole
is frequency-dependent
(see top-right panel of Fig. \ref{fig:FFeC}) 
and at low frequencies, that is, below $\approx 0.1$\,GHz, $\Delta R$ can be comparable in amplitude to $|R^{\rm BB} - 1|$ or even larger, mainly depending on the level of the FF diffuse emission.

\subsection{Bose-Einstein-like distortion}
\label{sec:BE}

Bose-Einstein-like distortions can be produced by a variety of early processes, including 
unconventional heating sources, which could occur before the end of the phase of kinetic 
equilibrium between radiation and matter.
Under near-equilibrium conditions, the stationary solution of the standard Kompaneets equation including only Compton scattering
is a Bose-Einstein (BE) photon distribution function \citep{1970Ap&SS...7...20S}:
\begin{equation}\label{eq:etaBE}
    \eta_{\rm BE}= {1\over e^{x_{\rm e}+\mu} -1} \, ,
\end{equation}
\noindent
with a frequency independent chemical potential, $\mu$; here $x_{\rm e} = x / \phi (z)$,
$\phi (z) = T_{\rm e}(z)/T_r = \phi_{\rm BE}(\mu)$, 
with $T_{\rm e}(z)$ 
the electron temperature.
For mechanisms intrinsically involving a negligible photon number density production or absorption,
$\mu$ is related to the fractional energy exchanged in the plasma during the
interaction, $\Delta \varepsilon/ \varepsilon_{\rm i}$, where the subscript ${\rm i}$ denotes the process initial time.
For small distortions, $\phi_{\rm BE} \simeq (1-1.11\mu)^{-1/4}$ 
and, for an almost instantaneous process, $\mu \simeq 1.4 \Delta \varepsilon/ \varepsilon_{\rm i}$ at the end of the dissipation phase. 
Photon production processes, such as bremsstrahlung and double (or radiative) Compton emission, are particularly efficient at low frequencies,
making the chemical potential dependent on the frequency, $\mu = \mu(x)$ \citep{1970Ap&SS...7...20S,1974AZh....51.1162I,1980A&A....84..364D}. In combination with photon diffusion by Compton scattering, they tend to decrease the value of $\mu$. 

At high frequencies, $x_{\rm e} \gsim 1$, the relaxation to a BE stationary solution can be achieved for processes that have occurred
at redshifts of $z \gsim z_{\rm p}$, corresponding to a time Comptonization parameter of $y_{\rm e} \gsim y_{\rm p} \simeq 1/4$ \citep{1980A&A....84..364D,1991A&A...246...49B}, where
$y_{\rm e} = \int_1^{1+z} [t_{\rm exp}/t_{\rm C}] [d(1+z') / (1+z')]$ and where $t_{\rm exp}$ is the cosmic expansion time and 
$t_{\rm C} = [1/(n_{\rm e} \sigma_{\rm T} c)] [m_{\rm e}  c^2 /(kT_{\rm e} )]$ is the timescale for the achievement of the kinetic equilibrium,
with $n_{\rm e}$ as the density of free electrons, $m_{\rm e}$ the electron mass, and $\sigma_{\rm T}$ the Thomson cross-section.
At $z \lsim z_{\rm p}$, if the dissipation mechanism is concluded, the evolution of the photon distribution function is mainly due to photon production processes that 
significantly affect the low-frequency spectral region up to the recombination epoch. 
Aside from this effect, a longer time is needed for the photon distribution function relaxation towards the final spectrum at low frequencies, 
$x_{\rm e} \lsim 1$; it can be achieved for processes occurred at redshifts $z \gsim z_1$ corresponding to
$y_{\rm e} \gsim y_{1}$, with $y_{1} \simeq 5$ for small distortions \citep{1991A&A...246...49B}.
Here, we define, using $\mu_0,$ the high frequency asymptotic value of $\mu$ 
at $z_1$, which substantially identifies the end of the kinetic equilibrium phase.
For the above reasons, the observational constraints on the chemical potential are typically referred to $\mu_0$, the constraints on $\mu = \mu(z)$ at higher redshifts 
being derived theoretically through (semi-)analytical formulas or numerical methods (see e.g. \cite{1991ApJ...379....1B}), according to the considered problem.
The limits on $\mu(z)$ can be significantly relaxed at increasing redshifts, and the constraints on $\Delta \varepsilon/
\varepsilon_{\rm i}$ before the thermalization redshift (when even large distortions can be
erased) are then set by cosmological nucleosynthesis.
The current upper limit on $\mu_0$ is mainly derived from FIRAS data at $\lambda \lsim 1$\,cm, $|\mu_0|<9 \times 10^{-5}$ at the 95\,\% confidence level
\citep{1996ApJ...473..576F}, although jointly recovering early and late spectral distortion parameters and 
including measurements at longer wavelengths can marginally change this constraint
(see e.g. \cite{1998astro.ph..5123N}, \cite{2002MNRAS.336..592S}, \cite{2008ApJ...688...24G} and \cite{2011ApJ...734....6S}).

While a value of $\mu_0\simeq$\,few\,$\times 10^{-5}$ cannot be excluded by current data,   
the existence of much smaller BE-like distortions is predicted as a consequence of two unavoidable processes. 
The dissipation of primordial perturbations at small scales~\citep{1994ApJ...430L...5H,2012MNRAS.425.1129C},
generates a positive chemical potential with values of $\mu_0$ between $\sim 10^{-9}$ and $10^{-7}$, 
mainly depending on the shape of spectrum of the primordial scalar perturbation, a wider range being achieved in some inflation models varying 
the amplitude of primordial perturbations at very small scales \citep{chlubaal12} that are not constrained by current CMB anisotropy data.
The faster decrease of the matter temperature relative to the radiation temperature in an expanding Universe
generates, instead, a negative chemical potential, 
because of the interaction of CMB photons with colder electrons, 
with an absolute value $\simeq 3 \times 10^{-9}$  \citep{2012MNRAS.419.1294C,sunyaevkhatri2013}.

\begin{figure*}[ht!]
\centering
         \includegraphics[width=18.5cm]{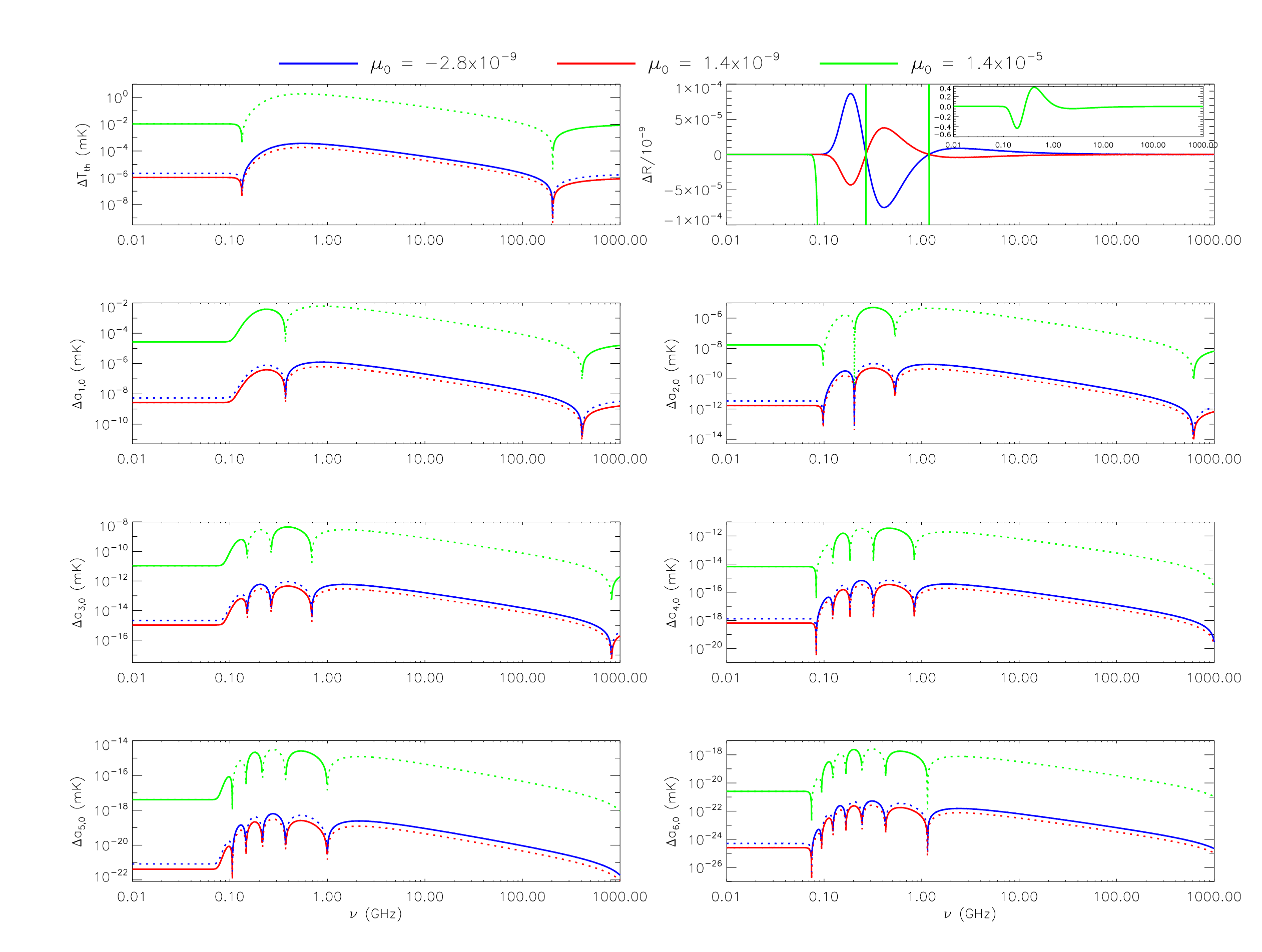}
    \caption{$\Delta T_{th}$, $\Delta R,$ and $\Delta a_{\ell,0}$ from $\ell = 1$ to $\ell_{\rm max} = 6$ for the considered BE-like distortion models. The inset in the top-right panel shows separately the case with the highest value of $\mu_0$, which cannot be appreciated
    in the main panel as it is targeted to much lower values of $\mu_0$, in order to appreciate the extremely similar spectral shape of the red and green lines.
      Except in the top-right panel where $\Delta R$ is shown in linear scale, solid lines (or dots) correspond to positive (or negative) values. See also the legend and the text.}
    \label{fig:BE}
\end{figure*}

According to the above discussion, we consider just three values of $\mu_0$: $1.4 \times 10^{-9}$, $1.4 \times 10^{-5}$, and $-2.8 \times 10^{-9}$. 
We adopt here an updated implementation of the semi-analytical representation of BE-like distortions, suitable also at low frequencies, as proposed by \cite{1980A&A....84..364D}
and described in detail in \cite{1995A&A...303..323B}.
We assume a cold DM plus cosmological constant ($\Lambda$CDM) model with the set of parameters 
based on the last {\it Planck} data release and derived in \cite{2020A&A...641A...6P} including 
CMB power spectra in combination with CMB lensing reconstruction
(see their Table 2, column labelled 'TT, TE, EE+lowE+lensing').
We adopt a universe with a Hubble constant, $H_0 = 67.36$\,km/s/Mpc, cosmological constant (or dark energy) and non-relativistic matter density parameters $\Omega_\Lambda = 0.6847$, $\Omega_{\rm m} = 0.3153$, $\Omega_{\rm b} [H_0 / (100\, {\rm km/s/Mpc})]^2 = 0.02237$ (implying a baryon density. $\Omega_{\rm b} = 0.0493017$), and, according to the standard model, an effective number of relativistic neutrinos $N_{\rm eff} = 3.046$. In principle, the fine accounting of the relativistic neutrinos contribution to the expansion rate in the presence of an energy injection should  also require the specification of the heating redshift, $z_{\rm h}$, particularly for $z_{\rm h} \gg z_1$ when $\mu$ could be significantly larger than $\mu_0$ \citep{1991A&A...246...49B}:
for simplicity, we treat this aspect in numerical estimates as in the case equivalent to $z_{\rm h} \simeq z_1$. We also assume $\phi (z_{\rm h}) \simeq \phi (z_1) = \phi_{\rm BE}(\mu_0)$. 
The cosmic expansion time, $t_{\rm exp}$, and the relevant rates depend on these parameters, that play the major role in determining the spectrum shape.
We compute the bremsstrahlung term according to \cite{1961ApJS....6..167K}, \cite{1979rpa..book.....R} and \cite{1991A&A...246...49B} but using, 
in its range of validity, the polynomial fitting formula for the non-relativistic exact Gaunt factor derived by \cite{2000ApJS..128..125I} (see also \cite{2020MNRAS.492..177C} for recent improvements). 
We separately compute the contributions from ionized hydrogen ($H^+$) and helium ($He^{++}$ and $He^+$), counting accordingly the overall fraction of free electrons,
given the helium mass fraction ($f_{He}=0.2454$). We calculate the double Compton rate in the elastic limit according to 
\cite{1981ApJ...244..392L} and \cite{1981MNRAS.194..439T}, and using the cross-section by \cite{1984ApJ...285..275G}. 

Considering the relevance of double Compton 
at high redshifts, we also include the correction factor, $C_{\rm mr}$, for mildly relativistic thermal plasma in the soft photon limit, $C_{\rm mr} \simeq 1/[1+14.6 k T_r / (m_{\rm e} c^2)]$,  
introduced by \cite{2007A&A...468..785C}. In Appendix \ref{app:gdc_integ}, we provide a fitting formula that, in the limit of very small distortions, 
can be used to compute the double Compton Gaunt factor at a precision level better than $\simeq 0.1-0.2$\,\% also at $x \gsim 1$.
Finally, replacing the simple approximation of full ionization up to the hydrogen recombination with the introduction of the redshifts 
($\sim 6 \times 10^3$ and $\sim 2 \times 10^3$) at which $He^{++}$ and $He^+$ 
disappear, resulting into a $\sim$ two-steps helium recombination, introduces only a small correction ($\lsim 0.2$\,\%) in the final spectrum computation. 
The above details enter in the computation of the key redshifts $z_1$ and $z_{\rm p}$ (respectively, $\simeq 5.38 \times 10^5$ and $\simeq 1.21 \times 10^5$ with the adopted parameters), in the frequency-dependent optical depth of the universe for absorption, 
$y_{\rm abs} = \int_1^{1+z} [t_{\rm exp}/t_{\rm abs}] [d(1+z') / (1+z')]$ (see \cite{1995A&A...303..323B}), and in the characteristic dimensionless frequency, $x_{\rm c}$,
which quantifies the low-frequency damping of the chemical potential, $\mu(x) = \mu_0 \, {\rm exp}\, (- x_{\rm c} / x_{\rm e})$, $x_{\rm c}$ being defined by 
$t_{\rm abs} (z_1) = t_{\rm C} (z_1)$, where $t_{\rm abs}$ is the absorption timescale for photon production processes. 
For our purposes, a simple Gaussian quadrature scheme is accurate and efficient enough for computing the relevant integrals over $z$ (we find advantageous to work with a logarithmic integration variable), while 
the NAG routine {\it D01AJF} can provide a better performance. The Brent's method \citep{press1992numerical}
is suitable to solve the equation for $x_{\rm c}$ (found to be $\simeq 4.86 \times 10^{-3}$ with the adopted parameters), given bracketing guesses based on the simple low frequency limit approximation.

The intrinsic monopole spectra so obtained for the three adopted values of $\mu_0$ are shown in Fig. \ref{fig:BE} (top-left panel) in terms of $\Delta T_{th}$. 
As expected, the amplitude of $|\Delta T_{th}|$ is proportional to $|\mu_0|$ and, for $\mu_0 < 0$, we find a spectrum shape, $\Delta T_{th}$, 
opposite in sign with respect to the case $\mu_0 > 0$. 
It is interesting to note the plateau at extremely low frequencies and the presence of two characteristic changes of sign in $\Delta T_{th}$, corresponding to the well-known excess (or decrement) of signal of the BE-like spectrum with respect to the blackbody at temperature $T_0$ at low and high frequencies and the remarkable decrement (or excess) at intermediate frequencies for positive (or negative) 
values of $\mu_0$. These sign changes also appear in the differences $\Delta a_{\ell,0}$
(see Fig. \ref{fig:BE}), but at two characteristic frequencies significantly
increasing at increasing $\ell$. We note that the sign change at the higher of the two characteristic frequencies occurs at $\nu > 1$\,THz for $\ell \ge 4$ and just for this reason it disappears in the corresponding plots. 
In addition, two further sign changes appear at each increase of an even multipole. They are located at frequencies between the smaller of the two above characteristic frequencies and the plateau at extremely low frequencies. 
Again, this pattern of sign changes is symmetric with respect to the sign of $\mu_0$. These are remarkable features of the BE-like spectrum:
they are almost independent of the value of $\mu_0$, while their behaviour at low frequencies depend on the underlying cosmological parameters.  

In Fig. \ref{fig:BE}, $\Delta R$ is displayed in the top-right panel. 
Over the whole frequency range, 
it is characterised by a
module 
proportional to $|\mu_0|$ 
and much smaller than $|R^{\rm BB} - 1|$ even for values of $|\mu_0|$ not far from FIRAS limits.
Again,
$\Delta R$ depends on frequency. For BE-like distortions, the shape of $\Delta R$ is again symmetric with respect to the sign of $\mu_0$,
with a well-defined maximum (or minimum) located between two minima (or two maxima), clearly defined at lower frequencies and less pronounced at higher frequencies, for $\mu_0 > 0$ (or $\mu_0 < 0$).

\begin{figure*}[ht!]
\centering
                  \includegraphics[width=18.5cm]{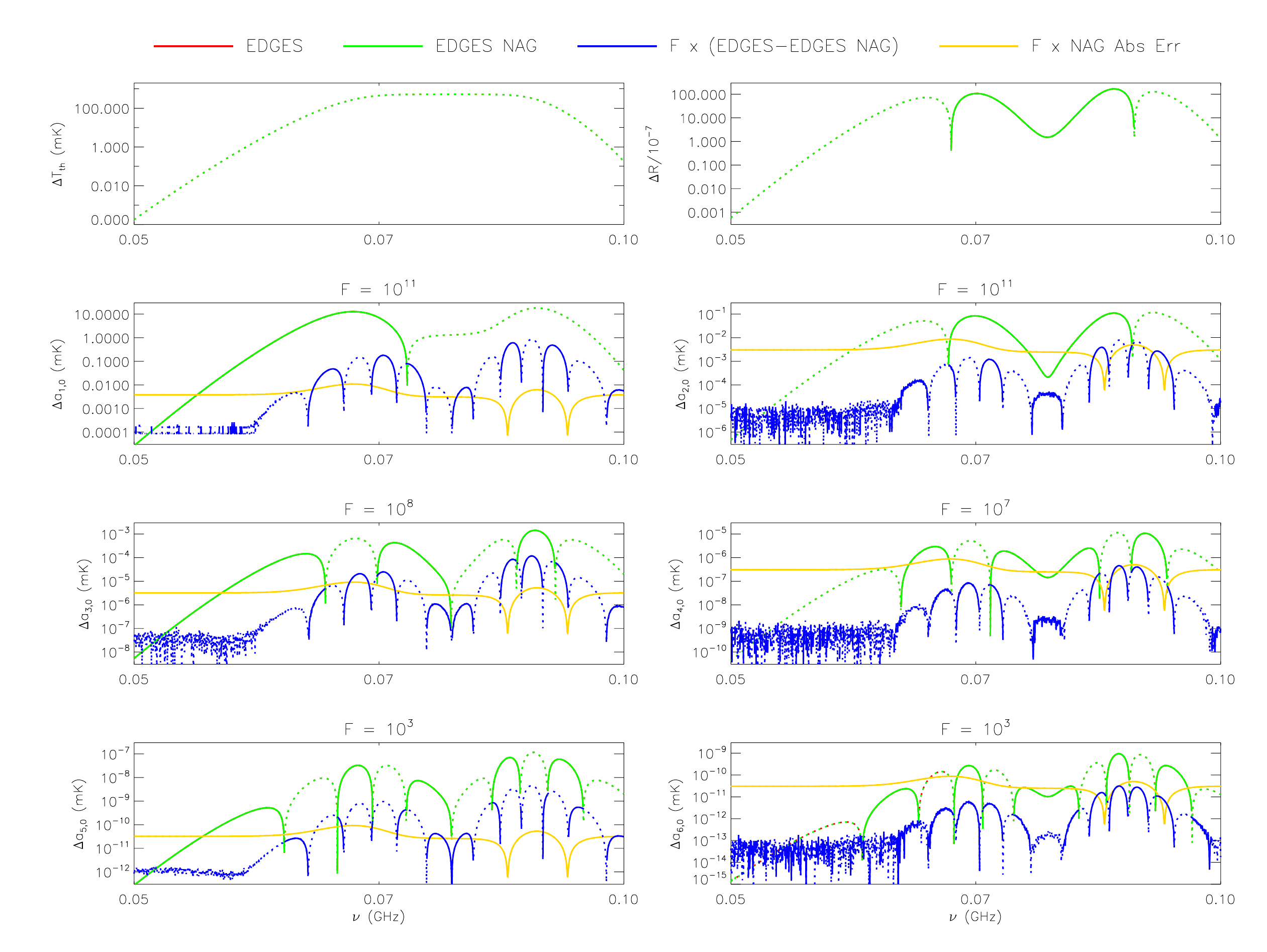}
    \caption{$\Delta T_{th}$, $\Delta R$ and $\Delta a_{\ell,0}$ from $\ell = 1$ to $\ell_{\rm max} = 6$ for the EDGES profile of redshifted 21cm line (summed in intensity with the CMB blackbody).
    Solid lines (or dots) correspond to positive (or negative) values. Red and green lines are essentially superimposed at any $\ell$.
    Their difference, multiplied by a factor F to have values compatible with the adopted range, is displayed by the blue lines.
    Yellow lines refer to the nominal integration error quoted by the routine {\it D01AJF}, again multiplied by the factor F. See also the legend and the text.}
    \label{fig:EDGES}
\end{figure*}

\subsection{21cm redshifted line}
\label{sec:21cm}

The 21cm line corresponds to the spin-flip transition in the ground state of neutral hydrogen. This signal is described as the offset of the 21cm brightness (i.e. antenna) temperature from the 
background temperature, $T_{\rm back}$, along the observed line of sight at a frequency $\nu$ that, because of cosmic expansion, is related to the rest frame frequency, $\nu_{21 {\rm cm}} = c / (21 {\rm cm})$, 
by $\nu = \nu_{21 {\rm cm}} / (1+z)$. 
$T_{\rm back}$ is usually assumed equal to $T_r$ but, 
in general, it could include potential distortions and other radiation backgrounds.
It depends on the evolution of the gas spin temperature, $T_{S}$, which represents the excitation temperature of the 21cm transition, 
of the fraction of neutral hydrogen, 
on the Hubble function, $H(z)$, on $\Omega_{m}$, 
on the matter density contrast, and on the comoving gradient of the line-of-sight component of the comoving velocity \citep{2006PhR...433..181F}.
If $T_{S} < T_{\rm back}$ (or $T_{S} > T_{\rm back}$), the gas is seen in absorption (or in emission). 
Since the signal detected at a given frequency corresponds
to a specific redshift, 
the 21cm line provides a
tomographic view of the cosmic evolution. A rich set of 21cm redshifted line models has been studied in \cite{2017MNRAS.472.1915C}, resulting in an wide envelope of predictions for $T_{\rm ant}^{\rm 21cm} (\nu)$.

In this work, we
consider only the 
pronounced absorption profile, with an almost symmetric U-shape centred at $(78 \pm 1)$ MHz, which was recently found by \cite{EDGESobs2018Nature}
based on an analysis of the data from the Experiment to Detect the Global EoR Signature (EDGES).
The absorption feature was found to have an amplitude of $0.5^{+0.5}_{-0.2}$\,K
and a spread of the profile with a full width at half-maximum of $19^{+4}_{-2}$ MHz. 
These data support the presence of an ionizing background by 180 million years after the Big Bang
and a phase of gas heating above the radiation temperature less than 100 million years later \citep{EDGESobs2018Nature}. 
The explanation of the 21cm redshifted line signal found by EDGES might require
a substantial cooling of the IGM gas 
or
an additional high-redshift extragalactic radio background, 
or a combination of them (see e.g. the references at the beginning of Sect. \ref{sec:NE}).
The authors provide a suitable analytical representation of the EDGES absorption profile in terms of a flattened Gaussian characterized
by a set of best-fit parameters, and we adopt here their expression of $T_{\rm ant}^{\rm 21cm} (\nu)$. 

The signals considered in Sects. \ref{sec:NE}--\ref{sec:BE}, referring to intrinsic CMB spectral distortions, already include the contribution of the unperturbed CMB spectrum.
In this section (as well as in Sects. \ref{sec:exback_radio}--\ref{sec:cib}), we are considering signals that are superimposed onto the CMB, assumed without spectral distortions,
and we then add (in terms of $\eta$) the CMB blackbody at the current temperature $T_0$ to $T_{\rm ant}^{\rm 21cm} (\nu)$ to construct the global signal to be studied
as in previous sections.  

We compute the coefficients $a_{\ell,0} (\nu,\beta)$ with the method described in Sect. \ref{sec:sol7}
and, for comparison, also on the basis of Eq. \eqref{eq:harminv}, as in Sect. \ref{sec:NE}. The results are shown in Fig. \ref{fig:EDGES}.

We note the consistency between the results based on the integral given by Eq. \eqref{eq:harminv}
and 
our approach:
the agreement is excellent up to $\ell_{\rm max} = 6$ (the green and red the lines are indistinguishable).
Again, their differences are compatible with a combination of higher order terms, that is, beyond $\ell = 6$, and integration errors, that are
only present in the numerical results. 
The two types of differences clearly appear, respectively, where the signal is higher and the relative integration error is lower, 
and where the signal is lower and the relative integration error is higher. The latter point is also evident from the comparison with the 
nominal integration error quoted by the routine {\it D01AJF}. Remarkably, except for the numerical integration uncertainty,
the spectral shape of the differences is very similar for all the odd,
as well as 
the even multipoles
because they mainly come from the contribution from $\ell = 7$ and from $\ell = 8$, respectively.
In Appendix \ref{app:beta_amplified}, we repeat this 
analysis 
adopting a much larger value of $\beta$, that implies much larger signals, relatively higher contributions from higher multipoles and
relatively lower numerical integration errors: the result clearly supports the above interpretation.

It is interesting to note the complexity, increasing with $\ell$, of the features displayed in Fig. \ref{fig:EDGES}: they 
include the alternation of increasing and decreasing behaviours, the number of relative minima and maxima, and the changes of the sign of $\Delta a_{\ell,0}$. 

Here, the result for $\Delta a_{1,0}$ corrects the dipole spectrum published in \cite{2019A&A...631A..61T}, expressed there in terms of $\Delta_{0,\pi/2} T_{\rm th}^{\rm BB/dist}$ (see Eq. \eqref{eq:DeltaTtherm_alm}), where, for EDGES, the profile $T_{\rm ant}^{\rm 21cm} (\nu)$ was accounted in equivalent thermodynamic (not in antenna) temperature. This mere oversight significantly affected only the very small values of $\Delta_{0,\pi/2} T_{\rm th}^{\rm BB/dist}$, 
making 
the wings
a bit 
steeper 
they are.

We note that the frequency, right above 70 MHz, corresponding to the change of sign of $\Delta a_{1,0}$ is shifted at slightly larger values in the case of $\Delta a_{3,0}$ and $\Delta a_{5,0}$, 
while for the even multipoles it corresponds to a well-defined minimum that falls in the middle of a positive interval of the $\Delta a_{\ell,0}$ profile. 
The size of this frequency interval decreases as $\ell$ increases; the same holds, in the case of odd multipoles,
for the interval identified by the two signs changes around the above minimum.
Also, the increasing with $\ell$ of the number of sign changes in $\Delta a_{\ell,0}$ implies that, at frequencies
outside the above interval, the size of each frequency range with unchanged sign of $\Delta a_{\ell,0}$ decreases with $\ell$.
These are remarkable features of the considered model.

Remarkably, 
$\Delta R$ (see top-right panel of Fig. \ref{fig:EDGES}), 
again exhibits a frequency dependence
related to the assumed intrinsic monopole spectrum: it is positive in the inner frequency range, almost corresponding to the plateau of the absorption feature, 
and negative in the profile wings. 
In particular, $|\Delta R|$ is comparable to or, typically, greater than $|R^{\rm BB} - 1|$, 
across most of the relevant frequency range.

\subsection{Extragalactic radio background}
\label{sec:exback_radio}

\begin{figure*}[ht!]
\centering
         \includegraphics[width=18.5cm]{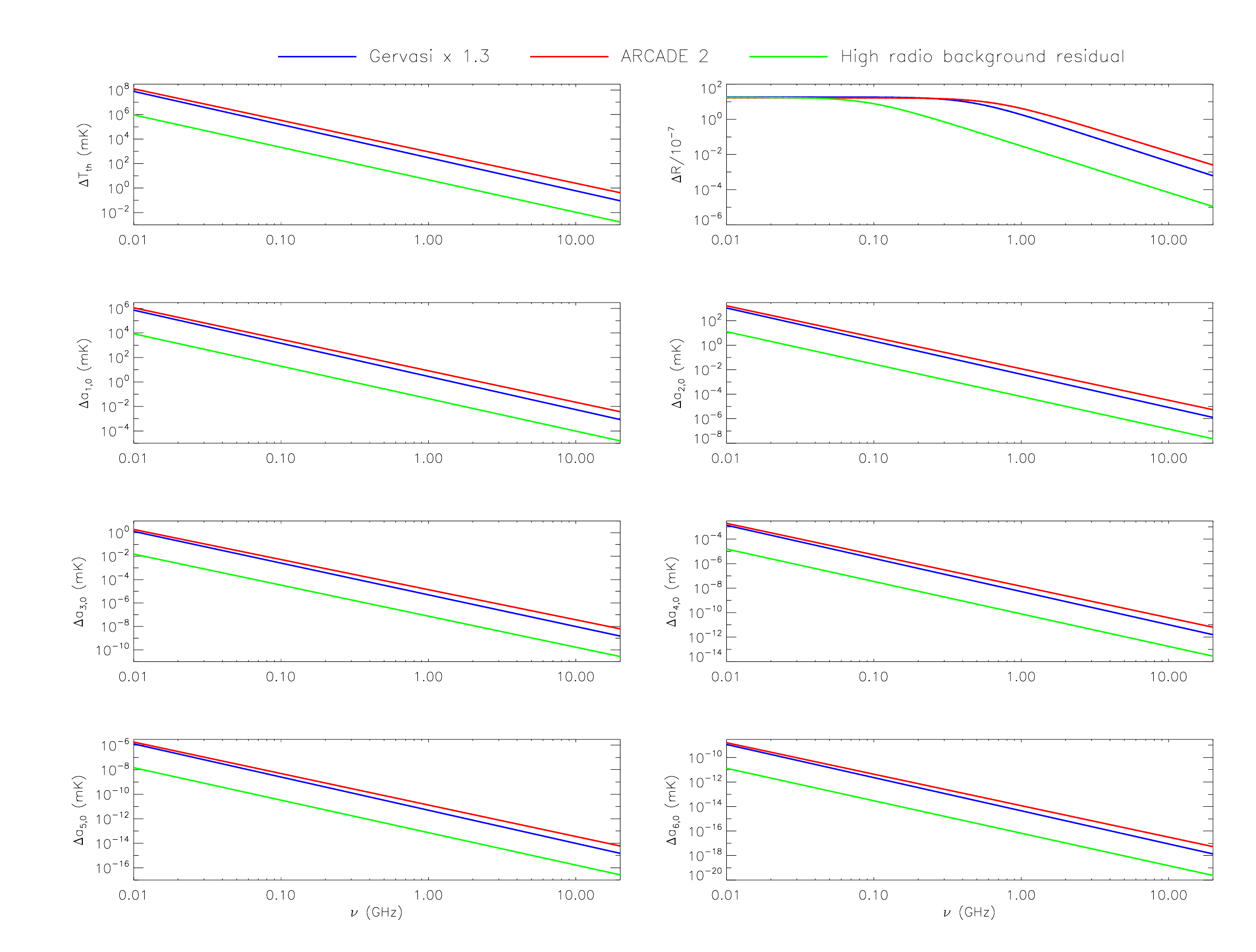}
    \caption{$\Delta T_{th}$, $\Delta R,$ and $\Delta a_{\ell,0}$ from $\ell = 1$ to $\ell_{\rm max} = 6$ for the two considered 
extragalactic radio background models and an estimate of extragalactic radio source background signal for
an assumption of source contribution subtraction (summed in intensity with the CMB blackbody).
    Solid lines (or dots) correspond to positive (or negative) values. See also the legend and the text.}
    \label{fig:BackBB}
\end{figure*}

An important radio background is produced by extragalactic sources.
Differently from most of the signals discussed in previous sections, which are of intrinsically diffuse origin,
this background, as well as those discussed in Sects. \ref{sec:exback_mm} and \ref{sec:cib}, results from the integrated contribution of discrete sources.
With galactic surveys able to reach increasingly deeper flux density levels, it is possible to resolve a large fraction of this background.
In spite of this, an observationally diffuse residual background comes from faint sources below the survey detection limits. 
Other than intrinsically interesting, this extragalactic background needs to be accurately known in order to
understand the reionization imprints correctly. Remarkably, these classes of signals may also be tightly related. 
A notable extragalactic radio background is evident in, for example, the radio data by \citep{Dowell:2018} and it was proposed by \cite{2011ApJ...734....6S} to explain the 
signal excess claimed by ARCADE 2. 

We exploit several simple analytical representations of the extragalactic radio background.
According to \cite{2019A&A...631A..61T}, we assume the best-fit power-law model,
\begin{equation}
T_{\rm ant}^{\rm Back} (\nu)  \simeq 18.4 \, {\rm K} \, (\nu/0.31 {\rm GHz})^{-2.57} \, ,
\label{eq:backseiff}
\end{equation}
\noindent
by \cite{2011ApJ...734....6S}.

A careful analysis and prediction of the extragalactic source radio background between 0.151 and 8.44 GHz, also including different source detection thresholds, was carried out by \cite{2008ApJ...682..223G}. We consider their best-fit single power-law model for the extragalactic source background signal, 
multiplied by a factor of 1.3 in order to approximately account for a larger contribution
that is, ultimately, likely ascribed to the emerging of star-forming galaxies and radio-quiet active galactic nuclei at fainter flux densities.
Indeed, the Lockman Hole Project and deep Low Frequency Array (LOFAR)
imaging of the Bo\"otes field support a certain flattening of differential number counts, $N'(\nu)$, at 1.4 GHz below $\approx 100$ $\mu$Jy \citep{2018MNRAS.481.4548P} and at 0.15 GHz below $\approx 1$ mJy \citep{2018A&A...620A..74R}. 
This may suggest an increase in $N'(\nu)$ of a factor of $\sim 2$ with respect to the estimate of \cite{2008ApJ...682..223G}
at the faint flux densities and 
a $\sim 30$\,\% increase in the extragalactic radio background,
which is proportional to $\int_{S_{\rm min}}^{S_{\rm max}} S N'(\nu) dS$.
We
then adopt
\begin{equation}
T_{\rm ant}^{\rm Back} (\nu)  \simeq 1.3 \times 0.88 \, {\rm K} \, (\nu/0.61 {\rm GHz})^{-2.707} \, .
\label{eq:backgerv}
\end{equation}

\cite{2008ApJ...682..223G} provided also an empirical analytical fit function of $N'(\nu)$ that can be used to estimate 
the remaining residual extragalactic radio background when a certain source detection threshold, $S_{\rm max}$, is assumed. 
According to \cite{2019A&A...631A..61T}, we exploit their differential number counts assuming 
$S_{\rm max} = 50$ nJy, 
which almost corresponds to typical detection limits of the ultra deep reference continuum surveys planned for the Square Kilometre Array 
(SKA)
\citep{2015aska.confE..67P}.
We consider the above factor of $\sim 2$ to be applied to $N'(\nu)$ by \citep{2008ApJ...682..223G} at faint flux densities
(we then label this case as a 'High-radio background residual') 
and fit the results found in the frequency range considered by the authors to find the corresponding estimate of the remaining residual extragalactic radio background,
\begin{equation}
T_{\rm ant}^{\rm Back} (\nu)  \simeq A \, (\nu/{\rm GHz})^{-2.65} \, ,
\label{eq:backgervres}
\end{equation}
\noindent
with $A \simeq 4.7$\,mK. Different choices of $S_{\rm max}$ mainly reflects into the value of $A$.

As in previous section, we add (in terms of $\eta$) the CMB blackbody at the current temperature $T_0$ to these radio background models to construct the global signal.
The intrinsic monopole spectra 
are shown in Fig. \ref{fig:BackBB} in terms of $\Delta T_{th}$ (top-left panel) with the derived coefficients $a_{\ell,0} (\nu,\beta)$. 

The typical power law shapes of the considered intrinsic monopole spectra, after the subtraction of the blackbody at the present temperature $T_0$,
are also maintained at higher multipoles and the same holds for their relative amplitudes, as already noted in \cite{2019A&A...631A..61T} for the dipole.

Again, $\Delta R$
exhibits a frequency dependence that is related to the assumed intrinsic monopole spectrum.
At low frequencies, below $\sim 1$\,GHz,
$\Delta R$ can have an amplitude comparable to $|R^{\rm BB} - 1|$ or even larger. 

\subsection{Extragalactic millimetre background}
\label{sec:exback_mm}

\begin{figure*}[ht!]
\centering
         \includegraphics[width=18.5cm]{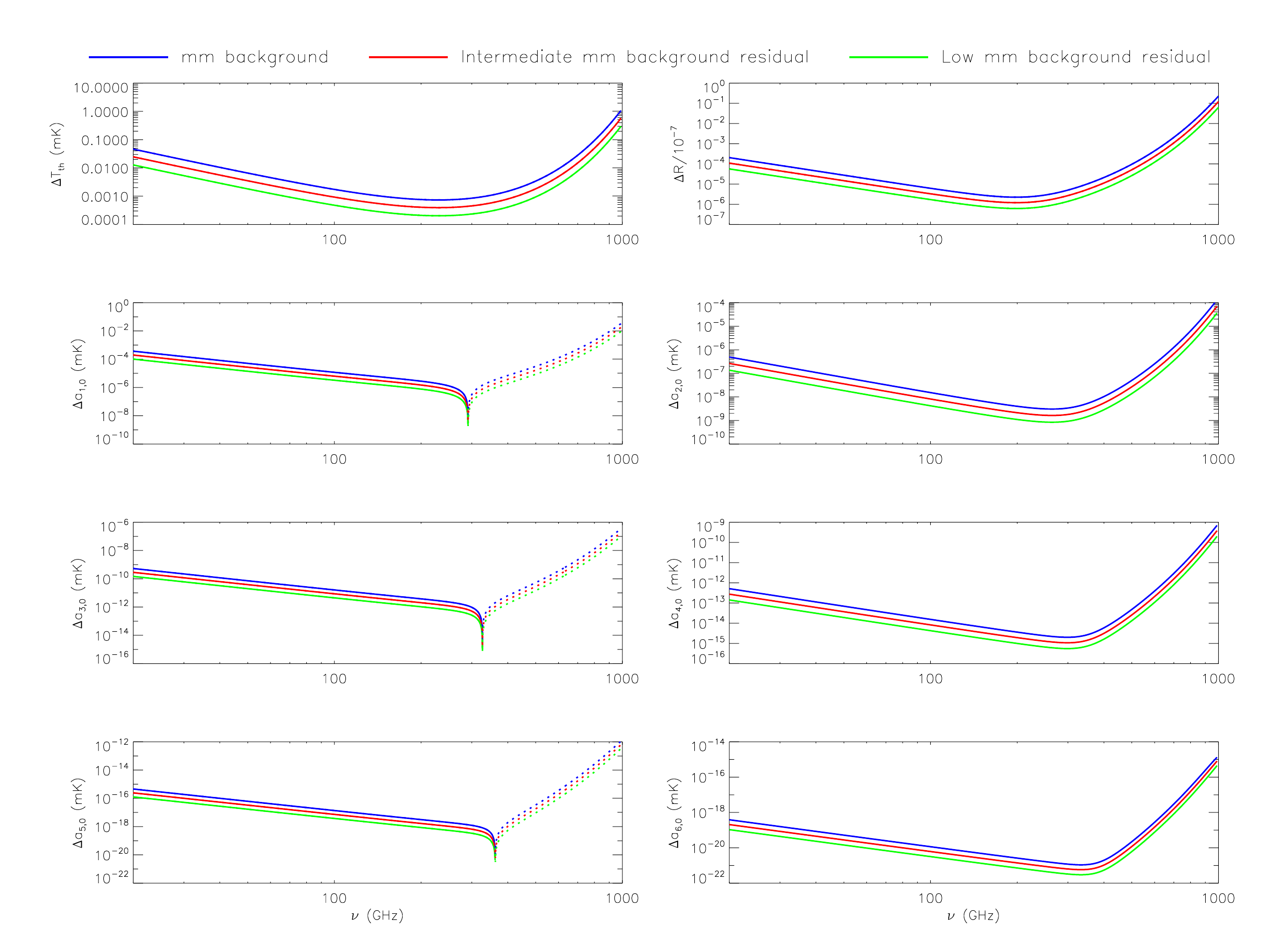}
    \caption{$\Delta T_{th}$, $\Delta R,$ and $\Delta a_{\ell,0}$ from $\ell = 1$ to $\ell_{\rm max} = 6$ for the considered 
millimetre background model from extragalactic radio sources and estimates of extragalactic source millimetre background signal for
two assumptions of source contribution subtraction (summed in intensity with the CMB blackbody).
    Solid lines (or dots) correspond to positive (or negative) values. See also the legend and the text.}
    \label{fig:BackMM}
\end{figure*}

The extragalactic radio source populations that mainly contribute to the radio and to the millimetre background are very different.
Whereas steep-spectrum radio sources, particularly at high flux densities, are most important
at radio frequencies,  
extragalactic compact sources with an almost flat, or possibly inverted, spectrum (primarily blazars, flat spectrum radio quasars, 
and BL Lacertae sources, 
where are not fully considered in Sect. \ref{sec:exback_radio}) become increasingly relevant at wavelengths 
shorter than a few centimetres.
They can be directly extracted by analyzing CMB maps. The products \citep{2013A&A...550A.133P,2016A&A...594A..26P,2016A&A...596A.106P,2018A&A...619A..94P} from the {\it Planck} mission, complemented by the 
available ground-based data (see e.g. \cite{2013ApJ...779...61M}) and, in the far-IR, via Herschel (see e.g. \cite{2013MNRAS.430.1566L}) observations, 
provide crucial information for the characterization of their number counts \citep{2015JCAP...06..018D} (see also \cite{2005A&A...431..893D} and \cite{2011A&A...533A..57T}),
while a substantial progress at fainter flux densities is expected from the next-generation space missions and deeper multi-frequency ground-based surveys (see e.g. \cite{2018JCAP...04..020D} and references therein).

\begin{figure*}[ht!]
\centering
         \includegraphics[width=18.5cm]{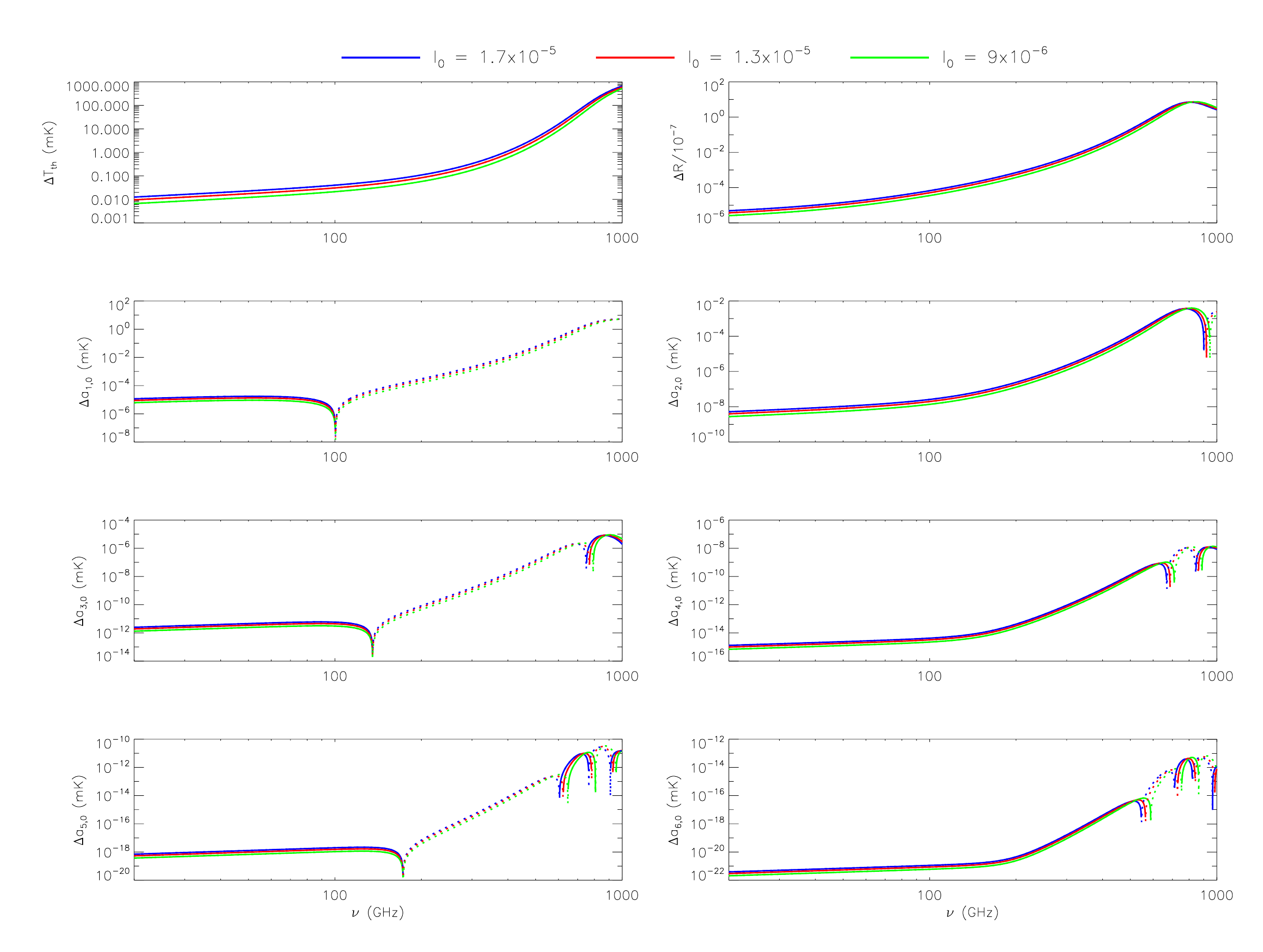}
    \caption{$\Delta T_{th}$, $\Delta R,$ and $\Delta a_{\ell,0}$ from $\ell = 1$ to $\ell_{\rm max} = 6$ for the considered CIB models (summed in intensity with the CMB blackbody).
    Solid lines (or dots) correspond to positive (or negative) values. See also the legend and the text.}
    \label{fig:CIB}
\end{figure*}

We exploit their differential number counts, $N'(\nu)$, including both steep and flat spectrum sources,
to estimate the corresponding millimetre background, $\propto \int_{S_{\rm min}}^{S_{\rm max}} S N'(\nu) dS$, for which we find a power-law approximation, as 
in Sect. \ref{sec:exback_radio}. We simply extrapolate  the power law behaviours of the differential number counts at the lowest available flux density ranges to fainter flux densities.
The (currently uncertain) characterization of $N'(\nu)$ at much fainter flux densities is relatively less relevant 
for the estimate of the global background than for the estimate of the residual millimetre background derived given a certain source detection threshold $S_{\rm max}$. 
We consider two assumptions of $S_{\rm max}$, namely:\ 1) 10 mJy (labeled as 'Intermediate millimetre background residual'),
a value that is similar to those
typically considered for next-generation space missions; 
and 2) 100\,$\mu$Jy (labelled as 'Low millimetre background residual'), 
representing
an estimate of the potential improvement achievable with SKA while considering frequency extrapolation uncertainties from deep radio surveys to millimetre bands.

As expected due to the spectral shapes of the sources that are more relevant at millimetre wavelengths, we find a background spectrum slightly flatter 
in the millimetre than in the radio, with slopes 
(of the spectrum expressed in antenna temperature) between $\simeq -2.10$ to $-2.62$, depending on the specific number counts model (with steeper spectra found for the number counts by
\cite{2011A&A...533A..57T}) and also on the adopted source detection threshold. 

For current numerical estimates we consider, for simplicity, a single slope 
based on
number counts consistent with {\it Planck} results from 30 GHz to 857 GHz. We adopt:
\begin{equation}
T_{\rm ant}^{\rm mm} (\nu)  \simeq A \, (\nu/{\rm GHz})^{-2.19} \, ,
\label{eq:backmm}
\end{equation}
\noindent
with $A \simeq 32.4$\,mK for the millimetre background spectrum and $\simeq 17.3$\,mK or $\simeq 8.93$\,mK for residual millimetre background derived for $S_{\rm max} = 10$ mJy or $S_{\rm max} = 100$ nJy, respectively. We construct the global signal by adding (in terms of $\eta$) the CMB blackbody at the current temperature $T_0$. 

Our results 
are shown in Fig. \ref{fig:BackMM}. The flattening of $\Delta T_{th}$ at $\nu \sim 200$\,GHz and its increasing at higher frequencies 
are due to
the representation in terms of equivalent thermodynamic temperature. We note the different behaviours of $\Delta a_{\ell,0}$ at odd and even multipoles, and the (minimal) increase with $\ell$ of the frequency where for odd $\ell$, the change of sign of $\Delta a_{\ell,0}$ occurs,
or for even $\ell$, the minimum of $\Delta a_{\ell,0}$ is located.

Figure \ref{fig:BackMM} reports the coefficients $a_{\ell,0} (\nu,\beta)$ for $\ell = 1, 6$, expressed in terms of $\Delta a_{\ell,0}$, derived using the solutions given in Sect. \ref{sec:sol7}.
The typical power law shapes of the considered intrinsic monopole spectra, after the subtraction of the blackbody at the present temperature, $T_0$, that are evident in the figure at low frequencies, 
are kept also at higher multipoles and the same holds for their relative amplitudes, as already noted in \cite{2019A&A...631A..61T} for the dipole.

$\Delta R$ exhibits a dependence on the frequency as well as on the assumed intrinsic monopole spectrum: 
it is positive and its frequency shape and minimum location 
are in line with the behaviours of $\Delta a_{\ell,0}$ at even multipoles.
On the other hand, the values of $|\Delta R|$ are less than $|R^{\rm BB} - 1|$ even at the 
highest frequencies.

Finally, we note that, despite the fact that the estimate of the extragalactic radio source background in the radio and at millimetre wavelengths, approximated respectively by 
Eq. \eqref{eq:backgerv} and Eq. \eqref{eq:backmm}, are based on different models,
the results found at $\sim 20$\,GHz, a frequency in the middle between the maximum and the minimum frequency of the approximations elaborated in the previous section and in this one, agree within a factor of two or better,
as shown by the comparison between Figs. \ref{fig:BackBB} and \ref{fig:BackMM}.

\begin{figure*}[ht!]
\centering
         \includegraphics[width=18.5cm]{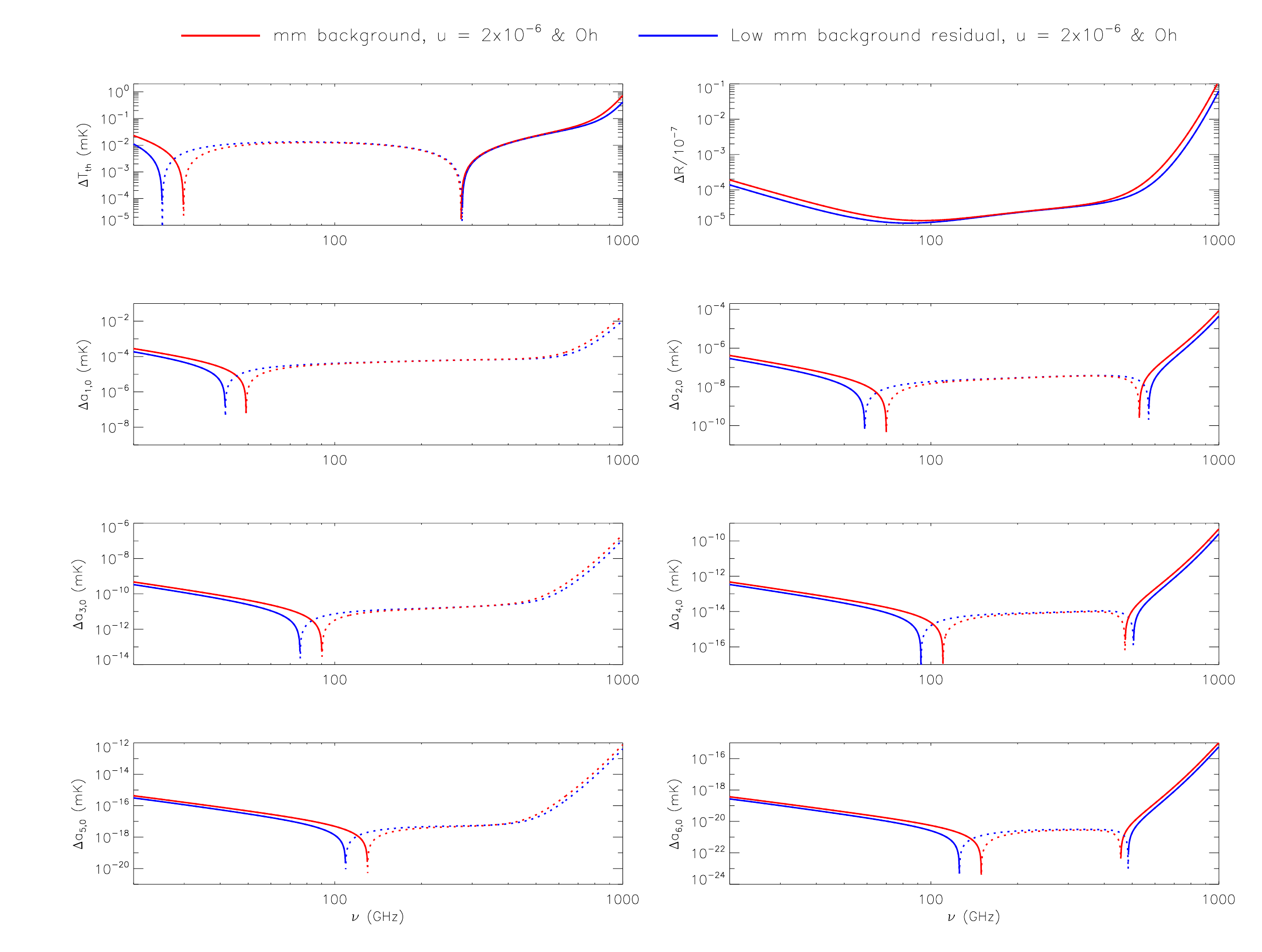}
    \caption{$\Delta T_{th}$, $\Delta R,$ and $\Delta a_{\ell,0}$ from $\ell = 1$ to $\ell_{\rm max} = 6$ for the sum of 
    the considered millimetre background model from extragalactic radio sources or an estimate of its residual signal for
an assumption of source detection threshold plus 
    a combined Comptonization and diffuse FF distortion model. Solid lines (or dots) correspond to positive (or negative) values. See also the legend and the text.}
    \label{fig:FFBackMM}
\end{figure*}

\begin{figure*}[ht!]
\centering
         \includegraphics[width=18.5cm]{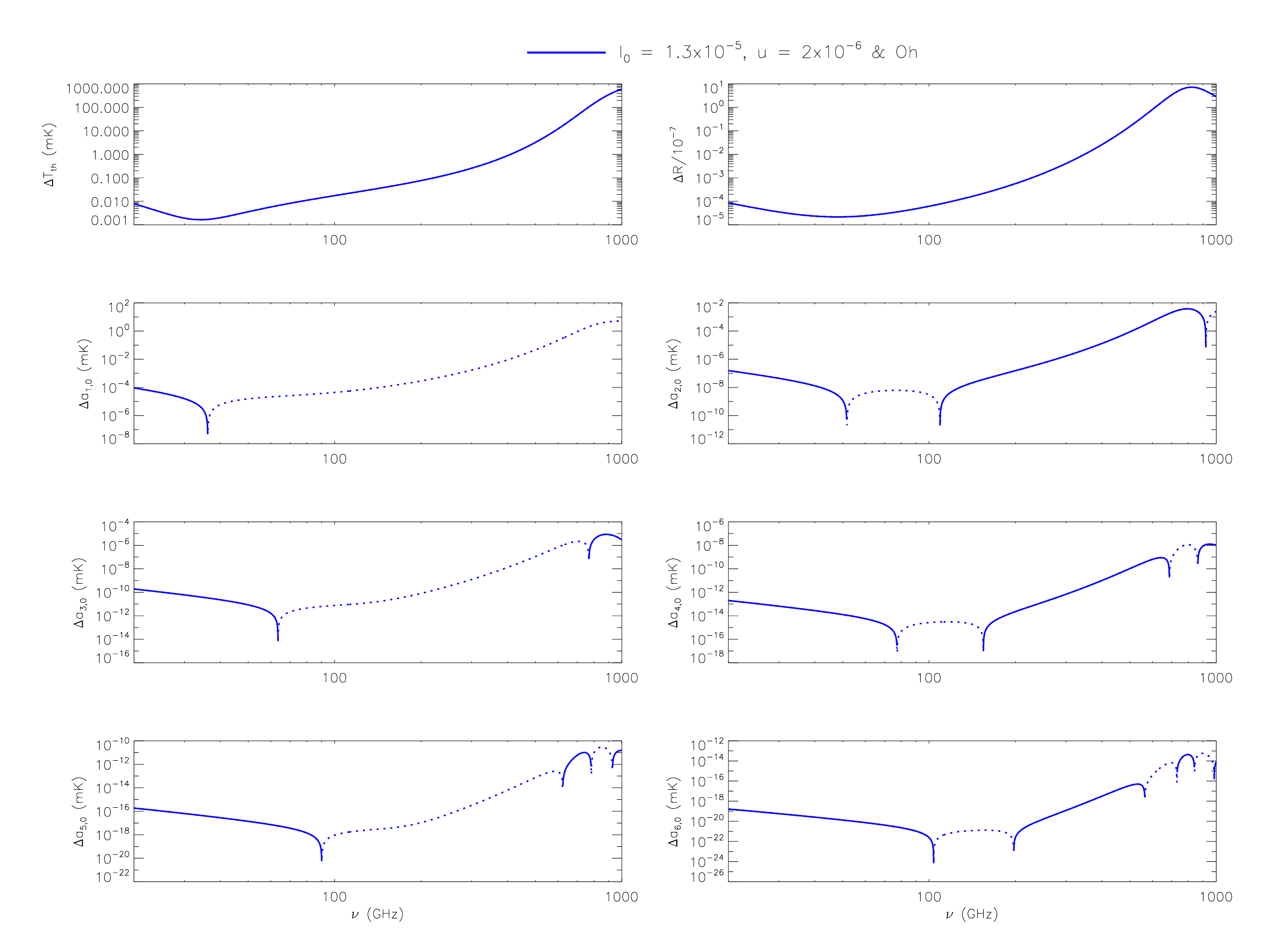}
    \caption{$\Delta T_{th}$, $\Delta R,$ and $\Delta a_{\ell,0}$ from $\ell = 1$ to $\ell_{\rm max} = 6$ for the sum of considered best-fit CIB model plus 
    a combined Comptonization and diffuse FF distortion model. Solid lines (or dots) correspond to positive (or negative) values. See also the legend and the text.}
    \label{fig:CIBFF}
\end{figure*}

\subsection{Cosmic infrared background}
\label{sec:cib}

A precise analysis of CIB spectrum, which is still not well known, can provide a better understanding of the dust-obscured star-formation phase of galaxy evolution.

In spite of its absolute calibration precision of 0.57 mK, 
the FIRAS characterization of CIB amplitude and shape still presents a substantial uncertainty. According to \cite{1998ApJ...508..123F},
a suitable analytic representation in terms of photon distribution function of the CIB spectrum at the present time, can be expressed by 
\begin{align}\label{eq:eta_CIB}
\eta_{\rm CIB} & = {c^2\over 2 h \nu^3} I_{\rm CIB}(\nu)
= {I_0 \, (\nu/\nu_0)^{k_{\rm F}} \over e^{x_{\rm CIB}}-1} = {I_0 \, (x_{\rm CIB} / x_{\rm 0,CIB})^{k_{\rm F}} \over e^{x_{\rm CIB}}-1} \, ,
\end{align}
\noindent
where $\nu_0 = c / (100\, \mu{\rm m}) \simeq 3\,$THz and $x_{\rm CIB}=h\nu/(kT_{\rm CIB})$.
The best-fit to FIRAS data gives $k_{\rm F} = 0.64 \pm 0.12$, $T_{\rm CIB} = (18.5\pm1.2)\,$K and $I_0 = (1.3 \pm 0.4) \times 10^{-5}$ \citep{1998ApJ...508..123F}, $I_0$ setting the CIB spectrum amplitude.
In the last equality of Eq. \eqref{eq:eta_CIB}, $x_{\rm 0,CIB}=h\nu_0/(kT_{\rm CIB}) \simeq 7.78$ 
(implying $x_{\rm CIB} \simeq 7.78 \times \nu/\nu_0)$.
The current uncertainty on CIB spectrum amplitude is quite high, with 1\,$\sigma$ accuracy of about 30\,\%.
Indeed, the direct determination of the CIB spectrum is hard to obtain as it requires absolute intensity measurements and is limited by foreground signals.

We construct the global signal by adding $\eta_{\rm CIB}$ with the photon distribution function of the CMB blackbody at the current temperature $T_0$ and consider
three simple cases corresponding to the above best-fit and 1\,$\sigma$ limits of $I_0$.
The results are shown in Fig. \ref{fig:CIB} in terms of $\Delta T_{th}$, $\Delta R,$ and $\Delta a_{\ell,0}$.
As shown by the comparison of Fig. \ref{fig:CIB} with Figs. \ref{fig:NEQ}-\ref{fig:BackMM} and anticipated in Sect. \ref{sec:CFF}, the signal associated to the CIB, strongly increasing with frequency, at $\nu \gsim 400$\,GHz dominates over the other extragalactic contributions at any multipole.

As has already been found for extragalactic millimetre background, we note the different behaviours of $\Delta a_{\ell,0}$ at odd and even multipoles and, in the frequency range between $\simeq 100$\,GHz and $\simeq 200$\,GHz, 
the increase with $\ell$ of the frequency where, for odd $\ell$, the change of sign of $\Delta a_{\ell,0}$ occurs or, for even $\ell$, the shape of $\Delta a_{\ell,0}$ shows a clear steepening. At the highest frequencies,
where the power law approximation of the intrinsic monopole spectrum breaks down and the spectrum changes its behaviour approaching its maximum, the shapes of $\Delta a_{\ell,0}$ show remarkable features.
They are located at frequencies that decrease as $\ell$ increases: for example, for $\ell =$  2 or 6 they occur at frequencies larger than $\sim 800$\,GHz or than $\sim 500$\,GHz. 

Here, $\Delta R$ (top-right panel of Fig. \ref{fig:CIB}) assumes positive values that are much smaller than $|R^{\rm BB} - 1|$ at $\nu \lsim 400$\,GHz, but 
significantly increase with $\nu$ at $\nu \gsim 400$\,GHz and
become comparable to or larger than $|R^{\rm BB} - 1|$ at $\nu \gsim 700$\,GHz, achieving a maximum at a frequency $\simeq 800-850$\,GHz that slightly increases as $I_0$ decreases.

\smallskip
\smallskip
\smallskip
\smallskip
\begin{figure*}
\minipage{0.32\textwidth}
\includegraphics[width=\linewidth]{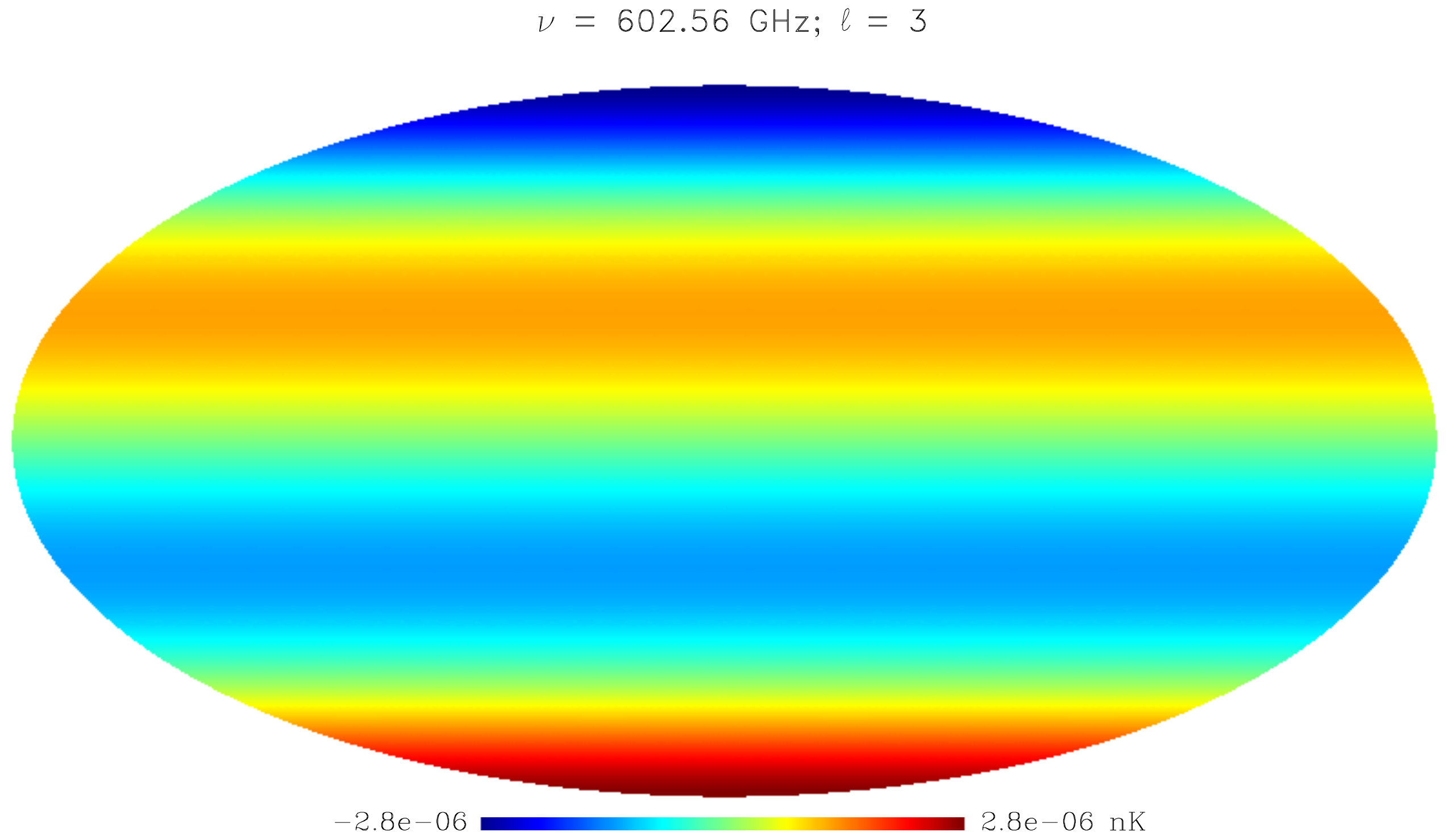}
\endminipage\hfill
\minipage{0.32\textwidth}
\includegraphics[width=\linewidth]{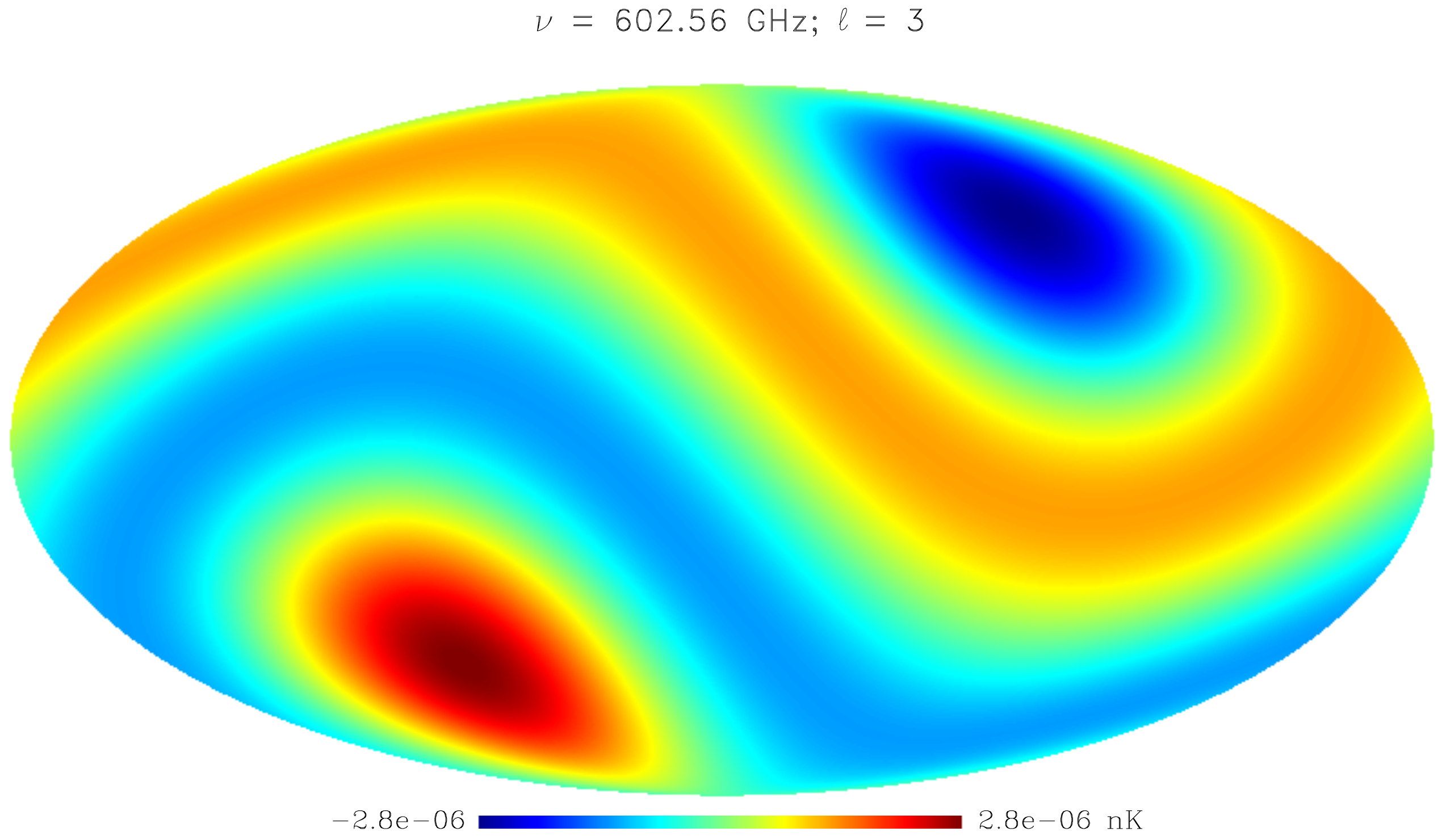}
\endminipage\hfill
\minipage{0.32\textwidth}
\includegraphics[width=\linewidth]{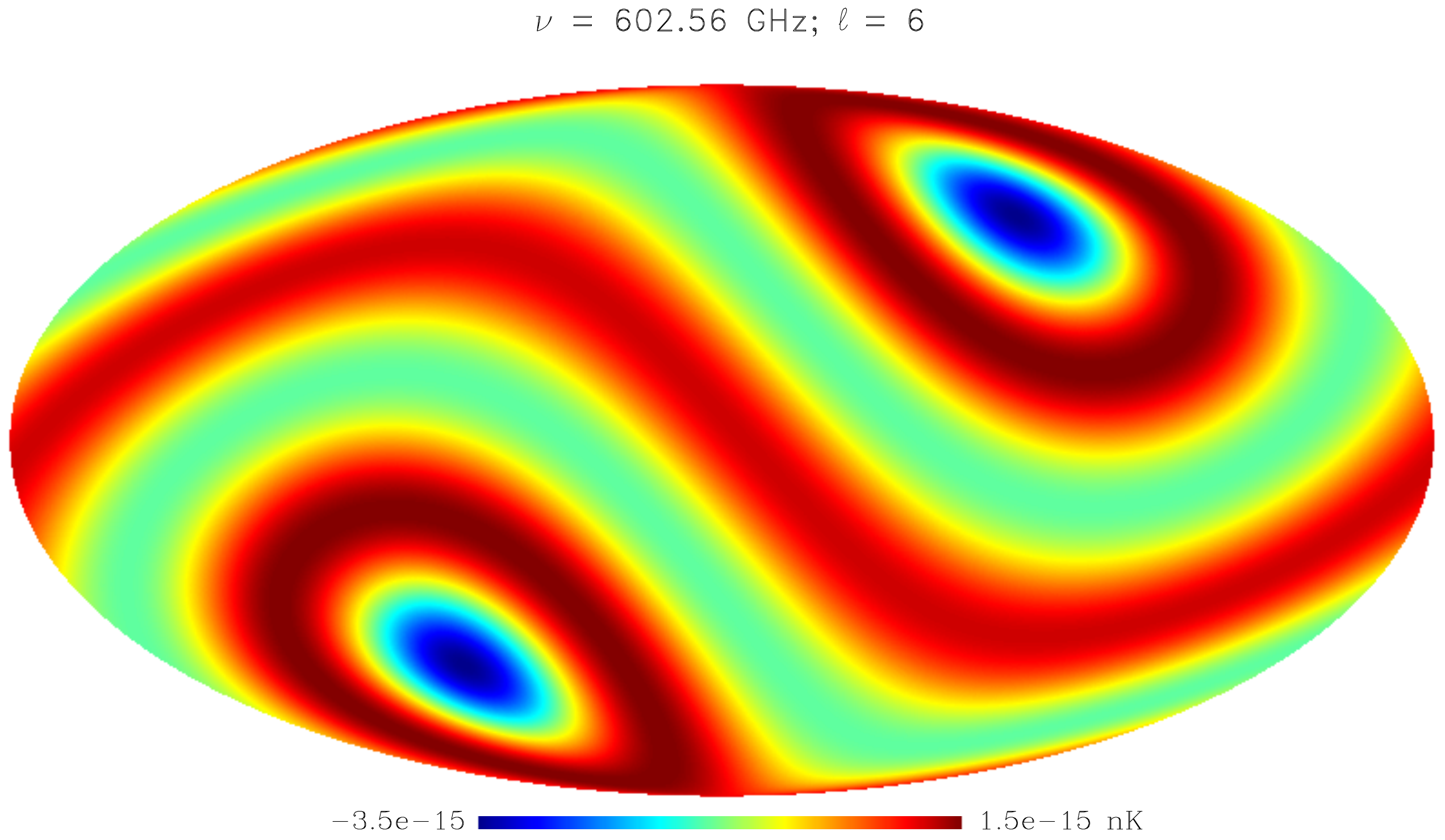}
\endminipage\hfill
\minipage{0.32\textwidth}
\includegraphics[width=\linewidth]{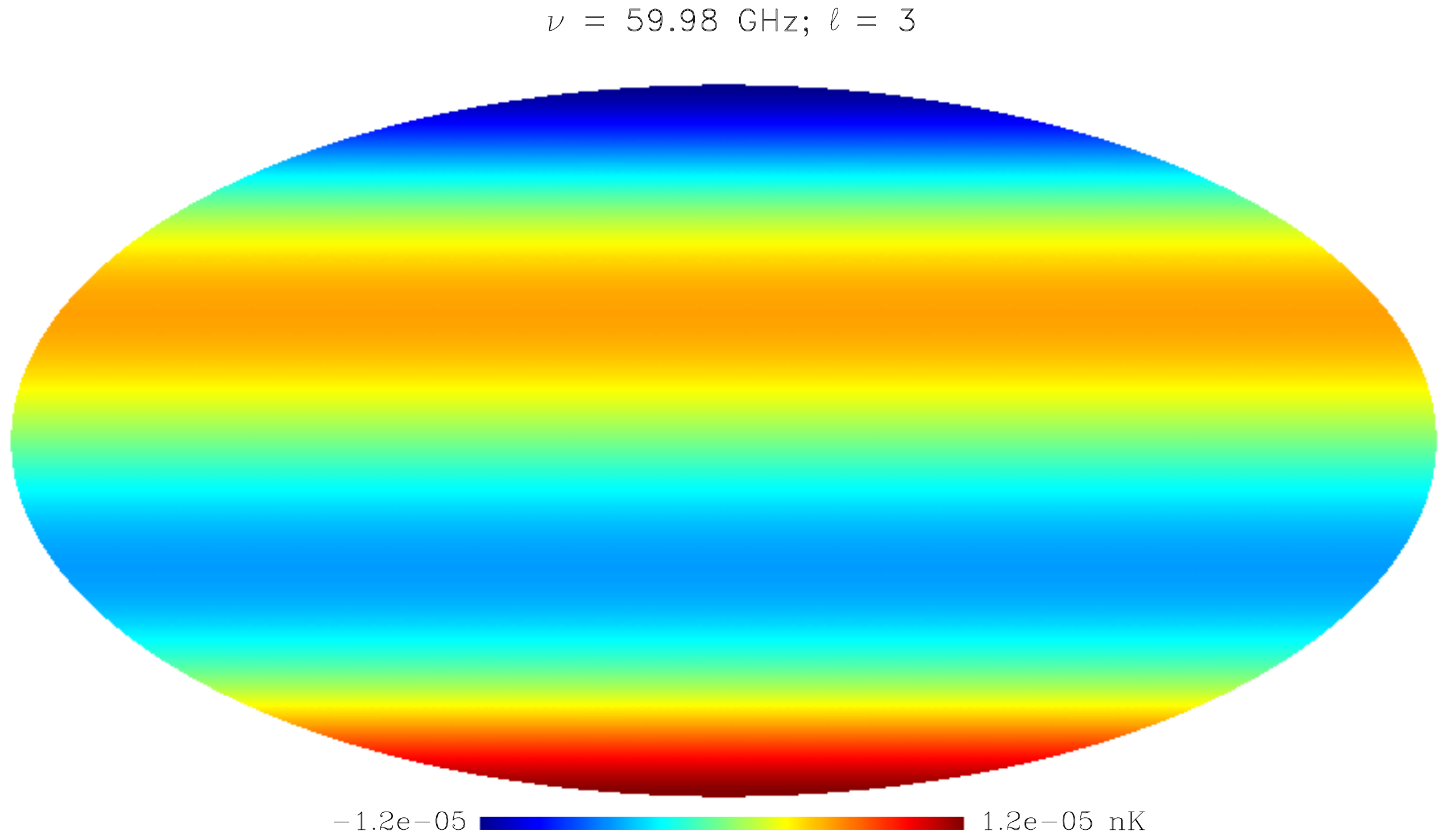}
\endminipage\hfill
\minipage{0.32\textwidth}
\includegraphics[width=\linewidth]{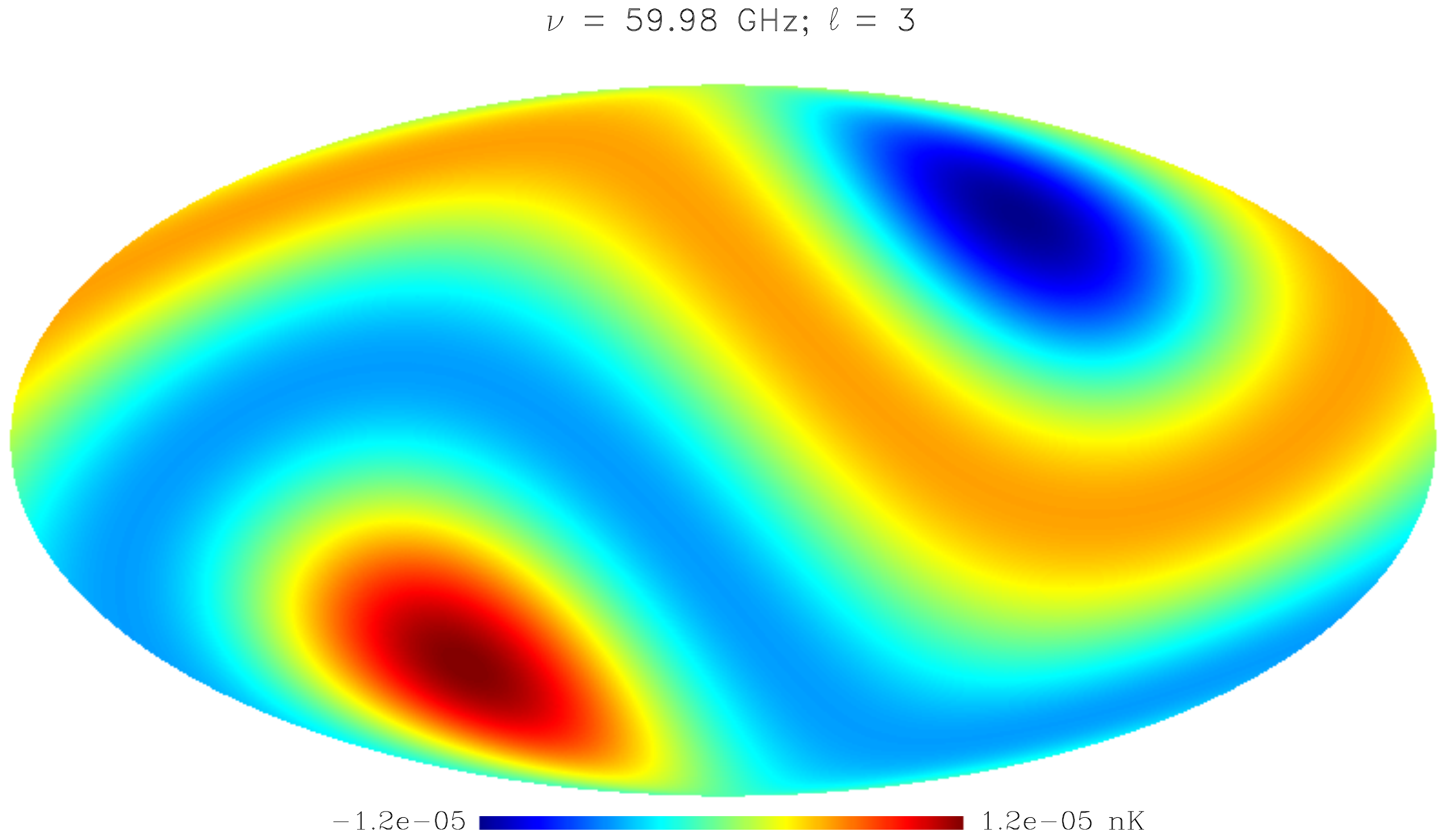}
\endminipage\hfill
\minipage{0.32\textwidth}
\includegraphics[width=\linewidth]{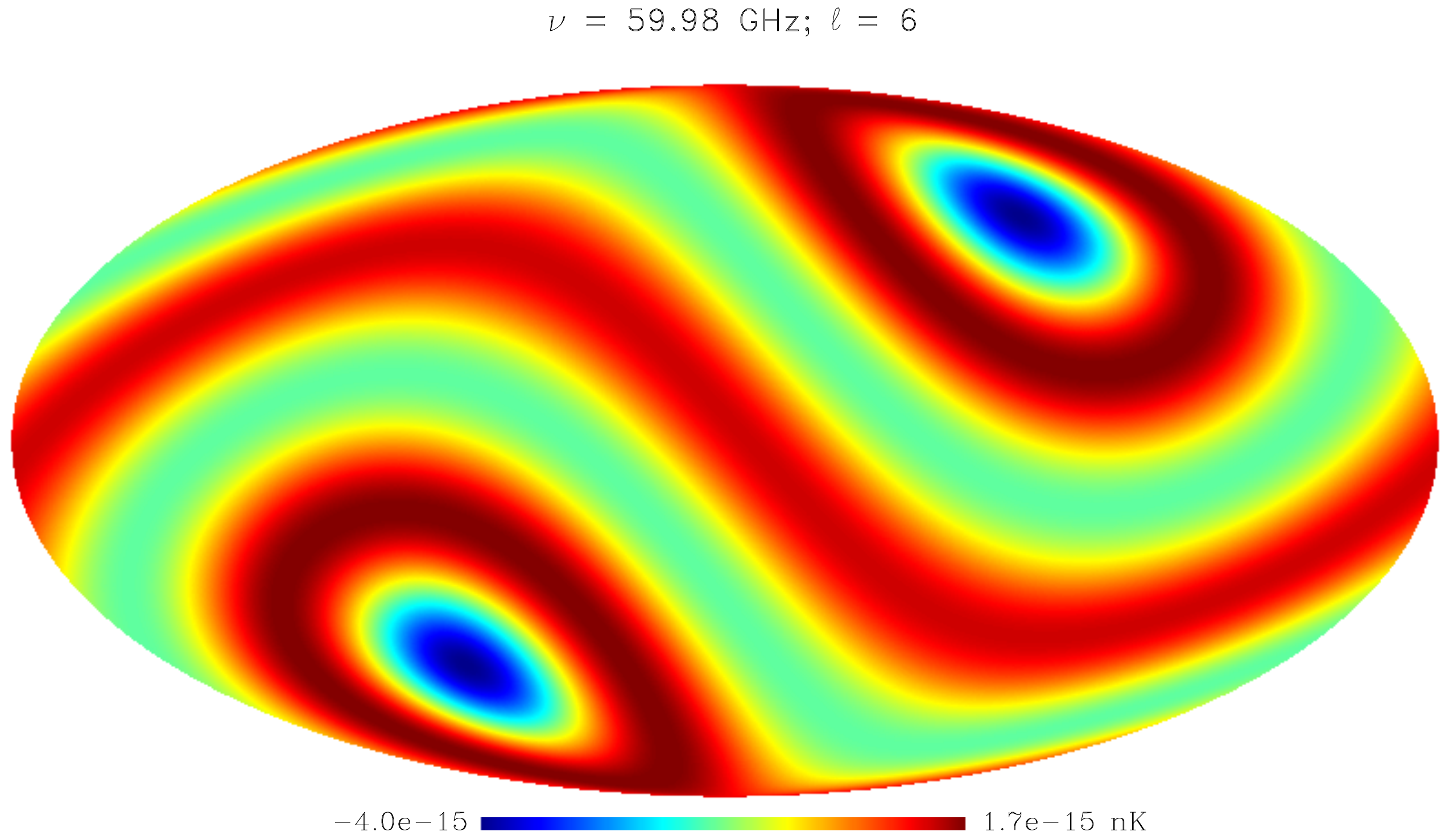}
\endminipage\hfill
\minipage{0.32\textwidth}
\includegraphics[width=\linewidth]{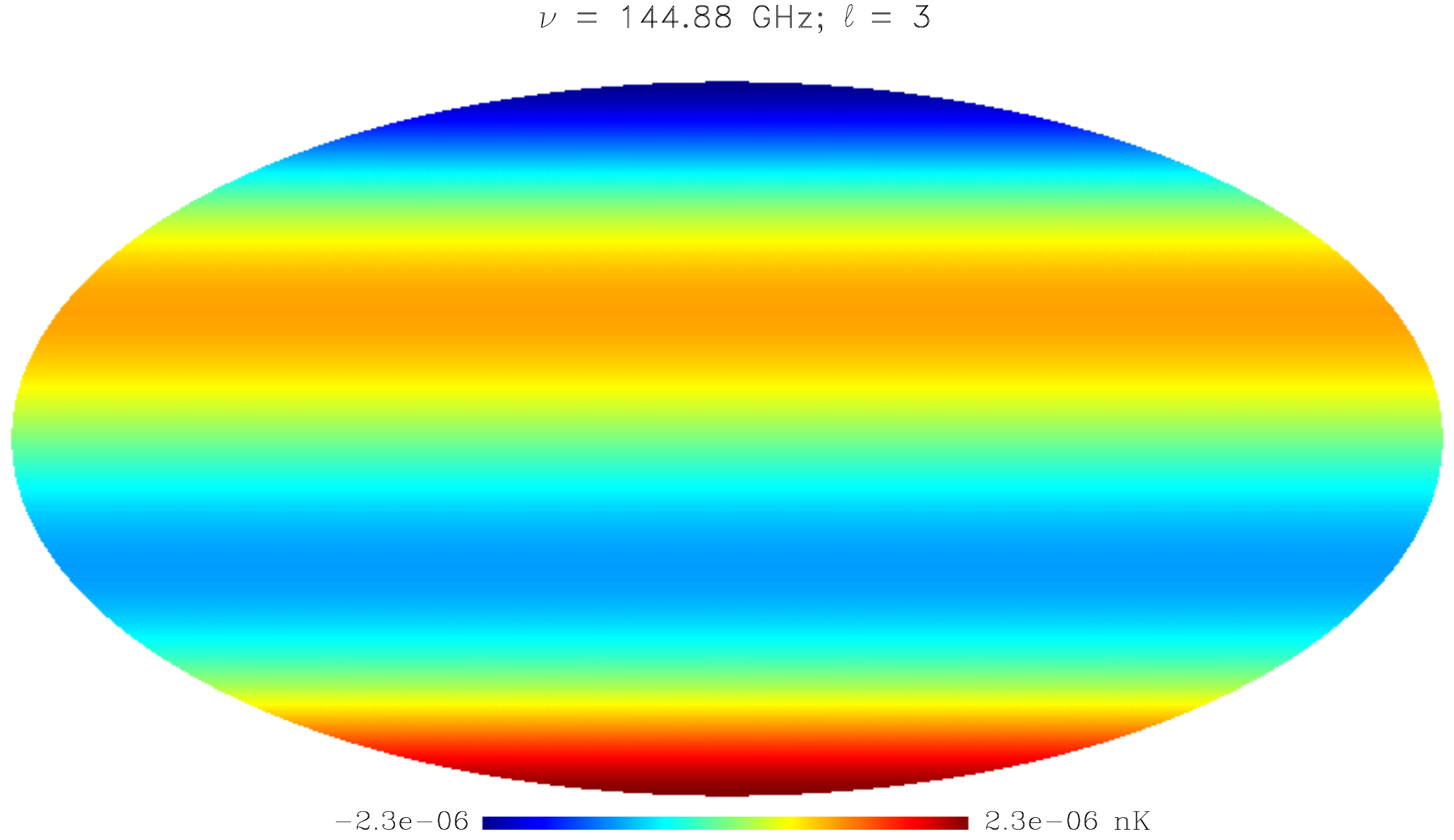}
\endminipage\hfill
\minipage{0.32\textwidth}
\includegraphics[width=\linewidth]{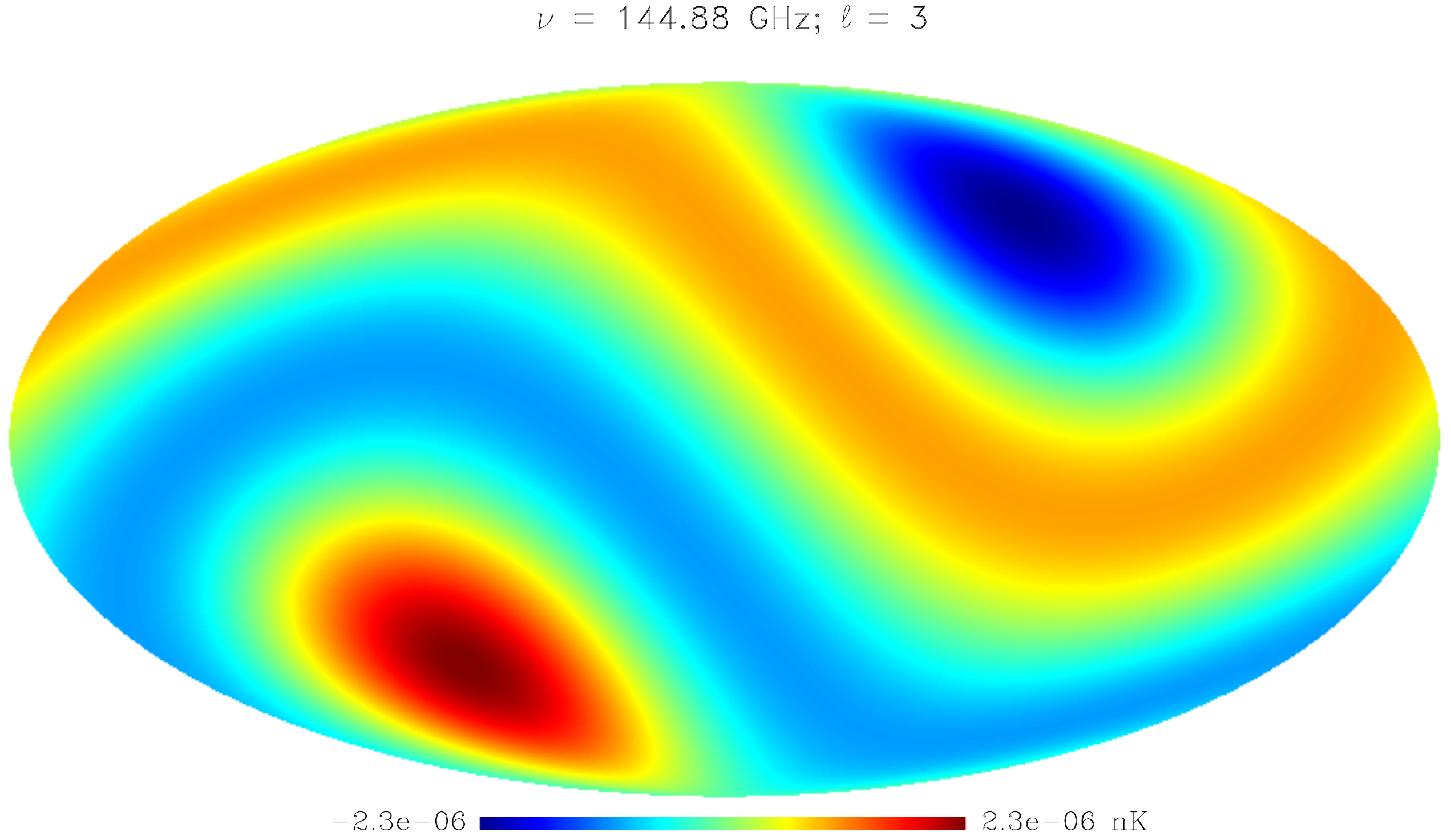}
\endminipage\hfill
\minipage{0.32\textwidth}
\includegraphics[width=\linewidth]{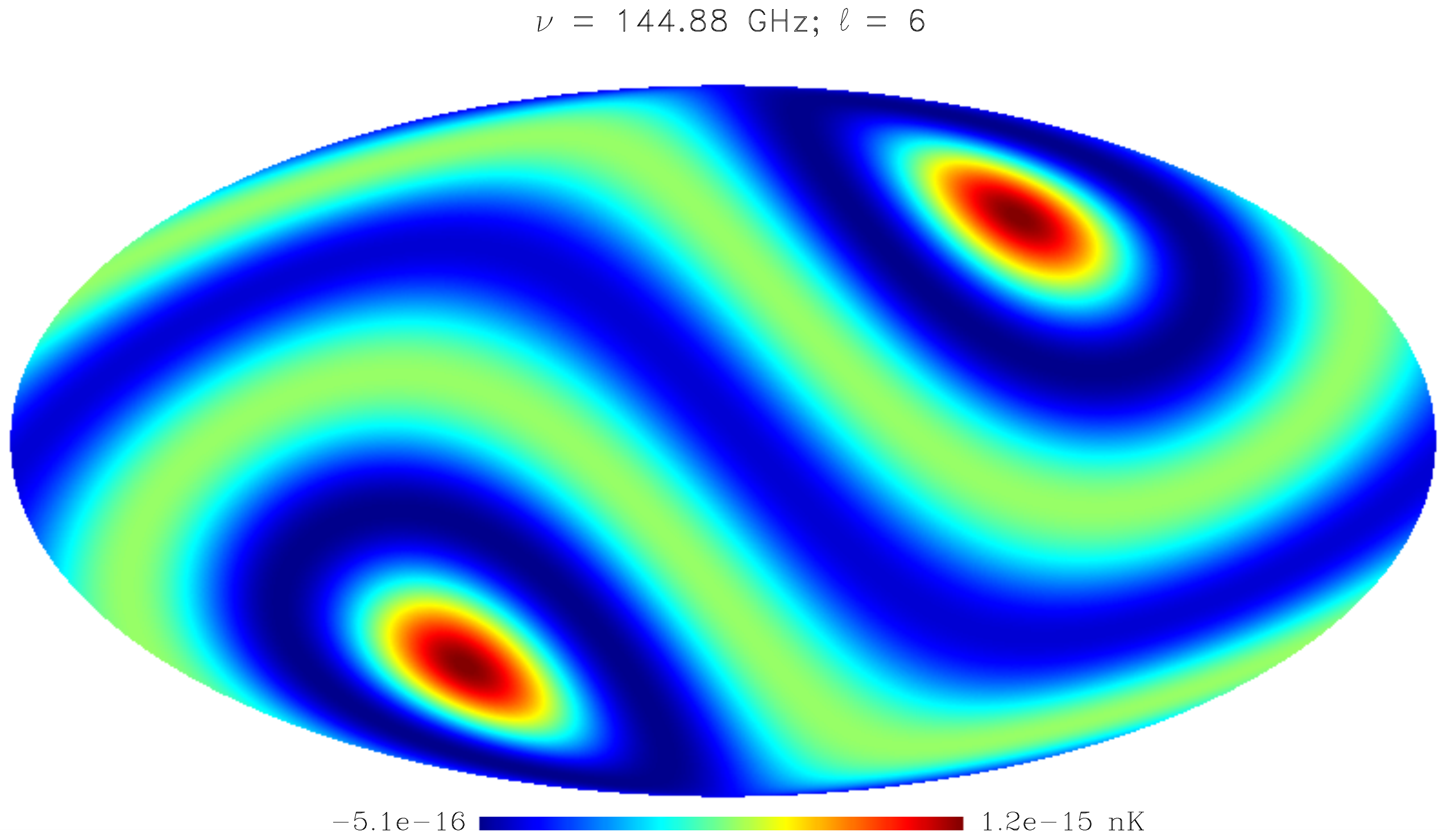}
\endminipage\hfill
\smallskip
\smallskip
\smallskip
\smallskip
\caption{Examples of maps at two specific multipoles in the reference system with the $z$ axis parallel to the observer velocity (left column)
and in Galactic coordinates (central and right columns). According to \cite{2020A&A...641A...1P},
we adopt $l = 264.021^\circ$ and $b = 48.253^\circ$ for the dipole direction in Galactic coordinates.
We display the temperature pattern for a BE-like distortion with $\mu_0 = 1.4 \times 10^{-5}$ (top row), a Comptonization distortion with $u = 2 \times 10^{-6}$ (central row) and 
the CIB distribution function with best-fit amplitude value $I_0 = 1.3 \times 10^{-5}$ added to the blackbody one (bottom row), minus the temperature pattern coming from the blackbody.
The considered frequency and multipole is given above each map. Maps are in equivalent thermodynamic (or CMB) temperature.}
\label{fig:maps}
\end{figure*}

\section{Signal combination results}
\label{sec:combi}

As already states previously in this paper, the method described in Sects. \ref{sec:theoframe}--\ref{sec:highl_derivs} and applied in 
Sect. \ref{sec:monmod}
to specific emissions 
can be also used to predict the signatures expected from the desired combinations of signals, provided that they are summed in terms of additive quantities.
A discussion of the imprints on the dipole spectrum left by combinations of backgrounds associated to cosmological reionization and 
relevant in the radio can be found in \cite{2019A&A...631A..61T}.
Of course, the number of the possible combinations of the models discussed in Sect. \ref{sec:monmod} is high. 
Here, we consider a couple of cases
that are relevant at millimetre and sub-millimetre wavelengths.

We combine the model of Comptonization plus diffuse FF distortion with the highest values of $u$ and $y_B$ with an astrophysical background, namely, the adopted
millimetre background model from extragalactic radio sources, an estimate of its residual signal given 
an assumption of source detection threshold (the 'Low millimetre background residual'), and the best-fit CIB spectrum.
The global signal is constructed by adding the photon distribution functions associated to the considered astrophysical background (Eq. \eqref{eq:backmm} or Eq. \eqref{eq:eta_CIB})
and to the Comptonization plus diffuse FF distortion (Eq. \eqref{eq:etaC}), that already contains the initial unperturbed CMB spectrum, $\eta_{\rm i}$. The results based on the solutions given in Sect. \ref{sec:sol7} are shown in Figs. \ref{fig:FFBackMM} and \ref{fig:CIBFF} in terms of $\Delta T_{th}$, $\Delta R,$ and $\Delta a_{\ell,0}$. 

From these figures (see also Figs. \ref{fig:FFeC}, \ref{fig:BackMM} and \ref{fig:CIB}), it is possible to appreciate that at the frequencies where a certain component is much stronger than the other, that component also dominates in the 
combined signal, but this simplification does not hold where the components have comparable amplitudes.

The comparison of Fig. \ref{fig:FFBackMM} with Figs. \ref{fig:FFeC} and \ref{fig:BackMM} shows that the millimetre background, as well as its residual for the assumed source detection threshold, 
dominates above a frequency of $\sim 500$\,GHz, with only a very little dependence on $\ell$. Below a frequency ranging from few tens of GHz to $\sim 100$\,GHz the millimetre background and 
the considered model of Comptonization plus diffuse FF distortion give comparable effects:
in particular, the millimetre background and the diffuse FF distortion have also similar power law behaviours (see Eqs. \eqref{eq:yBpowlaw} and \eqref{eq:backmm}).
As a result, the difference of considering the millimetre background or its residual is more evident
in the spectral shapes of their combinations, affecting, respectively, less or more the frequency where the spectra exhibit their sign change, that occurs at frequencies slightly higher  
than in the case of pure Comptonization plus diffuse FF distortion. 
The Comptonization distortion is, instead, clearly appreciable at intermediate frequencies, its characteristic plateau appearing in all the $\Delta a_{\ell,0}$.

The comparison of Fig. \ref{fig:CIBFF} with Figs. \ref{fig:FFeC} and \ref{fig:CIB} shows analogous results. The CIB dominates over the Comptonization distortion above a frequency that ranges 
from $\sim 100$\,GHz to $\sim 200$\,GHz, slightly increasing with $\ell$. Below that frequency, the Comptonization distortion emerges and the combined spectral shapes become flatter. They show also
a remarkable change of sign at even $\ell$ from Figs. 2 to 6, where $\Delta a_{\ell,0}$ is positive for the CIB and negative for the Comptonization distortion. This does not occur for $\Delta T_{\rm th}$ where the contributions 
from CIB is slightly larger than that from Comptonization, and for $\Delta R$, that is positive in both the cases.
At further decreasing frequencies, the FF diffuse emission emerges, and, consequently, the $\Delta a_{\ell,0}$ exhibit, at even and odd $\ell$, 
the typical shapes already found in that case at the lowest frequencies as well as the typical 
change of sign corresponding to the transition from the range dominated by the FF term to that dominated by the Componization term. The frequency of this transition is only slightly larger than that 
found in the case of pure Comptonization plus diffuse FF distortion. This frequency shift is significantly smaller than that found above, which combines the Comptonization plus diffuse FF distortion with the millimetre background, because of the smaller additional contribution from the CIB at low frequencies. 

Of course, the details in the above considerations depend also on the assumed model parameters. We discuss above some cases where the combined contribution from different signals may give not trivial
effects. In general, refinements in modelling specific signals and combinations or couplings of them can be included in our method to improve the quality of the results.

\section{Maps and angular power spectra}
\label{sec:compa}

\begin{figure*}[ht!]
\centering
         \includegraphics[width=18.5cm]{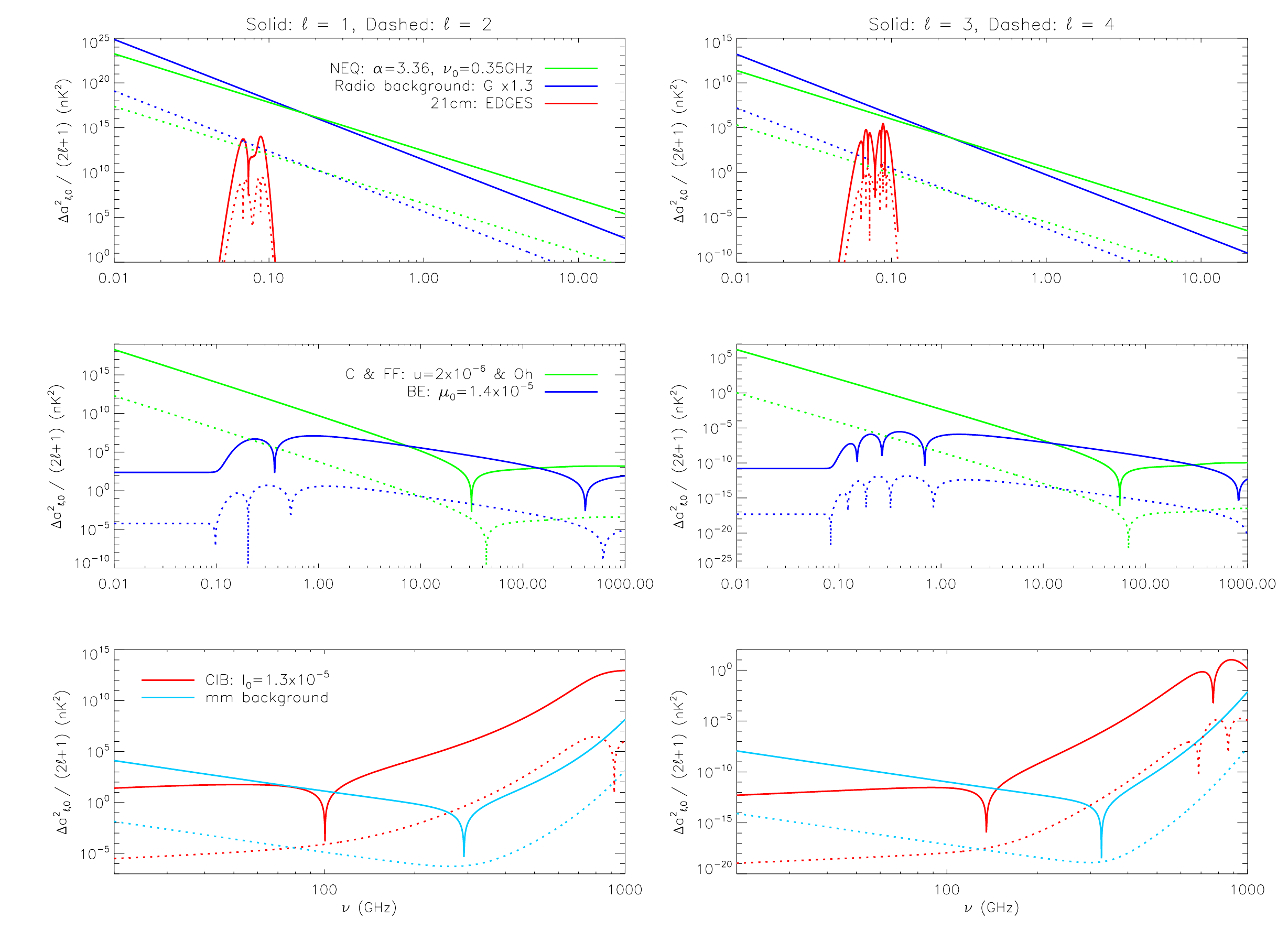}
    \caption{Angular power spectrum of the maps obtained from
    the difference between the maps produced in the various models
    and the map obtained for spectrum corresponding to the blackbody at the current temperature $T_0$.
    See also the legend and the text.}
    \label{fig:cl_dist}
\end{figure*}

Having evaluated the coefficients $a_{\ell,0}$,
it suffices then to compute the corresponding maps and angular power spectra. 
Here, we report  just a few examples, also for comparison with the results presented in \cite{2018JCAP...04..021B}, which are based on map generation and inversion performed with the 
great set of publicly available tools implemented in the 
Hierarchical Equal Area and isoLatitude Pixelization of the sphere (HEALPix)
\citep{2005ApJ...622..759G}. Here, we adopt the same pixelization scheme to generate the maps using the $a_{\ell,0}$ coefficients:
working in real space, based on Eq. \eqref{eq:harm6}, we immediately get the map for each multipole component in the reference system with the $z$ axis parallel to the observer velocity. 
They can be also simply computed in any other reference system (e.g. in Galactic coordinates) given the angle of a pixel direction with respect to the observer velocity direction.  
Here, we generate the maps with $n_{\rm side} = 1024$, corresponding to a pixel linear size of $\simeq 3.44$\,arcmin.

We show the maps for $\ell = 3$ and 6 in the case of a BE-like distortion with the maximum value of $\mu_0$ (see Sect. \ref{sec:BE}), 
of a pure Comptonization distortion with the maximum value of $u$ (see Sect. \ref{sec:CFF}) and of the best-fit CIB spectrum (see Sect. \ref{sec:cib}).
The three adopted frequencies, different for each type of signal, are selected to allow us to make an almost direct comparison with the maps displayed in \cite{2018JCAP...04..021B} and
to appreciate how the method presented here is suitable for a fast map computation,
very precise up to the desired order, 
even where the relevant signal is weak and, in principle, more sensitive to numerical uncertainty. 
The maps (see Fig. \ref{fig:maps}) are indeed very clean 
and without 
numerical artifacts up to the highest $\ell$ thanks to the adopted analytical approach;
each multipole pattern obviously reflects the corresponding scaling, see Eq. \eqref{eq:harm6}, and, in particular, for odd multipoles
the maximum and minimum values of each map are equal in module.

Given the map, we can use publicly available tools, such as the {\it anafast} facility of HEALPix, 
to compute the corresponding angular power spectrum and the $a_{\ell,m}$ coefficients in any reference system
(see \cite{1984JGR....89.4413G}, and also Appendix \ref{app:rotation}, for explicit formulas to transform the $a_{\ell,m}$ under rotation). 
Indeed, for the considered problem, the $a_{\ell,m}$ with $m \ne 0$ do not vanish in reference systems with the $z$ axis not parallel to the observer velocity direction. 
On the other hand, since the angular power spectrum, $C_\ell$, is an invariant under rotation of the reference system, we can compute it simply working in the reference system with the $z$ axis 
parallel to the observer velocity direction, where the $a_{\ell,m}$ with $m \ne 0$ vanish, and the coefficients $a_{\ell,0}$ are evaluated as in the previous sections 
\begin{equation}
C_\ell (\nu,\beta) = \sum_{m=-\ell}^{m=\ell} \frac{a_{\ell,m}^2(\nu,\beta)}{2\ell+1} =  \frac{a_{\ell,0}^2(\nu,\beta)}{2\ell+1} \, .
\label{eq:aps}
\end{equation}
In Fig. \ref{fig:cl_dist}, we only report
the result derived for some of the considered models for $\ell$ from 1 to 4. We plot the quantity:
\begin{equation}
\frac{\Delta a_{\ell,0}^2 (\nu,\beta)} {2\ell+1} = \frac{[a_{\ell,0}^{\rm dist}(\nu,\beta) - a_{\ell,0}^{\rm BB}(\beta)]^2}{2\ell+1} \, . 
\label{eq:deltaaps}
\end{equation}
Equation \eqref{eq:deltaaps} gives the angular power spectrum of the map obtained from the difference between the map produced in a given model
(i.e. for a CMB distorted spectrum or for an astrophysical background spectrum summed with the blackbody spectrum at the current temperature $T_0$)
and the map obtained for the blackbody spectrum at the current temperature $T_0$. 

There is a good
agreement
with the results reported by \cite{2018JCAP...04..021B} 
-- but in the frequency range between 60 and 600 GHz considered for the Cosmic Origins Explorer (COrE)
in Figure 14 (for the CIB) and in Figure 12 (for BE and Comptonization distortions).
As expected, the central panels of Fig. \ref{fig:cl_dist}, that span from the radio to the sub-millimetre, 
show that the replacement of the pure BE spectrum with a BE-like spectrum and the inclusion of FF diffuse emission,
which is not included in \cite{2018JCAP...04..021B}, is remarkable at lower frequencies 
(given the high FF model considered here, a little effect is already appreciable at the lowest frequencies of the COrE range, see also Fig. \ref{fig:FFeC}). 

The signals more relevant at low and high frequencies are displayed in terms of $\Delta a_{\ell,0}^2 / (2\ell+1)$  
in the top and bottom panels of Fig. \ref{fig:cl_dist}, respectively. All the well-defined minima of the various lines displayed in Fig. \ref{fig:cl_dist} correspond to the changes of sign of $\Delta a_{\ell,0}$
(see Fig. \ref{fig:FFeC}-\ref{fig:EDGES} and Figs. \ref{fig:BackMM}-\ref{fig:CIB}), but not in the case of the minimum at $\simeq 80$\,MHz  with respect to the EDGES profile at $\ell = 2$ and 4 
(appearing in the top panels of Fig. \ref{fig:cl_dist}, which correspond to the positive minima at the same $\ell$'s in Fig. \ref{fig:EDGES}).

Finally, we remark that the values of $\Delta a_{\ell,0}^2 / (2\ell+1)$ at the minima corresponding to the sign changes of $\Delta a_{\ell,0}$ 
(as well as the values of $\Delta a_{\ell,0}$) should, in principle, go to zero at the corresponding frequencies; typically,
this does not appear in the plots merely because of their frequency discretization.

\section{Varying observer velocity}
\label{sec:varybeta}

As discussed in Sect. \ref{sec:intro},
the observer velocity
$\vec{\beta} = \vec{\beta_{\rm C}} + \vec{\beta_{\rm V}}(t)$
is modulated 
by the time dependent component $\vec{\beta_{\rm V}}$ that, in real measurements, mainly comes from the revolution of the Earth or of L2 around the Solar System barycentre
($\beta_{\rm V} \simeq \beta_{\rm ES} \simeq \beta_{\rm L2} \simeq 10^{-4} \simeq 0.081 \, \beta$).
The formalism and the general properties of the solutions presented in previous sections hold for any choice of $\vec{\beta}$. 
Thus, we can consider a relatively small time interval, $\Delta t_i$, around a certain instant $t_i$,
in which the time variation of $\vec{\beta_{\rm V}}$ can be neglected and 
$\vec{\beta}$ can be considered as a constant.
For example, in a time interval $\Delta t_i \simeq 1$\,h, the relative change of 
$\beta$, which is given by $\Delta {\beta} / {\beta} \simeq (\Delta {\beta_{\rm V}} / {\beta_{\rm V}}) ({\beta_{\rm V}} / {\beta}) \simeq \omega \, \Delta t_i \, (\beta_{\rm V} / \beta) \simeq 0.0058$\,\%,
is about one order of magnitude smaller than the current relative uncertainty on $\beta_{\rm C}$, 
given the period of the main modulation $P=2\pi/\omega = 1$\,yr.

In a time interval, $\Delta t_i$, only a certain fraction of the sky can be observed. 
The detailed pattern of the sequence of the observed sky positions is defined 
by the so-called scanning strategy, or observational strategy, which is specific of each considered experiment.
Let us define, using $M_i(\nu, \theta'_i, \vec{\beta}_i),$ the all-sky map 
given by Eq. \eqref{eq:harm6},
where  $\vec{\beta}$ is replaced by $\vec{\beta}_i = \vec{\beta} (t_i) = \vec{\beta_{\rm C}} + \vec{\beta_{\rm V}}(t_i)$ 
and the colatitude $\theta$ is replaced by the angle $\theta'_i = \theta'_i(\vec{\beta}_i,\theta,\phi) = {\rm arccos}[({\hat{n}} \cdot \vec{\beta}_i)/\beta_i]$ between $\vec{\beta}_i$ and a sky direction 
corresponding to the unit vector ${\hat{n}}$ defined by the colatitude $\theta$ and the longitude $\phi$ in any adopted reference system, $S$.
The spherical harmonic coefficients to be used in Eq. \eqref{eq:harm6} are given in Sect. \ref{sec:sol7} for $\ell_{\rm max} = 6$ 
(or, alternatively, in Sect. \ref{sec:sol5} for $\ell_{\rm max} = 4$ or in Sect. \ref{sec:dip_3and2col} for $\ell_{\rm max} = 2$ or 1).

The experiment scanning strategy defines the sequence of pointing maps, $P_i(t_i,\theta,\phi)$, on a pixelized sky at the times $t_i$, 
with $i=1, n$ and $n = \tau_{\rm S}/\Delta t_i$ where $\tau_{\rm S}$ is the survey duration:
if the direction defined by $\theta$ and $\phi$ is observed in the interval $\Delta t_i$ then $P_i = 1$, otherwise $P_i = 0$.
The averaged map corresponding to a survey based on a set of observer velocities $\vec{\beta}_i$ is then given by:

\begin{equation}
M(\nu,\theta,\phi) = \sum_{i=1}^n \frac{M_i(\nu, \theta'_i, \vec{\beta}_i) \cdot P_i(t_i,\theta,\phi)} {N(\theta,\phi)} \, ,
\label{eq:allmap_varvel}
\end{equation}

\noindent
where $N(\theta,\phi)$ is the global number of observer velocities $\vec{\beta}_i$ for which the direction defined by $\theta$ and $\phi$ is observed.

As discussed in Sect. \ref{sec:theoframe}, we need to compute the signal in only $\ell_{\rm max}+1$ directions to predict the map up to $\ell_{\rm max}$, and the number of signal evaluations required by Eq. \eqref{eq:allmap_varvel}
is then $N_{\rm ev,\ell_{\rm max}} = (\ell_{\rm max}+1) \, n = (\ell_{\rm max}+1) \, \tau_{\rm S}/\Delta t_i$.  
Of course, the map $M(\nu,\theta,\phi)$ can be directly computed through Eqs. \eqref{eq:t_therm} and \eqref{eq:eta_boost}
using $\vec{\beta} = \vec{\beta_{\rm C}} + \vec{\beta_{\rm V}}(t)$ at each time, without  
applying the discretization
associated to the choice of the time interval $\Delta t_i$. In this case, since the receiver sampling time, $\tau_{\rm samp}$, is typically much smaller than $\Delta t_i$, 
the required number of signal evaluations,
$N_{\rm ev,d} = \tau_{\rm S}/\tau_{\rm samp}$, is much larger than $N_{\rm ev,\ell_{\rm max}}$.
The ratio, $r_{\rm ev} = N_{\rm ev,\ell_{\rm max}} / N_{\rm ev,d}$, between the required number of signal evaluations in the two cases is $r_{\rm ev} = \tau_{\rm samp} (\ell_{\rm max}+1)/\Delta t_i \ll 1$.
In order to properly compare the computing times in the two approaches, we evaluate this ratio discretizing $\vec{\beta} = \vec{\beta_{\rm C}} + \vec{\beta_{\rm V}}(t)$ 
according to the choice of $\Delta t_i$ also in the direct scheme based on Eqs. \eqref{eq:t_therm} and \eqref{eq:eta_boost}.
Thus, $N_{\rm ev,d}$ reduces to $N_{\rm ev,d} = N_{i,{\rm pix}} \tau_{\rm S}/\Delta t_i$, where $N_{i,{\rm pix}}$ is the number of sky pixels observed in the time interval $\Delta t_i$,
and we obtain $r_{\rm ev} = (\ell_{\rm max}+1) / N_{i,{\rm pix}}$ (again, $r_{\rm ev} \ll 1$).

The method based on Eq. \eqref{eq:allmap_varvel} is particularly advantageous when applied
to scanning strategies designed for future CMB missions that foresee the use of a huge number of receivers at the focal surface of a wide field of view telescope.
The scanning strategies typically foreseen for these missions involve fast spacecraft spin axis precessions,
in order to achieve in a short time a large sky coverage and to observe sky pixels with many orientations during the survey to improve the quality of map making results and of polarization analyses (see e.g. \cite{2018JCAP...04..014D} and \cite{2018JCAP...04..022N}). 

Since $\theta'_i$ depends on $\vec{\beta}_i$, $\theta$ and $\phi$, 
Eq. \eqref{eq:allmap_varvel} implies that the time varying component $\vec{\beta_{\rm V}}(t)$ will introduce
in the map $M(\nu,\theta,\phi)$ a further modulation along $\theta$ and $\phi$, 
which is superimposed onto the main pattern, even if we choose a reference frame $S$ with the $z$ axis parallel to the constant component $\vec{\beta_{\rm C}}$.
Although the details of this modulation depend on the experiment scanning strategy, its amplitude 
amounts to $\sim \beta_{\rm V} / \beta \sim 8.1$\,\% of the main signal due to the constant velocity component.

For a given model, fixing the other parameters and neglecting the terms in 
$\beta^{(\ell+2)}$ and beyond,
the differences of the spherical harmonic coefficients $\Delta a_{\ell,0}$
scale proportionally to $\beta^\ell$,
that is, $\Delta a_{\ell,0} (\beta) \simeq \Delta a_{\ell,0} (\beta_{\rm C}) \, (\beta / \beta_{\rm C})^\ell$ at the leading order.
This property comes from the separation of the system into even and odd multipoles (see Sects. \ref{sec:theoframe}--\ref{sec:dip_3and2col})
and is explicit in the formulas found for the blackbody spectrum (see Sects. \ref{sec:sol7BB} and \ref{sec:sol5BB}).
The comparison of the results displayed in Figs. \ref{fig:NEQ} and \ref{fig:EDGES} with those reported in Figs. \ref{fig:NEQper100} and \ref{fig:EDGESper10} where the velocity is multiplied by a factor 
100 and 10, respectively, also suggests this scaling, providing that the signal is sufficiently smooth in frequency and except for smearing effects where the signal rapidly changes.
We verified this property in our numerical results varying $\beta$ within a $\pm 10$\,\% because of the contribution of $\beta_{\rm V}$, namely, for realistic deviations of $\beta$ from $\beta_{\rm C}$.   
This feature can further speed up the computation of $M_i(\nu, \theta'_i, \vec{\beta}_i)$ because it allows a fast estimation of the contribution of the time varying component of the velocity
over two consecutive multipole patterns, for example, in the case of the dipole and quadrupole, as useful in many applications.

\section{Global pattern}
\label{sec:all}

In general, the global sky pattern is a combination of intrinsic anisotropies and of the anisotropies induced by the peculiar observer motion. We focus here on the
diffuse cosmic signals more relevant in the microwaves, 
where the intrinsic (mainly of primordial nature) anisotropies are better studied and the background frequency spectrum can be modelled in terms of small deviations from a blackbody.  

Working within a reference frame at rest with respect to the background, in a given sky direction identified by $\theta$ and $\phi$ the frequency dependent equivalent thermodynamic temperature
can be seen as a function of an effective temperature and of a set of $P$ distortion parameters $p_j$ with $j=0,P-1$

\begin{equation}\label{Tthlocal}  
T_{\rm th} (\nu, \theta, \phi) = T_{\rm th} (\nu, T_{\rm BB}(\theta, \phi) , p_j(\theta, \phi)) \, . 
\end{equation}

\noindent
Since both fluctuations and distortions are small, $T_{\rm BB}(\theta, \phi) \sim T_0$ and $p_j (\theta, \phi) \sim 0$ and we can expand $T_{\rm th}$ in Taylor's series around these values. At linear order

\begin{align}\label{Tthlocal_lin}
T_{\rm th} (\nu, \theta, \phi) & = (T_{\rm th})_0 + \left({ \frac{\partial{T_{\rm th}}}{\partial{T_{\rm BB}}} }\right)_0 \, (T_{\rm BB} - T_0) \, + \, \sum_{j=0}^{P-1} \, \left({ \frac{\partial{T_{\rm th}}}{\partial{p_j}} }\right)_0 \, p_j  \nonumber
\\ & = T_0 + \left({ \frac{\partial{T_{\rm th}}}{\partial{\eta}} \frac{\partial{\eta}}{\partial{T_{\rm BB}}} }\right)_0 \, (T_{\rm BB} - T_0) \nonumber
\\ & + \sum_{j=0}^{P-1} \, \left({ \frac{\partial{T_{\rm th}}}{\partial{\eta}} \frac{\partial{\eta}}{\partial{p_j}} }\right)_0 \, p_j \, , 
\end{align}

\noindent
where $()_0$ denotes that the quantities are evaluated at $T_{\rm BB}(\theta, \phi) = T_0$ and $p_j (\theta, \phi) = 0$, that is, for a blackbody with effective temperature $T_0$. Expanding the temperature fluctuation $(T_{\rm BB}(\theta, \phi)  - T_0)$ and 
the distortion parameters $p_j(\theta, \phi)$ in spherical harmonics with coefficients $a_{\rm BB,\ell,m}$ and $a_{j,\ell,m}$, we have

\begin{align}\label{Tthlocal_linSphArm}
T_{\rm th} (\nu, \theta, \phi) & = a_{\rm BB,0,0} Y_{0,0} + \sum_{j=0}^{P-1} \, a_{j,0,0} Y_{0,0} \left({ \frac{\partial{T_{\rm th}}}{\partial{\eta}} \frac{\partial{\eta}}{\partial{p_j}} }\right)_0
\\ & + \sum_{\ell=1}^{\ell_{\rm max}} \sum_{m=-\ell}^{\ell} a_{\rm BB,\ell,m} Y_{\ell,m} \left({ \frac{\partial{T_{\rm th}}}{\partial{\eta}} \frac{\partial{\eta}}{\partial{T_{\rm BB}}} }\right)_0 \nonumber
\\ & + \sum_{j=0}^{P-1} \sum_{\ell=1}^{\ell_{\rm max}} \sum_{m=-\ell}^{\ell} a_{j,\ell,m} Y_{\ell,m} \, \left({ \frac{\partial{T_{\rm th}}}{\partial{\eta}} \frac{\partial{\eta}}{\partial{p_j}} }\right)_0 \, , \nonumber 
\end{align}

\noindent
where $a_{\rm BB,0,0} Y_{0,0} = T_0$ and $a_{j,0,0} Y_{0,0} = \bar{p}_j$ are the average of $T_{\rm BB}(\theta, \phi)$ and $p_j(\theta, \phi)$ over the full sky. 
Although not necessary, 
we 
adopt here for simplicity a reference system with the $z$ axis parallel to the observer velocity in order to avoid rotations in the following considerations
(see Eq. \eqref{eq:alm_glob}).
In the right-hand side of Eq. \eqref{Tthlocal_linSphArm},
the first line represents the (possibly distorted) monopole spectrum, the second line the intrinsic temperature fluctuations, the third line the intrinsic fluctuations of the distortion parameters and
(see Eq. \eqref{eq:t_therm})

\begin{equation}\label{DevTth}
\frac{\partial{T_{\rm th}}}{\partial{\eta}} = \frac{h\nu}{k} \frac{1}{\eta \, (1+\eta) \, {\rm ln}^2 (1+1/\eta)} \, ,
\end{equation}

\noindent
implying  that (see Eq. \eqref{eq:etaBB}):

\begin{equation}\label{DevTth}
 \left({ \frac{\partial{T_{\rm th}}}{\partial{\eta}} }\right)_0 = T_0 \frac{(e^{x} -1)^2}{x\,e^{x}}  \, ,
\end{equation}

\noindent
and the functions $({\partial{\eta}}/{\partial{p_j}})_0$ ($j=0,P-1$) depend on the type of distortion. Eq. \eqref{Tthlocal_linSphArm} shows that the spectral shape of each distortion term in the monopole
spectrum (added to $T_0$) and of the fluctuations of the corresponding distortion parameter is the same, but they are  weighted differently depending
on the coefficients $a_{j,\ell,m}$.

Next, we consider, in the case of distorted spectra, the derivatives ${\partial{\eta}}/{\partial{p_j}}$ and ${\partial{\eta}}/{\partial{T_{\rm BB}}}$.
In the case of a BE distortion (see Eq. \eqref{eq:etaBE}) with a frequency independent chemical potential $\mu_0$ (i.e. neglecting for simplicity 
the spectrum modifications introduced by considering a BE-like distortion that are relevant at lower frequencies), and for small values of $\mu_0$,
with the approximation $\phi_{\rm BE} \simeq (1-1.11 \mu_0)^{-1/4}$ we have

\begin{equation}\label{DevBE}
\frac{\partial{\eta}}{\partial{\mu_0}} \simeq \frac{e^{h\nu/(kT_{\rm BB})/\phi_{\rm BE} + \mu_0} } {(e^{h\nu/(kT_{\rm BB})/\phi_{\rm BE} + \mu_0} -1)^2}
 \cdot \left (\frac{1.11}{4}\frac{h\nu}{kT_{\rm BB}\phi_{\rm BE}}\,\phi_{\rm BE}^4-1 \right ) \, ,
\end{equation}

\begin{equation}\label{Dev_BE_TBB}
\frac{\partial{\eta}}{\partial{T_{\rm BB}}} \simeq \frac{1}{T_{\rm BB}} \frac{e^{h\nu/(kT_{\rm BB})/\phi_{\rm BE} + \mu_0} \, h\nu/(kT_{\rm BB})/\phi_{\rm BE}}{(e^{h\nu/(kT_{\rm BB})/\phi_{\rm BE} + \mu_0} -1)^2} \, ,
\end{equation}

\noindent
implying

\begin{equation}\label{DevBE0}
\left({ \frac{\partial{\eta}}{\partial{\mu_0}} }\right)_0 \simeq \frac{e^{x} [(1.11/4)x-1]}{(e^{x} -1)^2} \, ,
\end{equation}

\begin{equation}\label{Dev_BE0_TBB}
\left({ \frac{\partial{\eta}}{\partial{T_{\rm BB}}} }\right)_0 \simeq \frac{1}{T_0} \frac{x \, e^{x}}{(e^{x} -1)^2} \, .
\end{equation}

We derive now ${\partial{\eta}}/{\partial{p_j}}$ and ${\partial{\eta}}/{\partial{T_{\rm BB}}}$ in 
the case of a Comptonization distortion with a small Comptonization parameter $u$ and an initial Planckian spectrum, $\eta_{\rm i}$ with $\phi_{\rm i} \simeq 1-u$ 
(see Eq. \eqref{eq:etaC}), combined with
a FF distortion with $y_B$ approximated 
by $y_B \simeq {\tilde A}_{FF} x^{-\zeta}$ 
with ${\tilde A}_{FF} = A_{FF} [(h/k)\,({\rm GHz}/T_{\rm BB})]^\zeta \simeq 0.5456 \, A_{FF}$ (see Eq. \eqref{eq:yBpowlaw}).
In principle, also the slope parameter $\zeta$ or, for the more general description in Eq. \eqref{eq:expweight},
an alternative set of five parameters could be included in the set of distortion parameters $p_j$.
For simplicity, we include in the $p_j$ only the most relevant emission amplitude parameter in the power law approximation.
In the derivatives below, we report only the terms that will not vanish when specified at $p_j = 0$.
The terms that are not multiplied by $u$ in ${\partial{\eta}}/{\partial{u}}$ give 

\begin{align}\label{DevC}
\frac{\partial{\eta}}{\partial{u}} & = - \frac{[h\nu/(kT_{\rm BB})]\,e^{h\nu/(kT_{\rm BB})/\phi_{\rm i}}\,/\,\phi_{\rm i}^2}{(e^{h\nu/(kT_{\rm BB})/\phi_{\rm i}} - 1)^{2}} \nonumber
\\ & + {h\nu/(kT_{\rm BB}) / \phi_{\rm i} \, e^{h\nu/(kT_{\rm BB})/\phi_{\rm i}} \over (e^{h\nu/(kT_{\rm BB})/\phi_{\rm i}} - 1)^{2}} \left ( {h\nu/(kT_{\rm BB})/\phi_{\rm i} \over { {\rm tanh}[h\nu/(kT_{\rm BB})/(2\phi_{\rm i})] }} - 4 \right) \, ,
\end{align}

\noindent
and we get

\begin{equation}\label{DevC0}
\left({ \frac{\partial{\eta}}{\partial{u}} }\right)_0 = - \frac{x\,e^{x}}{(e^{x} - 1)^{2}}
+ {x \, e^{x} \over (e^{x} - 1)^{2}}  \left ( {x \over { {\rm tanh}(x/2) }} - 4 \right) \, ,
\end{equation}

\noindent
while

\begin{equation}\label{DevFF}
\frac{\partial{\eta}}{\partial{{\tilde A}_{FF} }} \simeq \left({ \frac{h\nu}{kT_{\rm BB}} }\right)^{-(3+\zeta)} \, ,
\end{equation}

\noindent
implying

\begin{equation}\label{DevFF0}
\left({ \frac{\partial{\eta}}{\partial{{\tilde A}_{FF} }} }\right)_0 \simeq x^{-(3+\zeta)} \, .
\end{equation}

\noindent
The terms that are not multiplied by $u$ nor by ${\tilde A}_{FF}$ in ${\partial{\eta}}/{\partial{T_{\rm BB}}}$ give 

\begin{align}\label{Dev_CFF_TBB}
\frac{\partial{\eta}}{\partial{T_{\rm BB}}} & \simeq \frac{1}{T_{\rm BB}} \frac{e^{h\nu/(kT_{\rm BB})/\phi_{\rm i}} \, h\nu/(kT_{\rm BB})/\phi_{\rm i}} {(e^{h\nu/(kT_{\rm BB})/\phi_{\rm i}} -1)^2} \, , 
\end{align}

\noindent
implying, again,

\begin{align}\label{Dev_CFF0_TBB}
\left({ \frac{\partial{\eta}}{\partial{T_{\rm BB}}} }\right)_0 \simeq \frac{1}{T_0} \frac{x \, e^{x}}{(e^{x} -1)^2} \, .
\end{align}

Formally, in Eqs. \eqref{DevBE}, \eqref{Dev_BE_TBB}, \eqref{DevC}, \eqref{DevFF}, and \eqref{Dev_CFF_TBB},  $T_{\rm BB}$, $\mu_0$, $\phi_{\rm BE}(\mu_0)$ and
$\phi_{\rm i}(u)$ refer to  
a given sky direction identified by $\theta$ and $\phi$. However, their relations with $\theta$ and $\phi$ vanish when they are specified for $T_{\rm BB} = T_0$ and $p_j = 0$ as in Eqs. 
\eqref{DevBE0}, \eqref{Dev_BE0_TBB}, \eqref{DevC0}, \eqref{DevFF0}, and \eqref{Dev_CFF0_TBB}, where indeed $x = h\nu/(kT_0)$.

In general, we have $\left({ ({\partial{T_{\rm th}}}/{\partial{\eta}}) ({\partial{\eta}}/{\partial{T_{\rm BB}}}) }\right)_0 = 1$, as evident also from Eqs. \eqref{DevTth},  \eqref{Dev_BE0_TBB},  \eqref{Dev_CFF0_TBB};
the coefficients $a_{\rm BB,\ell,m}$ are then the usual spherical harmonic expansion coefficients defining the CMB temperature anisotropies. Considering both the effects induced by the observer peculiar motion on the monopole and the intrinsic anisotropies, 
working in the observer reference system with the $z$ axis parallel to the observer velocity, 
we can combine Eq. \eqref{Tthlocal_linSphArm} and Eq. \eqref{eq:harm} to derive 
a global anisotropy pattern

\begin{align}\label{eq:alm_glob}
T_{\rm th} (\nu, \theta, \phi, \beta) & = 
\sum_{\ell=1}^{\ell_{\rm max}} \sum_{m=-\ell}^{\ell} a_{\rm BB,\ell,m} Y_{\ell,m} \left({ \frac{\partial{T_{\rm th}}}{\partial{\eta}} \frac{\partial{\eta}}{\partial{T_{\rm BB}}} }\right)_0
\\ & + \sum_{j=0}^{P-1} \sum_{\ell=1}^{\ell_{\rm max}} \sum_{m=-\ell}^{\ell} a_{j,\ell,m} Y_{\ell,m} \, \left({ \frac{\partial{T_{\rm th}}}{\partial{\eta}} \frac{\partial{\eta}}{\partial{p_j}} }\right)_0 \nonumber 
\\ & + \sum_{\ell=0}^{\ell_{\rm max}} \sum_{m=-\ell}^{\ell} a_{\ell,m} (\nu, \beta) Y_{\ell,m}(\theta, \phi) \, , \nonumber
\end{align}

\noindent
where the coefficients $a_{\ell,m} (\nu, \beta)$ are given in Sect. \ref{sec:sol7} (or in Sect. \ref{sec:sol5}); we omitted here the first line 
of the right-hand side of Eq. \eqref{Tthlocal_linSphArm} to obviously avoid a double counting of the monopole that is already included in the term with $\ell = 0$ of the last line, where the 
effect of the observer motion with respect to a frame at rest with the CMB is also taken into account. The focus in this work on
the analysis at low multipoles of 
the effects caused by the observer motion on the isotropic monopole component, neglects the Doppler and aberration effects
on the anisotropies, that is, on the first and second lines of the right-hand side of Eq. \eqref{eq:alm_glob}. These effects couple multipoles $\ell$ to $\ell \pm n$, particularly in the correlation between $\ell$ and $\ell \pm 1$
\citep{Challinor:2002zh,Burles:2006xf,Kosowsky:2010jm,Amendola:2010ty,2011MNRAS.415.3227C,2014PhRvD..89l3504D}, a property that has been used to independently constrain $\beta$.
The effects are indeed more important at high multipoles: their main information comes from $\ell \gsim 100$, where many modes can be exploited \citep{2018JCAP...04..021B}
(see \cite{2020arXiv200312646P} for a recent application to {\it Planck} data based on the modulation of the thermal Sunyaev-Zel'dovich effect \citep{2005A&A...434..811C,2015arXiv151008793N}).

We can rewrite the coefficients $a_{\ell,m} (\nu, \beta)$ as

\begin{equation}\label{eq:Dalm}
a_{\ell,m} (\nu, \beta) = \Delta a_{\ell,m} (\nu, \beta) + a_{\ell,m}^{\rm BB} (\beta) \, , 
\end{equation}

\noindent
where 
$a_{\ell,m}^{\rm BB} (\beta)$
refer to the case of a blackbody spectrum and
$\Delta a_{\ell,m} (\nu, \beta) = a_{\ell,m} (\nu, \beta) - a_{\ell,m}^{\rm BB} (\beta)$ depend on the type of considered distortion.

\section{Intrinsic dipole versus kinematic dipole}
\label{sec:dip_all}

We focus here on the dipole anisotropy. Including both the effect induced by the observer peculiar motion on the monopole and the intrinsic anisotropies, 
and using a reference system with the $z$ axis parallel to the observer velocity, the global dipole pattern is characterized by the coefficients:

\begin{equation}\label{eq:a1m_glob}
a_{1,m}^{\rm glob} (\nu, \beta) = \Delta a_{1,m}(\nu, \beta) + a_{1,m}^{\rm BB} (\beta) + a_{\rm BB,1,m} + \sum_{j=0}^{P-1} {\tilde a}_{j,1,m} (\nu) \, ,
\end{equation}
\noindent
where 

\begin{equation}\label{eq:tilde_aj10}
{\tilde a}_{j,1,m} (\nu) = a_{j,1,m} \left({ \frac{\partial{T_{\rm th}}}{\partial{\eta}} \frac{\partial{\eta}}{\partial{p_j}} }\right)_0 \, 
\end{equation}

\noindent
and $\Delta a_{1,m}(\nu, \beta) = 0$, $a_{1,m}^{\rm BB}(\beta) = 0$ for $m \ne 0$. 
In these equations, the superscript `BB' refers to observer peculiar motion effects while the subscript `BB' refers to intrinsic anisotropies.
The coefficients $\Delta a_{1,0}(\nu, \beta)$ and ${\tilde a}_{j,1,m} (\nu)$ (through $({ ({\partial{T_{\rm th}}}/{\partial{\eta}}) ({\partial{\eta}}/{\partial{p_j}}) })_0$), 
that are related to the type of distortion, do not vanish
only in the presence of deviations from a Planckian spectrum.

The typical amplitude of the coefficients $\Delta a_{1,0}(\nu, \beta)$ is 
of the order of 
$a^{\rm BB}_{1,0}(\beta)$, that is, of $\approx \beta T_0$, 
multiplied by the amplitude of the monopole spectral distortion, $\Delta T_{th}$ (see the first two left panels from the top in Figs. \ref{fig:FFeC} and \ref{fig:BE}).
The coefficients $a_{j,1,m}$ characterize the fluctuations of the different types of distortion at $\ell = 1$ and then
depend significantly on the specific mechanism and not only on the corresponding average distortion parameter and spectral shape.
Global spectral distortions are still unobserved, as well their fluctuations (obviously except for the Sunyaev-Zel'dovich effect on galaxy clusters),
and it is then reasonable to assume ${\tilde a}_{j,1,m} (\nu) < a_{\rm BB,1,m}$. 

Assuming Gaussian random temperature fluctuations, the coefficients $a_{\rm BB,1,m}$ are expected to have zero mean and variance given by the angular power spectrum $C_\ell$ at $\ell = 1$. Currently,
$C_1$ is  unknown, but it is typically predicted to be of the order of the temperature anisotropy intrinsic quadrupole $C_2$. 
Constraining the intrinsic anisotropy power at $\ell = 1$ is difficult, but very interesting 
in the context of future CMB surveys (see \cite{2017PhRvL.119v1102Y} for a method based on the exploitation the leakage 
of the intrinsic dipole into the CMB monopole and quadrupole and \cite{2017PhRvD..96h3519M} for an analysis
based on the observation of the small scale temperature fluctuations that result from gravitational lensing). 
This topic is related to the power at low multipoles, and in particular, to the low power of the quadrupole
discovered by the COBE Differential Microwave Radiometer (DMR) \citep{1996ApJ...464L..21W,1996ApJ...464L..25H} and then confirmed by WMAP
\citep{2013ApJS..208...20B,2013ApJS..208...19H} and {\it Planck} \citep{2020A&A...641A...5P}.
Reconstructing the intrinsic anisotropy power at very low multipoles is very important for inflationary models (see \cite{2020A&A...641A..10P} for recent constraints) 
predicting power suppression at large scales
(see e.g. \cite{1982PhRvD..26.1231V},  \cite{1992JETPL..55..489S}, \cite{2003MNRAS.342L..72B}, \cite{2003JCAP...07..002C},
 \cite{2003MNRAS.343L..95E} and \cite{2006PhRvD..74d3518S})
and for their connection with universe geometry and topology
(see e.g.
 \cite{1995PhLB..351...99L}, \cite{2002PhRvD..65d3513G}, \cite{2002GReGr..34.1461E}, \cite{2003JCAP...05..002L}, \cite{2004AIPC..736...53L}
and \cite{2002PhR...365..251L}).

In general, the amplitudes of the coefficients $a_{\rm BB,1,m}$ are significantly smaller than 
the amplitude of the coefficient $a_{1,0}^{\rm BB}(\beta)$, the observed dipole being dominated by the Doppler effect associated to our peculiar motion with respect to the CMB.
In the absence of deviations from a Planckian spectrum, ${\tilde a}_{j,1,m} (\nu) = 0$ and $\Delta a_{1,0}(\nu, \beta) = 0$, 
and any (relatively minor) frequency independent contribution from $a_{\rm BB,1,0}$ is degenerate with $a_{1,0}^{\rm BB}(\beta)$.
The same holds for $a_{\rm BB,1,m}$ with $m \ne 0$ because the sum of dipole terms is still a dipole and, for a Planckian spectrum, 
it is possible to find a rotation of the reference system that jointly drops the terms with $m \ne 0$ at all frequencies (see Appendix \ref{app:rotation}). Thus, the peculiar motion 
Doppler effect alone does not allow to distinguish the intrinsic dipole from the kinematic dipole, at least in the absence of
a very accurate measure of $\vec{\beta}$ provided by other methods.

Now, we consider the presence of deviations from a Planckian spectrum. 
The only frequency dependent terms in Eq. \eqref{eq:tilde_aj10} are $\Delta a_{1,0}(\nu, \beta)$ and ${\tilde a}_{j,1,m} (\nu)$.

Let us assume, as a first simple case (A), that the very large angular scale fluctuations of the spectral distortion parameters are very small in amplitude,
namely ${\tilde a}_{j,1,m} (\nu) \ll \Delta a_{1,0}(\nu, \beta)$.
Thus, in the adopted reference system, 
the only relevant frequency dependence in the dipole pattern is for $m=0$ and comes from $\Delta a_{1,0}(\nu, \beta)$, while any different choice of 
the $z$ axis, that is, not parallel (or not antiparallel) to the observer velocity, would imply that the same frequency dependence is polluted in the dipole at $m \ne 0$.

In a more general case (B), when we relax the above assumption, that is, for a non-negligible frequency dependent contribution from ${\tilde a}_{j,1,m} (\nu)$, 
the situation is a bit more complex, but conceptually not so different (in particular for given prescriptions of the coefficients ${\tilde a}_{j,1,m} (\nu)$). 
In the adopted reference system, a frequency dependence 
related only to the fluctuations of the spectral distortions would appear at $m \ne 0$, 
and the combination of the different frequency dependencies related to the peculiar motion effect and to the fluctuations of the spectral distortions, properly weighted, 
would appear at $m=0$. For a reference system 
with the $z$ axis not parallel (or not antiparallel) to the observer velocity these frequency dependencies are polluted in the dipole at any $m$, in a way related to the underlined frequency spectra.

Let us assume that the intrinsic dipole and the kinematic dipole are not aligned (the opposite is possible by chance, but it is very unlikely).

In case (A), one can search for a reference system that drops (or, more realistically, minimize in a statistical sense) the frequency dependence of the coefficients, $a_{1,m}^{\rm glob}$ for 
$m \ne 0$, implying that its 
$z$ axis is parallel to $\vec{\beta}$. The components, $a_{1,m}^{\rm glob}$, with $m \ne 0$ are then to be ascribed only to the intrinsic temperature fluctuation terms, $a_{\rm BB,1,m}$,
and although the contribution from $a_{\rm BB,1,0}$ remains hidden in the larger term, $a_{1,0}^{\rm BB}(\beta)$, 
the squares of the components, $a_{1,m}^{\rm glob}$, with $m \ne 0$, allow us to provide an estimate of the intrinsic dipole angular power spectrum, $C_1$,
although with a slightly larger cosmic variance because this estimate is based on only two, instead than three, coefficients.
We note that (see Eq. \eqref{eq:alm_glob}) similar considerations apply in the limit of that approximation, also for $\ell > 1$, allowing us to exploit a larger number of modes $m$. 
On the other hand, because $\Delta a_{\ell,m} (\nu, \beta)$ (see Eq. \eqref{eq:Dalm}) decreases as $\beta^\ell$, the information from $\ell > 1$ does not add relevant constraints in this scheme.

In case (B),
relaxing the assumption ${\tilde a}_{j,1,m} (\nu) \ll \Delta a_{1,0}(\nu, \beta)$, it is possible to search for a reference system that drops, or minimize, 
the difference of the frequency dependence of the coefficients, $a_{1,m}^{\rm glob}$, with $m \ne 0$, with the behaviour expected from the terms, ${\tilde a}_{j,1,m} (\nu)$, related to the fluctuations of the distortion parameters
(see Eqs. \eqref{eq:a1m_glob} and \eqref{eq:tilde_aj10}), added with the intrinsic temperature fluctuation terms, $a_{\rm BB,1,m}$. 
We also note that the frequency dependencies of 
${\tilde a}_{j,1,m} (\nu)$ and $\Delta a_{1,0}(\nu, \beta)$, although different, are physically connected,
being related to the types of involved distortions, and this property can be exploited in the joint analysis of the modes with $m=0$ and $m \ne 0$, helping, at $m \ne 0$, the discrimination 
between ${\tilde a}_{j,1,m} (\nu)$ and $a_{\rm BB,1,m}$.  

Therefore, in the presence of spectral distortions, a very careful multifrequency analysis of the dipole pattern, 
namely, of the frequency behaviour of its spherical harmonic expansion coefficients,
can be used, at least in principle, to set constraints on the intrinsic dipole embedded in the kinematic dipole.

We note that the inclusion of the time variation of $\vec{\beta}$ (see Sect. \ref{sec:varybeta}), although requiring specific implementations,
does not conceptually modify the above considerations since it is possible to split the sky area observed in a survey into a proper set of sky areas,
each one observed in a shorter time interval and, hence, with a negligible variation of $\vec{\beta}$, without reducing the global statistical information contained in the survey.

\section{Discussion and conclusion}
\label{sec:conclu}

The peculiar motion of an observer with respect to the cosmic background in a certain frequency band
produces boosting effects in the background anisotropy pattern. In this work, we studied how the frequency spectrum of the background isotropic monopole emission
is modified and transferred to the frequency spectra of the patterns at higher multipoles.
We performed the analysis in terms of spherical harmonic expansion for various models of background radiation, ranging from the radio to the far-infrared.

Adopting a reference frame with the $z$ axis parallel to the observer motion direction allows us to simplify the problem since it is thanks to this choice that only 
the spherical harmonic coefficients $a_{\ell,m}$ with $m = 0$ do not vanish. 
We derive the system of linear equations to obtain the $a_{\ell,m}$ up to a desired value of $\ell_{\rm max}$. For each observational frequency, the $a_{\ell,m}$ are written 
as linear combinations of the signals at the set of frequencies corresponding to the chosen $N = \ell_{\rm max} +1$ colatitudes $\theta_i$ (Sect. \ref{sec:theoframe}).
We explicitly write the system and provide the solutions up to $\ell_{\rm max} = 6$, as well as for other smaller values of $\ell_{\rm max}$.  
The symmetry property of the associated Legendre polynomials with respect to $\theta = \pi/2$ is used to separate the system into two subsystems, 
one for $\ell=0$ and even multipoles and the other for odd multipoles, improving the solutions accuracy with respect to an arbitrary colatitudes choice 
(Sects. \ref{sec:sol7}--\ref{sec:dip_2col}).
We apply these solutions to the case of a blackbody and verify their agreement with the exact analytical solutions for $\ell = 0$ and $\ell =1$
at the order in $\beta$ related to $\ell_{\rm max}$ (Sects. \ref{sec:sol7BB} and \ref{sec:sol5BB}). 

The structure of the solutions is discussed and compared with respect to the general properties of monopole spectrum integration and differentiation (Sect. \ref{sec:highl_derivs}). 
The coefficients of the solutions can be regarded as 
sets of 
weights
assigned to a small number of function evaluations, according to adopted order of accuracy,
to compute
the integrals that define the $a_{\ell,m}$ coefficients.
The coefficients of these solutions exhibit remarkable symmetry properties. 
Some of these properties are the same of the ones shown by the weights used in finite difference formulas
to compute numerical derivatives.
The implicit mixing of the spectrum derivatives in the solutions
is reflected by the absence of that sign alternation property of the weights which appears in finite difference formulas.
Indeed, the frequency behaviours of the $a_{\ell,m}$ coefficients 
are particular sensitive to the local
monopole spectrum variation in a way characterized by derivative orders increasing with $\ell$. 

We applied the method to some models for different types of monopole spectra that can be represented in terms of analytical or semi-analytical functions
(Sects. \ref{sec:monmod} and \ref{sec:combi}), namely: four types of 
CMB distorted photon distribution functions, that is, a non-equilibrium imprint at low frequencies, Comptonization, FF, and BE-like distortions;
four types of extragalactic background signals, that is, 21cm redshifted line, radio, and millimetre backgrounds from extragalactic sources, CIB, superimposed onto the CMB Planckian spectrum;
some combinations of signal relevant at millimetre and sub-millimetre wavelengths, that is, Comptonization and FF distortions combined with millimetre background from extragalactic sources or with CIB.

For each model, we show the intrinsic monopole spectrum 
in terms of the difference, $\Delta T_{th} (\nu)$, 
between the equivalent thermodynamic temperature for the model and the present CMB temperature $T_0$.
Our results are presented in terms of:
(i) the difference $\Delta R (\nu,\beta) = R(\nu,\beta) - R^{\rm BB}(\beta)$, where 
$R (\nu,\beta) = (a_{0,0} (\nu,\beta) / \sqrt{4\pi}) / T_{th} (\nu)$ and $R^{\rm BB} (\beta)$ are the ratios between the equivalent thermodynamic temperature of observed 
and intrinsic monopole for the model and for the blackbody at the present temperature $T_0$;
(ii) the differences $\Delta a_{\ell,0}(\nu,\beta)$ between 
the spherical harmonic coefficients computed for the model, $a_{\ell,0} (\nu,\beta)$, 
and 
for the
blackbody, $a_{\ell,0} (\beta)$;
we also present (iii) all-sky maps and (iv) angular power spectra (Sect. \ref{sec:compa})
directly derived from the spherical harmonic coefficients, for some representative cases.

The
results are in excellent agreement
with those based on more computationally demanding numerical integrations or map generation and inversion (Sects. \ref{sec:monmod} and \ref{sec:compa}; see also Appendix \ref{app:beta_amplified}),
and even more accurate.
The method could be obviously implemented for any $\ell_{\rm max}$, however,
since $\beta$ is of the order of $10^{-3}$ and the coefficients $a_{\ell,0} (\nu,\beta)$ scale approximately as $\beta^\ell$,
the solutions presented for $\ell_{\rm max} = 6$ allows us to achieve an extremely high accuracy that is sufficient for any application even in the very distant future. We provide also explicit solutions for
$\ell_{\rm max} = 4$ that are fully adequate for the analysis of forthcoming and proposed surveys.
The only accuracy limitation of the proposed scheme derives from neglecting the contributions
from
higher multipoles, the largest relative errors appearing
at $\ell_{\rm max}$ from $\ell_{\rm max} + 2$ and at $\ell_{\rm max} -1$ from $\ell_{\rm max} + 1$: the relative errors are very small even at $\ell \lsim \ell_{\rm max}$,
while, remarkably, they are fully negligible at the lowest multipoles. 

The high number of possible model combinations may result in a variety of signatures in the global signal expected by cosmological plus astrophysical backgrounds.
The simplicity and computational efficiency of the proposed method can significantly alleviate the computational effort needed for theoretical predictions and for the comparison with data from future 
projects for a plethora of cases of interest. These include, for example, signal combinations or possible differences between vectors $\vec{\beta}$ that could refer to specific backgrounds
or that appear when the observer velocity variation is taken into account (Sect. \ref{sec:varybeta}).

We discuss the main features found for the considered background models at the various multipoles in wide frequency ranges. 
All the patterns at different multipoles are related to the observer peculiar motion, although with a signal amplitude that decreases with $\ell$.
Thus, they exhibit, both in the whole sky and in different sky regions, spatial correlations at different angular length scales
and well-defined geometrical properties, that are related each other, and this would improve their joint analysis that
can be optimized considering the observational method and specifications of a given project. As discussed in \cite{2019A&A...631A..61T} for the dipole in the radio domain, 
the analysis does not necessarily require the mapping of the entire sky or of a very large fraction of it.
Future radio surveys, and in particular the excellent resolution and sensitivity offered by the SKA \citep{dewdney2016} for a variety of scientific themes (see e.g. \cite{2020PASA...37....2W}) can be used to 
investigate the spectra of the multipoles patterns at low frequencies.
In general, the ultra-precise comprehension and subtraction of the Galactic foreground emission is likely to act as the most difficult problem in cosmological analyses at large angular scales.

Finally, we move on to a discussion of 
the superposition of the CMB intrinsic anisotropies and of the effects induced by the observer peculiar motion
focussing on their different frequency behaviours in the presence of CMB spectral distortions (Sect. \ref{sec:all}).
We find that they can be used, at least in principle, to set constraints on the intrinsic dipole embedded in the kinematic dipole,
through a very careful multifrequency analysis of the corresponding spherical harmonic expansion coefficients (Sect. \ref{sec:dip_all}; see also Appendix \ref{app:rotation}).
Detailed studies, object of future works, can clarify the feasibility of this approach and, possibly, the required specifications.

In general, the above considerations do not strictly rely on the direct absolute determination of the monopole spectrum.
Thus, although they come with challenges, these studies can, in principle, be pursued with an 
anisotropy project,
such as the Lite (Light) satellite for the studies of B-mode polarization and Inflation from cosmic background Radiation Detection 
(LiteBIRD) \citep{2014JLTP..176..733M} or a
mission like 
COrE \citep{2018JCAP...04..014D,2018JCAP...04..015D} or
the Probe of Inflation and Cosmic Origins (PICO) \citep{2019BAAS...51g.194H}, provided that an extremely accurate relative and inter-frequency calibration 
and suppression of systematic effects are achieved \citep{2018JCAP...04..022N}.

Clearly, the most advantageous observational chance is offered by the next generation of CMB space missions designed to perform ultra-accurate temperature measurements of the whole sky with 
a highly precise absolute calibration and a relatively high resolution mapping over a wide set of frequencies, as for example, in 
the Primordial Inflation Explorer (PIXIE) \citep{2011JCAP...07..025K,2016SPIE.9904E..0WK}, in the
Polarized Radiation Imaging and Spectroscopy Mission (PRISM) \citep{2014JCAP...02..006A},
and, as in more recent proposals, to the National Aeronautics and Space Administration (NASA) \citep{2019BAAS...51c.184C}, 
to the Indian Space Research Organisation (ISRO) (CMB Bharat\footnote{http://cmb-bharat.in/.}), and to the 
European Space Agency (ESA) \citep{2019arXiv190901591D,2019arXiv190901593C}. 
These projects have the advantage of measuring the frequency spectrum of the relevant multipole patterns -- starting from the monopole.

\begin{acknowledgements}
We gratefully acknowledge financial support
from the research program RITMARE SP3 WP3 AZ3 U02 and the research contract SMO at CNR/ISMAR,
from the INAF PRIN SKA/CTA project FORmation and Evolution of Cosmic STructures (FORECaST) with Future Radio Surveys
and from the ASI/Physics Department of the University of Roma--Tor Vergata agreement n. 2016-24-H.0 for study activities of the Italian cosmology community.
We gratefully acknowledge the use of the NAG Numerical Library. Some of the results in this paper have been derived using the HEALPix \citep{2005ApJ...622..759G} package.
We also thank the anonymous referee for comments that helped improve the paper.
\end{acknowledgements}


%%%   \bibliographystyle{aa} % style aa.bst
%%%      \bibliography{references_monop2highl} % your references Yourfile.bib

\appendix

\section{Numerical tests with amplified observer velocity}
\label{app:beta_amplified}

\begin{figure*}[ht!]
\centering
         \includegraphics[width=18.5cm]{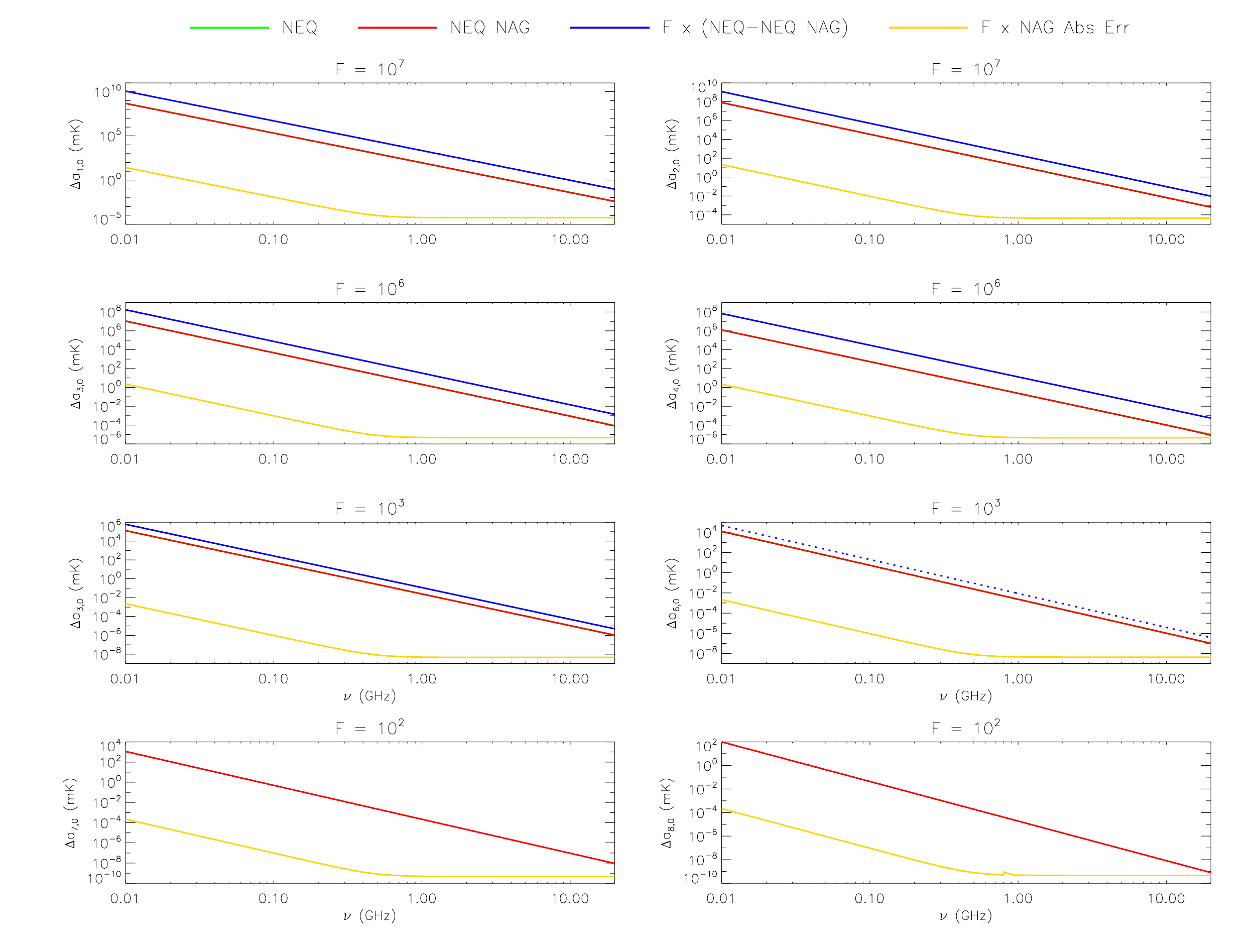}
    \caption{$\Delta a_{\ell,0}$ for the considered non-equilibrium model ($\nu_0 \simeq 0.35$\,GHz and $\alpha \simeq 3.36$), assuming a value of $\beta$ multiplied by a factor 100; 
    $\ell$ ranges from $1$ to $\ell_{\rm max} = 6$ in the case of the solutions described in Sect. \ref{sec:sol7} and up to $8$ in the case of numerical integration (with the routine {\it D01AJF}) 
    based on Eq. \eqref{eq:harminv}.
    Solid lines (or dots) correspond to positive (or negative) values. Green and red lines are essentially superimposed up to $\ell =6$.
    Their difference, multiplied by a factor F to have values compatible with the adopted range, is displayed by the blue lines. Yellow lines refer to the nominal integration error quoted by 
    the routine {\it D01AJF}, again multiplied by the factor F. See also the legend and the text.}
    \label{fig:NEQper100}
\end{figure*}

In Sects. \ref{sec:NE} and \ref{sec:21cm}, we compare, for two very different cases, the results based on the solutions described in Sect. \ref{sec:sol7} and on a direct numerical integration (see Eq. \eqref{eq:harminv}) to
find differences that are compatible with a combination of higher order terms, that is, beyond $\ell = 6$, and integration errors, the latter becoming more relevant at increasing $\ell$ and decreasing signals, in connection with the low value of the observer velocity, $\beta$.

To better clarify this aspect, we perform the same type of comparison, but adopting a much larger value of $\beta$, in order to exploit much larger signals and relatively higher contributions from higher multipoles and to deal 
with relatively lower numerical integration errors. Here, we reconsider the cases of the non-equilibrium model and of the EDGES profile of redshifted 21cm line, but assuming a value of $\beta$ arbitrarily amplified respectively by a factor 100 and 10, this choice being motivated by their different 
signal amplitudes in their relevant frequency ranges. 

\begin{figure*}[ht!]
\centering
         \includegraphics[width=18.5cm]{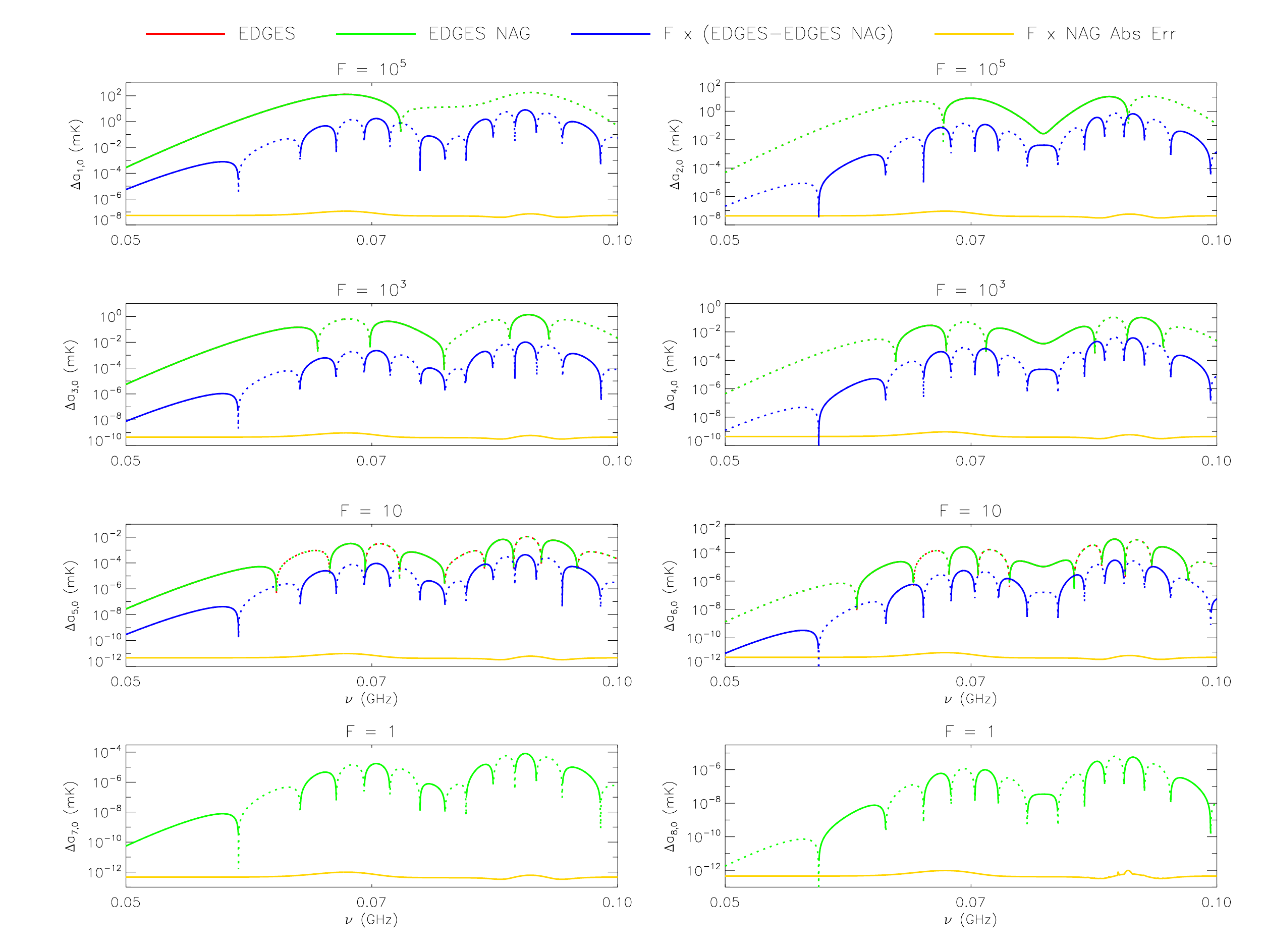}
    \caption{Same as in Fig. \ref{fig:NEQper100}, but for the EDGES profile of redshifted 21cm line (summed in intensity with the CMB blackbody) and assuming a value of $\beta$ multiplied by a factor 10. See also the legend and the text.}
    \label{fig:EDGESper10}
\end{figure*}

We report the results for $\Delta a_{\ell,0}$ found using the equations in Sect. \ref{sec:sol7} and the Eq. \eqref{eq:harminv} for $\ell = 1, 6$, as well as the results based on Eq. \eqref{eq:harminv} for $\ell = 7$ and 8. They are shown 
in Figs. \ref{fig:NEQper100} and \ref{fig:EDGESper10}. As expected, and as evident from the figures, such higher values of $\beta$ imply a strong reduction of the relative numerical integration error, making feasible an accurate computation of 
$\Delta a_{\ell,0}$ at larger $\ell$.
Thus, the differences at $\ell = 1, 6$ between the results found with the two methods come essentially only from higher order terms, neglected in the equations of Sect. \ref{sec:sol7},
that are obviously dominated by 
the contributions from $\ell = 7$ at odd multipoles and from $\ell = 8$ at even multipoles. 

We note that the order of magnitude of $\Delta a_{7,0}$ (or of $\Delta a_{8,0}$) is equal to the order of magnitude of the differences between the results found with the two methods at odd (or even) multipoles.
The spectral shapes of $\Delta a_{\ell,0}$ in the case of the non-equilibrium model are featureless, reflecting the original power law shape of the monopole spectrum when expressed in terms of $\Delta T_{th}$.
Conversely, the spectral shapes of $\Delta a_{\ell,0}$ in the case of the EDGES profile are rich in features, increasing in number at increasing $\ell$. Remarkably, the spectral shape of the differences between the results found with the two methods 
reflects the spectral shape of $\Delta a_{7,0}$ (or of $\Delta a_{8,0}$) at odd (or even) multipole.

As discussed in Sect. \ref{sec:theoframe}, the above properties derive from the separation of the system in two subsystems, one at odd $\ell$ the other at even $\ell$, associated to the adoption of 
a set of colatitudes symmetric with respect to $\pi/2$ (plus $\pi/2$). This suppresses in each $a_{\ell,0}$ at odd (or even) $\ell$ the 
contribution of the multipole immediately larger than maximum odd (or even) multipole considered in the system solution (see also, for comparison, the discussion in Sect. \ref{sec:dip_3and2col}).
This is particularly important in practice for the 'robustness' of the accuracy of the solutions presented in this work, suitable for the real (low) value of $\beta$,
and especially in the case of spectra rich in features for which changes of sign of $\Delta a_{\ell,0}$ and extremely low values of $\Delta a_{\ell,0}$ may occur at different frequencies
for different $\ell$, possibly enhancing the relative contribution of the missing higher order terms. 

\section{Double Compton Gaunt factor}
\label{app:gdc_integ}

\begin{figure}[ht!]
\centering
         \includegraphics[width=9.cm]{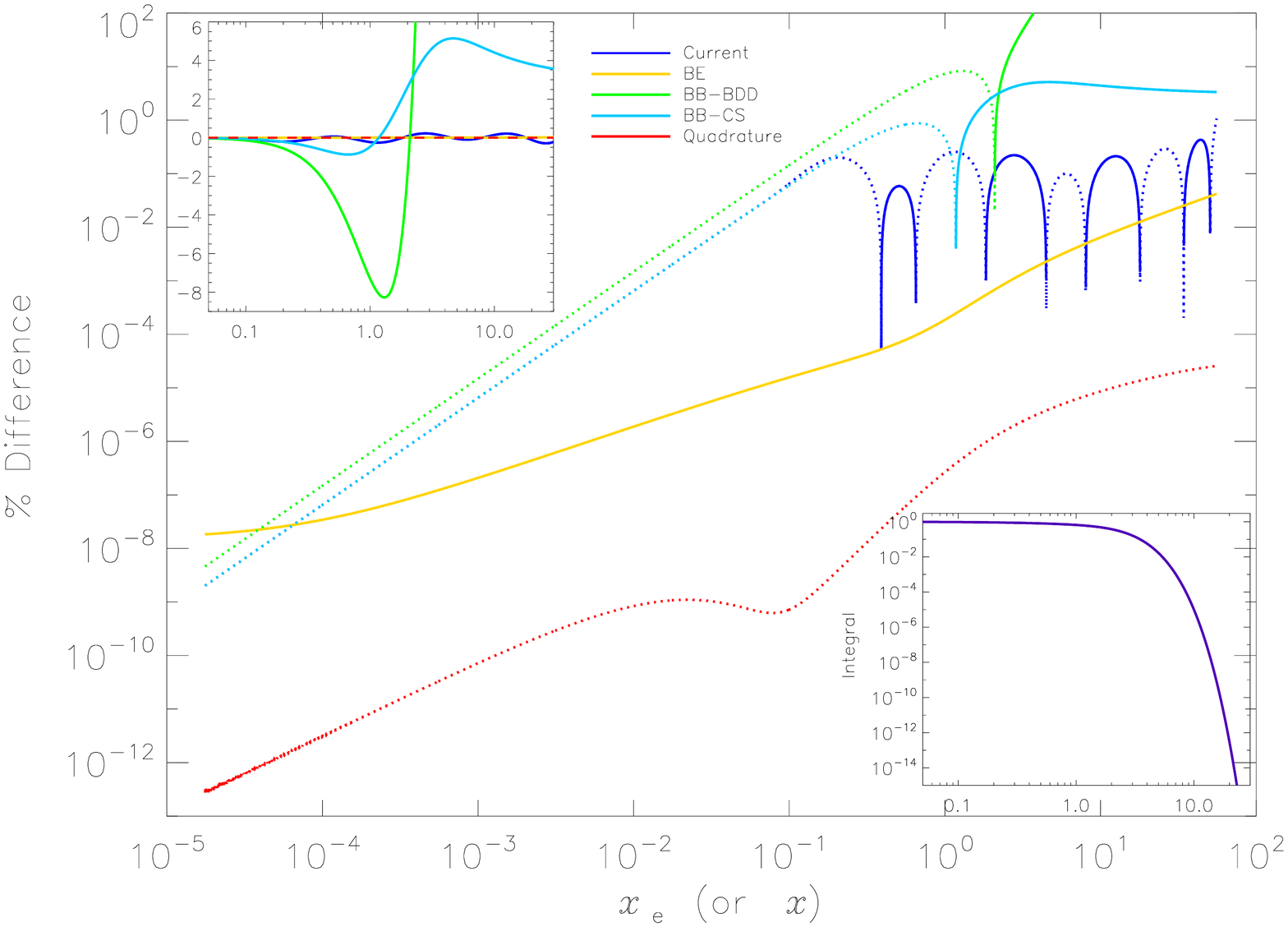}
    \caption{Comparison between the results of various approximations and accurate numerical integrations of Eq. \eqref{eq:intgdc}. The bottom-right inset gives the numerical result
    derived with the NAG routine {\it D01AJF} with a relative accuracy of $10^{-12}$ in the case of the blackbody spectrum.
    The main plot reports the per cent difference of various estimations of Eq. \eqref{eq:intgdc} 
    with respect to the NAG {\it D01AJF} result for the blackbody spectrum: the approximation presented here (Current, see Eq. \eqref{eq:intgdc_currentapp}); the NAG {\it D01AJF} result 
    but for a pure BE spectrum with $\mu_0 = 1.4 \times 10^{-5}$ (BE); the approximations reported in \cite{1991A&A...246...49B} (BB-BDD, see Eq. \eqref{eq:BDD}) 
    and in \cite{2012MNRAS.419.1294C} (BB-CS, see Eq. \eqref{eq:CS}); the numerical result
    derived with a Gaussian quadrature scheme \citep{press1992numerical} with the accuracy parameter ({\it EPS}) set to $10^{-9}$ and 2048 points
    in the case of the blackbody spectrum. Solid lines (or dots) correspond to positive (or negative) values.
    The top-left inset displays the same results of the main plot, but in a restricted frequency range and with a linear scale on the $y$-axis. 
    See also the legend and the text.}
    \label{fig:gDC}
\end{figure}

As anticipated in Sect. \ref{sec:BE}, here we provide an improved approximation for the double Compton Gaunt factor 
(see also \cite{2020JCAP...10..025R} and \cite{2020MNRAS.498..959C} for recent studies)
in the elastic limit, suitable for small BE-like distortions.  
Writing the double Compton term according to \cite{1995A&A...303..323B}, and accounting separately in the rate for the correction factor, $C_{\rm mr}$,
for mildly relativistic thermal plasma in the soft photon limit \citep{2007A&A...468..785C}, the double Compton Gaunt factor can be written as \citep{1991A&A...246...49B}
\begin{equation}\label{eq:intgdc}
g_{DC} (x_{\rm e}) = \frac{ \int_{2x_{\rm e}}^\infty x'^4_{\rm e} \, [1+\eta(x'_{\rm e} - x_{\rm e})] \, \eta(x'_{\rm e}) \, \left[{ w { F(w) \over 2 } }\right] \, dx'_{\rm e} }
{ \int_0^\infty [1+\eta(x'_{\rm e})] \, \eta(x'_{\rm e}) \, x'^4_{\rm e} dx'_{\rm e} } \, ,
\end{equation}
\noindent
where
\begin{align}\label{eq:Fw}
w { F(w) \over 2 }  & = \frac{1}{2} (1-w) 
\\ & \cdot \left[ { 1 + (1-w)^2 + \frac{w^2(1+w^2)}{(1-w)^2} + w^4 +w^2(1-w)^2 } \right] \, , \nonumber
\end{align}
\noindent
with $w$ the ratio between the frequencies of the created and incident photons, $0 \le w \le 1/2$ and $w F(w) / 2 \rightarrow 1$ as $w \rightarrow 1$ \citep{1984ApJ...285..275G}. For a (pure) BE spectrum with a frequency independent chemical potential $\mu$ and 
equilibrium temperature $T_{BE} = T_{\rm eq} = [h/(4k)] \int_0^\infty (1+\eta) \eta \nu^4 d\nu \, / \, \int_0^\infty  \eta \nu^3 d\nu$
\citep{refId0,1969JETP...28.1287Z}, the integral at denominator in Eq. \eqref{eq:intgdc} is simply $4I_3 f(\mu)$, 
with $f(\mu) = \int_0^\infty  \eta_{\rm BE}(x_{\rm e}) x_{\rm e}^3 dx_{\rm e} / I_3$, $I_3 = \pi^4/15$ 
and 
$f(\mu) \rightarrow 1$ in the Planckian limit $\mu \rightarrow 0$. Thus, for a pure BE spectrum only the integral at numerator of Eq. \eqref{eq:intgdc} needs a numerical computation. 
We have performed this calculation using both a Gaussian quadrature scheme \citep{press1992numerical}
and the NAG routine {\it D01AJF}, adopting a very large upper integration limit (set to $500$, to have a good estimation also at very high frequencies) and working with a dimensionless frequency in logarithmic space as integration variable.
We first consider the case of a blackbody spectrum and compare the results of two methods (see Fig. \ref{fig:gDC}): clearly, they are in excellent agreement. We then consider the case of a (pure)
BE spectrum with $\mu_0 = 1.4 \times 10^{-5}$, the largest value considered in this work.
The relative differences between $g_{DC}$ computed for this case and in the case of a blackbody spectrum are less than few\,$\times 10^{-2}$\,\% in the whole frequency range, 
and obviously it decreases for decreasing values of $\mu_0$. Although the very low frequencies give only a little relative contribution to the integral, a similar comparison
for a BE-like spectrum 
is in principle a bit more difficult, because the spectrum at low frequencies depends also on $g_{DC}$. On the other hand, the above difference represents a good upper limit estimation also when applied to a BE-like spectrum with 
the same high frequency asymptotic value of $\mu_0$, because a BE-like spectrum differs from the blackbody less than a pure BE spectrum.

An approximation for $g_{DC} (x_{\rm e})$ which, as explicitly stated by the authors, works well only at low frequencies, where the rate is higher, was found by \cite{1991A&A...246...49B}

\begin{equation}\label{eq:BDD}
{\tilde g}_{DC} (x_{\rm e}) \simeq e^{ -x_{\rm e}/2} \, .
\end{equation}

\noindent
Of course, for very small distortions $x_{\rm e} \simeq x$. In the limit $\mu \ll 1$, \cite{2012MNRAS.419.1294C} found the approximation 
\begin{align}\label{eq:CS}
{\hat g}_{DC} (x) & \simeq e^{ -2x} 
\\ & \cdot [1+ (3/2) x + (29/24) x^2 + (11/16) x^3 + (5/12) x^4]   \, . \nonumber
\end{align}
\noindent
Figure \ref{fig:gDC} shows the relative accuracies of these two formulas.
The latter works significantly better than the former at $x \gsim 0.3$, providing a little improvement at smaller frequencies (the two approximations clearly 
agree to first order in Taylor's series for small $x$). On the other hand, an error at some \% level still remains when using Eq. \eqref{eq:CS} at $x > 1$.

We then search for a better description of the numerical result. Considering the very good accuracy of Eq. \eqref{eq:CS} at low frequencies, we consider
the following expression:
\begin{equation}\label{eq:intgdc_currentapp}
g_{DC} (x) \simeq \,\, {\hat g}_{DC} (x) \, 10^{P(X)} \, e^{-({\hat x}/x)^m}
 +  {\hat g}_{DC} (x) [1 - e^{-({\hat x}/x)^m}] \, 
,\end{equation}
\noindent
where $X = {\rm log_{10}} x$, $P(X)$ is a polynomial of a certain degree $d$, ${\hat x}$ and $m$ are a dimensionless frequency and an exponent. $P(X)$ modifies the approximation represented by ${\hat g}_{DC} (x)$ to better describe 
the numerical results at high frequencies, while ${\hat x}$ and $m$ define the exponential weights to assure a smooth transition from low to high frequencies, around ${\hat x}$.
We fit our numerical results (for instance, the ones derived
with the {\it D01AJF} routine) for $g_{DC} (x)$ in the Planckian limit to find the values of ${\hat x}$, $m$, $d$, varying them on a simple three dimensional grid, and of the best-fit polynomial coefficients.
We find $m=1.6$, ${\hat x} = 0.23$, $d = 17$ and the following polynomial coefficients, from the power of order 0 to the power of order $d$:

\noindent
$+1.137720\times10^{-3}$, $-2.949735\times10^{-2},$
$-5.578415\times10^{-2}$, \\$+3.212930\times10^{-2},$
$+8.794246\times10^{-2}$, $-4.492392\times10^{-3},$ \\
$-5.746682\times10^{-2}$, $-1.277239\times10^{-2},$
$+1.573671\times10^{-2}$, \\$+7.068248\times10^{-3},$
$-9.391122\times10^{-4}$, $-1.073426\times10^{-3},$\\
$-1.716636\times10^{-4}$, $+8.114468\times10^{-6},$
$-4.003431\times10^{-7}$, \\$-1.886616\times10^{-6},$
$-3.587674\times10^{-7}$, $-2.055208\times10^{-8}$.

\noindent
The accuracy of this approximation is also displayed in Fig. \ref{fig:gDC}: it is always better than $\simeq 1$\,\% 
and better than $\simeq 0.1-0.2$\,\% in the whole relevant frequency range.

\section{Reference system rotation and dipole terms with $m \ne 0$}
\label{app:rotation}

Let us consider a real field, $T(\theta,\phi)$, defined by the spherical harmonics coefficients $A_{\ell,m}$ in a given reference system, $S$, with polar coordinates $\theta$ and $\phi$, that is:\ 

\begin{equation}\label{eq:T_S}
T (\theta, \phi) =
\sum_{\ell=0}^{\ell_{\rm max}} \sum_{m=-\ell}^{\ell} A_{\ell,m} Y_{\ell,m}(\theta, \phi) \, ,
\end{equation}

\noindent
and another reference system $S'$, with coordinates $\theta'$ and $\phi'$, that is defined, with respect to $S$, by the Euler angles $\alpha_{\rm E}, \beta_{\rm E}, \gamma_{\rm E}$. 
The range of $\alpha_{\rm E}$ and $\gamma_{\rm E}$ is defined modulo $2\pi$ radians: we adopt the range $[0,2\pi]$ (but another widely adopted choice is, for example, $[-\pi,\pi]$).
The range of $\beta_{\rm E}$ covers $\pi$ radians: we adopt the range $[0,\pi]$ (but it could be, for example, $[-\pi/2,\pi/2]$). In the reference system $S'$

\begin{equation}\label{eq:T_Sp}
T (\theta', \phi') =
\sum_{\ell=0}^{\ell_{\rm max}} \sum_{j=-\ell}^{\ell} D_{\ell,j} Y_{\ell,j}(\theta', \phi') \, ,
\end{equation}

\noindent
where $D_{\ell,j}$ are the corresponding (rotated) spherical harmonics coefficients. According to \cite{1984JGR....89.4413G}
 
\begin{equation}\label{eq:D_lj}
D_{\ell,j} = \sum_{m=-\ell}^{\ell} Q_{\ell,m,j} A_{\ell,m} \, ,
\end{equation}

\noindent
where 

\begin{align}\label{eq:Q_lmj}
Q_{\ell,m,j} & = (-1)^{\ell-m} { \binom{2 \ell}{\ell + m} }^{1/2} { \binom{2 \ell}{\ell + j} }^{-1/2}  \, i^{j-m} \, e^{i (m \alpha_{\rm E} + j \gamma_{\rm E})} \nonumber
\\ & \cdot {\rm cos}^{m+j}(\beta_{\rm E}/2) \, {\rm sin}^{j-m}(\beta_{\rm E}/2) \, J_{l-j}^{m+j,j-m}(-{\rm cos} \, \beta_{\rm E})  \, ;
\end{align}

\noindent
here $i^2 = -1$ and
$J_{r}^{s,t}(z)$ are the Jacobi polynomials

\begin{equation}\label{eq:JacExplExpr}
J_{r}^{s,t}(z) = \frac{1}{2^{r}} \sum_{k=0}^{r}  \binom{r+s}{k} \, \binom{r+t}{r-k} \, (z-1)^{r-k} \, (z+1)^{k} \, 
\end{equation}

\noindent
expressed in explicit polynomial form. For $r=0,$ we simply have 
$J_{0}^{s,t}(z) = 1$.

\noindent
Defining $A_{1,0} = a_{1,0}$,  $A_{1,1} = b_{1,1} + i \, c_{1,1}$ (and, for symmetry, $A_{1,-1} = -b_{1,1} + i \, c_{1,1}$), 
after some algebra one gets

\begin{align}\label{eq:RE_D_11}
Re (D_{1,1}) & =  \frac{\sqrt{2}}{2} \, a_{1,0} \, {\rm sin} \, \beta_{\rm E} \, {\rm sin} \, \gamma_{\rm E} 
\\ & + \frac{1+{\rm cos} \, \beta_{\rm E}}{2} \, [ b_{1,1}  \, {\rm cos} \, (\alpha_{\rm E} + \gamma_{\rm E}) \, - \, c_{1,1} \, {\rm sin} \, (\alpha_{\rm E} + \gamma_{\rm E}) ] \nonumber
\\ & + \frac{1-{\rm cos} \, \beta_{\rm E}}{2} \, [ b_{1,1} \, {\rm cos} \, (\alpha_{\rm E} - \gamma_{\rm E}) \, - \, c_{1,1} {\rm sin} \, (\alpha_{\rm E} - \gamma_{\rm E})  ] \nonumber \, ,
\end{align}

\begin{align}\label{eq:IM_D_11}
Im (D_{1,1}) & =  - \frac{\sqrt{2}}{2} \, a_{1,0} \, {\rm sin} \, \beta_{\rm E} \, {\rm cos} \, \gamma_{\rm E} 
\\ & + \frac{1+{\rm cos} \, \beta_{\rm E}}{2} \, [ b_{1,1} \, {\rm sin} \, (\alpha_{\rm E} + \gamma_{\rm E}) \, + \, c_{1,1} \, {\rm cos} \, (\alpha_{\rm E} + \gamma_{\rm E}) ] \nonumber
\\ & -  \frac{1-{\rm cos}  \, \beta_{\rm E}}{2} \, [ b_{1,1} \, {\rm sin} \, (\alpha_{\rm E} - \gamma_{\rm E}) \, + \, c_{1,1} \, {\rm cos} \, (\alpha_{\rm E} - \gamma_{\rm E})  ] \, . \nonumber
\end{align}

A rotation of the reference system $S \rightarrow S'$, that is, a set of Euler angles $\alpha_{\rm E}, \beta_{\rm E}, \gamma_{\rm E}$, that makes $D_{1,1}=0=D_{1,-1}$ or, equivalently, for which the whole dipole signal 
is along the $z'$ axis of the $S'$ reference system,
does not depend on the choice of the angle $\gamma_{\rm E}$ that specifies the (last) rotation around the $z'$ axis. Thus, setting for simplicity $\gamma_{\rm E} = 0$, the condition $D_{1,1}=0$ is satisfied for

\begin{equation}\label{eq:RE_D_11a}
b_{1,1} \, {\rm cos} \, \alpha_{\rm E}  \, - c_{1,1} \, {\rm sin} \, \alpha_{\rm E}  = 0 
\end{equation}

\noindent
and

\begin{equation}\label{eq:IM_D_11a}
- \frac{\sqrt{2}}{2} \, a_{1,0} \, {\rm sin} \, \beta_{\rm E} \, 
+ {\rm cos} \, \beta_{\rm E} [ c_{1,1} \, {\rm cos} \, \alpha_{\rm E} \, + \, b_{1,1}\,  {\rm sin} \, \alpha_{\rm E}  ] = 0 \, .
\end{equation}

\noindent
Equations \eqref{eq:RE_D_11a} and \eqref{eq:IM_D_11a} imply

\begin{align}\label{eq:sinalpha}
{\rm sin} \, \alpha_{\rm E}  & = \frac{\sqrt{2}}{2} \, {\rm tan} \, \beta_{\rm E} \, \frac{a_{1,0} \, b_{1,1}}{b_{1,1}^2+c_{1,1}^2} = \pm \frac{b_{1,1}}{\sqrt{b_{1,1}^2+c_{1,1}^2}} \, ,
\end{align}

\begin{align}\label{eq:cosalpha}
{\rm cos} \, \alpha_{\rm E}  & = \frac{\sqrt{2}}{2} \, {\rm tan} \, \beta_{\rm E} \, \frac{a_{1,0} \, c_{1,1}}{b_{1,1}^2+c_{1,1}^2} = \pm \frac{c_{1,1}}{\sqrt{b_{1,1}^2+c_{1,1}^2}} \, ,
\end{align}

\noindent
where the last equalities of Eqs. \eqref{eq:sinalpha} and \eqref{eq:cosalpha} come from the obvious condition
${\rm sin}^2 \, \alpha_{\rm E} + {\rm cos}^2 \, \alpha_{\rm E} = 1$ that gives

\begin{equation}\label{eq:tanbeta}
{\rm tan} \, \beta_{\rm E}  = \pm \frac{\sqrt{2 \, (b_{1,1}^2+c_{1,1}^2) } }{a_{1,0} } \, ;
\end{equation}

\noindent
the sign is related to the orientation of the (starting) reference system, $S$, with respect to the dipole pattern.

Finally, simply considering that the angular power spectrum, $C_\ell$, is invariant under rotation, for $\ell = 1, $ we have

\begin{align}\label{eq:RE_D_10}
\sqrt{C_1} & = \frac{  \sqrt{D_{1,0}^2 + D_{1,1}^2 + D_{1,-1}^2} } {\sqrt{3} } = 
\frac {D_{1,0}}{\sqrt{3}}  = \frac {Re (D_{1,0})} {\sqrt{3}} 
\\ &  = \frac{  \sqrt{A_{1,0}^2 + A_{1,1}^2 + A_{1,-1}^2} } {\sqrt{3} } = \frac{\sqrt{a_{1,0}^2 + 2 \, (b_{1,1}^2+c_{1,1}^2)}}{\sqrt{3}} \, . \nonumber
\end{align}

Let us consider now, see Eq. \eqref{eq:a1m_glob}, $A_{1,m} = a_{1,m}^{\rm BB} (\beta) + a_{\rm BB,1,m}$, as in the case 
of frequency independent $a_{1,m}^{\rm glob}$ coefficients evaluated for a Planckian spectrum, but in a reference system $S$ with the $z$ axis only roughly aligned 
with the observer peculiar velocity $\vec{\beta}$, that is easily identified from the sky area where the observed (large scale pattern) temperature is not far from its maximum
(this implies $A_{1,0} = a_{1,0} > 0$ and, with the adopted ranges for the Euler angles, this also 
defines a unique, positive choice of the sign 
in Eq. \eqref{eq:tanbeta} as well as in 
Eqs. \eqref{eq:sinalpha} and \eqref{eq:cosalpha}, 
since $\beta_{\rm E}$ should be relatively small). 
Thus, the above expressions allow to simply find a reference system $S'$ where $D_{1,1}=0=D_{1,-1}$ or, equivalently, the global dipole signal is along the $z'$ axis for all the frequencies
(and, in Eq. \eqref{eq:RE_D_10}, the positive sign of $D_{1,0}$ means that the $z'$ axis points towards the direction of the maximum of the 
dipole pattern signal).

Considering, instead, $A_{1,m} = a_{1,m}^{\rm glob} (\nu, \beta)$ with $a_{1,m}^{\rm glob} (\nu, \beta)$ given by Eq. \eqref{eq:a1m_glob} but 
in the presence of deviations from a Planckian spectrum. The coefficients $a_{1,0}$, $b_{1,1}$ and $c_{1,1}$ in the above expressions are then frequency dependent, 
it is no longer possible to find a reference system $S'$ satisfying $D_{1,1}=0=D_{1,-1}$ for all the frequencies, and, in principle, this property can be used to constrain the intrinsic dipole, 
as discussed in Sect. \ref{sec:dip_all}.

\end{document}